\documentclass[final,3p,times,twocolumn]{elsarticle}

\usepackage{amssymb}

\usepackage{booktabs}
\usepackage{multirow}
\usepackage{array}
\usepackage{subfig}
\usepackage{makecell}
\usepackage{soul}
\usepackage{xcolor}
\usepackage{url}

\journal{Journal of Information Security and Applications}

\begin{document}

\begin{frontmatter}



\title{On the Cross-Dataset Generalization of Machine Learning for Network Intrusion Detection}


\author[inst1]{Marco Cantone\corref{cor1}}
\cortext[cor1]{Corresponding author}
\ead{marco.cantone@unicas.it}

\affiliation[inst1]{organization={Department of Electrical and Information Engineering, University of Cassino and Southern Latium},
            addressline={Via Gaetano Di Biasio 43}, 
            city={Cassino},
            postcode={03043}, 
            state={FR},
            country={Italy}}

\author[inst1]{Claudio Marrocco}
\ead{c.marrocco@unicas.it}
\author[inst1]{Alessandro Bria}
\ead{a.bria@unicas.it}


\begin{abstract}
Network Intrusion Detection Systems (NIDS) are a fundamental tool in cybersecurity. Their ability to generalize across diverse networks is a critical factor in their effectiveness and a prerequisite for real-world applications. In this study, we conduct a comprehensive analysis on the generalization of machine-learning-based NIDS through an extensive experimentation in a cross-dataset framework. We employ four machine learning classifiers and utilize four datasets acquired from different networks: CIC-IDS-2017, CSE-CIC-IDS2018, LycoS-IDS2017, and LycoS-Unicas-IDS2018. Notably, the last dataset is a novel contribution, where we apply corrections based on LycoS-IDS2017 to the well-known CSE-CIC-IDS2018 dataset. The results show nearly perfect classification performance when the models are trained and tested on the same dataset.  However, when training and testing the models in a cross-dataset fashion, the classification accuracy is largely commensurate with random chance except for a few combinations of attacks and datasets. We employ data visualization techniques in order to provide valuable insights on the patterns in the data. Our analysis unveils the presence of anomalies in the data that directly hinder the classifiers capability to generalize the learned knowledge to new scenarios. This study enhances our comprehension of the generalization capabilities of machine-learning-based NIDS, highlighting the significance of acknowledging data heterogeneity.
\end{abstract}




\end{frontmatter}




\section{Introduction}
\label{sec:Introduction}
The rapid expansion of network interconnections has led to a corresponding growth in the cyber threat landscape, attracting the interest of an increasing number of cyber attackers. This has resulted in the disruption of essential services with significant economic consequences. Globally, cybercrime is estimated to have an impact of around 1 trillion dollars in 2020, with an increase of more than 50\% compared to 2018~\cite{cremer2022cyber}. It is therefore necessary to use systems and strategies to counter this phenomenon~\cite{martinez2019machine, dasgupta2022machine, sarker2020cybersecurity}.

Network Intrusion Detection Systems (NIDS) are specifically designed to identify intrusions analyzing network traffic, enabling targeted entities to take timely actions against potential threats~\cite{bhuyan2013network}. They can be grouped into three categories: statistics-based, knowledge-based, and Machine-Learning-based (ML-based)~\cite{alkasassbeh2023intrusion, khraisat2019survey}. The statistics-based paradigm concerns the meticulous examination of each record within a dataset, with the goal of building a statistical model that encompasses the established norms of user behavior within a network~\cite{tan2013system, camacho2016pca}. In contrast, the knowledge-based approach is based on attempting to identify the requested actions by leveraging existing system data, including protocol specifications and network traffic instances. Knowledge-based NIDS, guided by human-defined rules, deduces typical system activities and identifies deviations from this benchmark as potential intrusions~\cite{sakilaannarasi2014secure, hendry2008intrusion}. A knowledge-based method to attain traffic classification is Deep Packet Inspection. This technique involves examining the content and structure of individual packets, allowing for detailed inspection of packet headers, payload data, and protocol-specific information~\cite{finsterbusch2013survey}.

ML-based NIDS acquire intricate pattern-matching capabilities from extensive training data either with a supervised or unsupervised approach. In the context of supervised IDS, each data instance is characterized by a pair comprising a source of network data and an associated output value, typically denoting intrusion or normal behavior. Utilizing the training dataset, a supervised learning technique is then employed to train a classifier in discerning and modeling the inherent correlation that exists between the input data and the labeled output values~\cite{raikar2020data, sarhan2022evaluating, verkerken2023novel}.
In contrast, unsupervised learning is employed to extract patterns from input datasets that lack predefined class labels. In this approach, the input data points are typically treated as a set of random variables, and a joint density model is constructed for the dataset to identify intrusions and anomalous patterns~\cite{choi2019unsupervised, verkerken2020unsupervised}.
ML-based-NIDS leverage a number of different methodologies including decision trees, neural networks, Support Vector Machines (SVM), and k-nearest neighbors~\cite{liu2019machine, saranya2020performance}.
For instance, Zhou et al.~\cite{zhou2020building} introduced an intrusion detection framework that employs feature selection and ensemble learning. They first utilized a heuristic algorithm, CFS-BA, for dimensionality reduction based on feature correlations. Next, an ensemble approach combined C4.5, Random Forest, and Forest by Penalizing Attributes algorithms. A voting technique was finally used to merge base learners probability distributions for attack recognition.
The study of Stawan et al.~\cite{stiawan2020cicids} investigated feature selection, aimed at improving the accuracy and reducing execution time of traffic anomaly detection in a large network dataset. Employing Information Gain, the authors identified and grouped pertinent features by considering minimum weight values. Subsequently, the selected features underwent empirical evaluation through diverse classifier algorithms, including Random Forest, Bayes Net, Random Tree, Naive Bayes, and J48, applied to the CICIDS-2017 dataset.
The research by Gu et al.~\cite{gu2021effective} introduced a novel intrusion detection framework that integrates SVM with naïve Bayes feature embedding. The methodology involves applying naïve Bayes feature transformation to the original dataset and subsequently training an SVM classifier on the transformed data to construct the intrusion detection system.

More recently, also Deep Learning (DL) approaches have been adopted in the design of NIDS. DL is a subfield of ML in which deep artificial neural networks automatically learn increasingly abstract and hierarchical data representations~\cite{mahdavifar2019application, shone2018deep, marin2018rawpower, saba2022anomaly, mushtaq2022two}. 
Duan et al.~\cite{duan2022application} proposed to transform network traffic data into spatiotemporal graphs and used a dynamic line graph neural network (DLGNN) with semisupervised learning to capture evolving communication patterns.
The paper by Li et al.~\cite{li2020building} proposed an effective deep learning method, namely AE-IDS (AutoEncoder Intrusion Detection System) based on Random Forest and autoencoder. This method constructs the training set with feature selection and feature grouping using the random forest algorithm. After training, the model predicts the results with an autoencoder, which greatly reduces the detection time and effectively improves the prediction accuracy compared to traditional ML-based-IDS.

The popularity of DL for image analysis has prompted researchers to use these models for network traffic classification as well, applying them to synthetic images generated from the network traffic or a set of extracted network flows features. The literature includes research that employs both convolutional networks~\cite{wang2017end, wang2017malware, jo2020packet, agrafiotis2022image} and vision transformers~\cite{ho2022network, agrafiotis2022image}.
The complexity of the transformer model combined with its ability to merge heterogeneous information has paved the way for a strand of research that uses it as a feature extractor and/or classifier~\cite{li2022extreme, li2022mfvt, wu2022rtids}.

In this work, we extensively analyze the generalization capabilities of ML-based-NIDS including a variety of training strategies, data sources and evaluation metrics. Generalization refers to the ability of a model to perform well on data it has never seen during training. For this investigation, we design two types of experiments: a within-dataset scenario in which we split a dataset in train and test sets, and a cross-dataset scenario where train and test sets are taken from different datasets. The datasets we utilize include CIC-IDS-2017, CSE-CIC-IDS2018~\cite{sharafaldin2018toward}, LycoS-IDS2017~\cite{rosay2022network}, and LycoS-Unicas-IDS2018 generated for this research.
These datasets are available both as raw network traffic and features-processed format computed on bidirectional flows.
We utilize the available extracted features for training the models, which is the prevailing approach adopted in the literature~\cite{verkerken2023novel, duan2022application, sarhan2022evaluating, yang2022intrusion, gu2021effective, li2020building, zhou2020building, stiawan2020cicids, ustebay2018intrusion, shone2018deep, javaid2016deep}.
The selection of CIC-IDS2017 and CSE-CIC-IDS2018 was based on the presence of similar attack classes, making them suitable for cross-dataset evaluation. We also employed LycoS-IDS2017 to investigate generalization as it offers a distinct feature set resulting from the refinement of specific features from CIC-IDS2017.

Our main contributions are:
\begin{enumerate}
\item \textbf{Extensive experimentation}: we conducted a total of $1,728$ experiments encompassing a variety of ML models, feature sets, and training configurations including the discrimination between benign and malicious instances and single-attack analysis.
\item \textbf{Cross-dataset analysis}: we evaluated all models generalization capabilities when trained and tested across $4$ different datasets.
\item \textbf{New dataset}: we introduced LycoS-Unicas-IDS2018, an enhanced CSE-CIC-IDS2018 dataset with refined feature extraction techniques, specifically addressing the concerns highlighted by Rosay et al. [7].
\item \textbf{Explainability}: We employed ad hoc visualization techniques to enable a deeper understanding of the datasets structure and class distributions, as well as to provide visual explanations for the obtained results.
\end{enumerate}

The rest of this work is organized as follows. In Section \ref{sec:Related work} we revise the literature regarding the datasets used and the generalization capabilities of ML-based NIDS. In Section \ref{sec:Materials and methods} we describe the ML classifiers employed and the characteristics of the datasets. We also detail the experimental methodology and how we evaluate the performance. Section \ref{sec:Results and discussion} discusses the results obtained and shows some insights about the patterns present in the data. Finally, in Section \ref{sec:Conclusions}, we conclude with a summary and a critical discussion.

\section{Related work}
\label{sec:Related work}
In this section, we explore research on generalization and literature works that question the construction and labeling of the CIC datasets. The discussion of generalization is enriched by an examination of federated learning (FL) approaches. FL, contrary to centralized learning, enables organizations to work together without sharing sensitive data, ensuring data privacy, compliance, and enhanced security against attacks. This decentralized approach also supports real-time adaptability and robust intrusion detection in evolving threat landscapes~\cite{agrawal2022federated}.

\subsection{Generalization in centralized IDS}
D’Hooge et al.~\cite{d2020inter} conducted a study on CIC-IDS2017 and CSE-CIC-IDS2018 to evaluate the performance of different ML classifiers in a cross-dataset scenario only for DoS and Botnet attacks. They concluded that the tested classifiers were not able to work properly on data from unseen networks at train time and further analysis is necessary.
Verkerken et al.~\cite{verkerken2022towards} trained four unsupervised ML models on the two CIC datasets and evaluated their performance in a cross-dataset manner. Their results indicate an average decrease of 30.45\% in Area Under the Receiver Operating Characteristic curve (AUROC) during cross-dataset experiments compared to within-dataset experiments. Notably, even in the most favorable scenario, a consistent performance decrease of 17.85\% in AUROC is observed for the One-Class SVM. The authors conclude that the models exhibit challenges in transferring learned information across datasets, yet they still perform better than a random classifier.
In a very recent study~\cite{sivcic2023evaluation}, unsupervised training using PCA and autoencoders was performed on four NetFlow datasets~\cite{sarhan2022towards}. The research evaluated generalization performance resulting in 16 experimental combinations: 4 within-dataset and 12 cross-dataset scenarios.
Autoencoders demonstrated some potential for generalization since they achieved an average AUROC of 61.08\% in cross-dataset evaluation, with respect to a within-dataset performance of 96.62\%. In contrast, PCA obtained an average AUROC of 51.56\% for cross-dataset evaluation with respect to a within-dataset performance of 95.76\%. However, both methods performed worse than a random classifier in some cross-dataset configurations.

The abovementioned research sheds light on the limitations of machine learning classifiers in adapting to unseen network data, highlighting the challenges of generalization across datasets.

\subsection{Generalization in federated IDS}
The study conducted by Popoola et al.~\cite{popoola2021federated} aimed to enhance generalization performance by utilizing a FL approach on four NetFlows datasets. These datasets originated from different sources but shared the same set of features~\cite{sarhan2022towards}. The centralized models demonstrated robust performance in the within-dataset configuration but challenges arose when applied to other datasets. For instance, after training on NF-ToN-IoT-v2, the centralized model achieved an F1-score of 94.91\% when tested in a within-dataset scenario. In contrast, its performance varied significantly in cross-dataset modality, with values of 78.96\%, 29.19\%, and 62.83\% observed when tested on NF-UNSW-NB15-v2, NF-BoT-IoT-v2, and NF-CSE-CIC-IDS2018-v2, respectively. Introducing the Federated Learning approach significantly improved performance across all datasets while at the same time achieving better or comparable performance with respect to centralized models.
In a study involving FL with two clients holding respectively the NF-UBSW-NB15-v2 and NF-BoT-IoT-v2 datasets, the authors explored three training scenarios~\cite{sarhan2023cyber}. They considered localized training, where only one dataset was used, centralized training, which involved training on the merged dataset, and federated training, utilizing both datasets in an FL manner. Localized training demonstrated effectiveness when trained on NF-UBSW-NB15-v2 and assessed on NF-BoT-IoT-v2 for binary-class classification, yielding an average F1-score of 87.9\%. Conversely, when the models were trained on NF-BoT-IoT-v2 and tested on NF-UBSW-NB15-v2, the average F1-score was 9.36\%, indicating poor generalization. The centralized approach demonstrated the highest performance, achieving an F1-score of 95.63\% averaged across the various models and datasets employed. In comparison, the federated approach exhibited slightly lower performance at 91.08\% F1-score.
In the study of de Carvalho Bertoli et al.~\cite{de2023generalizing}, a stacked-unsupervised federated learning approach was employed in the context of a flow-based NIDS to enhance cross-silo generalization, referring to generalization across different splits of the dataset. The authors introduced the stacked Energy Flow Classifier and an autoencoder, both trained using federated learning techniques on the 4 NetFlow datasets. Each client in this setup possessed a distinct dataset. The results were promising, demonstrating good performance on each dataset.

In all of the abovementioned literature works the FL approach involved training on data from all the available data sources, contributing to its superior cross-dataset generalization. Further, in none of these works the federated learning models have been tested on an external dataset to fully characterize their generalization capability.

\subsection{CIC dataset integrity}
\label{sec:CIC dataset integrity}
Recently, some research investigated the integrity of the CIC datasets.
Engelen et al.~\cite{engelen2021troubleshooting} conducted a rigorous evaluation of the CIC-IDS2017 dataset and its data collection procedures, uncovering significant issues in traffic generation, flow construction, feature extraction, and labeling. They implemented an improved data processing methodology, resulting in the reconstruction and relabeling of over 20 percent of original traffic flows. Machine learning benchmarks on the refined dataset yielded performance improvements, highlighting the impact of data collection issues on model evaluation.
Rosay et al.~\cite{rosay2022network} analyzed the raw traffic data of CIC-IDS2017 and the software used for feature extraction and labeling. They found different issues related to features duplication, features miscalculation, wrong protocol detection, inconsistent TCP termination flows, and error in labeling. To overcome these problems, they developed a tool  to re-extract features from bidirectional flows. This yielded a corrected version of the CIC-IDS2017 dataset called LycoS-IDS2017.
The study of Lanvin et al.~\cite{lanvin2022errors} addresses deficiencies in the CIC-IDS2017 dataset labeling and traffic acquisition, including duplication and misclassified attacks. The authors contributed by presenting a refined dataset, introducing a CICFlowMeter patch, and assessing corrections on intrusion detection models. Their evaluation revealed nuanced challenges in achieving high precision without overfitting. The study underscores the critical role of dataset quality in NIDS evaluation, advocating for transparent labeling practices.
Another conference paper criticizing the construction of the CIC-IDS2017 and CSE-CIC-IDS2018 datasets is that of Liu et al.~\cite{liu2022error}. Their work mainly focused on labeling issues and they concluded that labeling errors totaled $6.67\%$ and $7.53\%$ on CIC-IDS2017 and CSE-CIC-IDS2018 respectively, with a corruption rate that exceeded $75\%$ for some attacks.
The research by Catillo et al.~\cite{catillo2023machine} critically examined the limitations of public intrusion datasets and associated ML experiments, challenging the reliability of conclusions drawn in existing literature despite perceived progress. Although academic enthusiasm and increasing complexity in machine and deep learning, the practical effectiveness of public datasets for real-world network intrusion detection is questioned. The paper emphasizes the necessity for caution in handling the collection, release, and utilization of existing intrusion datasets in research. It also asserts that genuine advances in NIDS are hampered by numerous flaws and limitations, supported by scientific insights from existing datasets, related literature, and direct field experience.

\section{Materials and methods}
\label{sec:Materials and methods}
\subsection{Datasets}
In this study, we employ four different datasets: CIC-IDS2017, CSE-CIC-IDS2018, LycoS-IDS2017 e LycoS-Unicas-IDS2018. The first two are from the Canadian Institute for Cybersecurity (CIC) and have the same features and similar attack classes, whereas the other two are constructed from the same raw traffic of the CIC datasets but have slightly different features and correct some of the problems that we have mentioned in section~\ref{sec:CIC dataset integrity}.

In Table~\ref{tab:datasets_composition} is reported the class composition for all the datasets. It is worth noting that DDoS LOIC-UDP and DDoS HOIC attacks are absent in the 2018 datasets, whereas PortScan and Heartbleed are not present in the 2017 datasets. Furthermore, there is a significant class imbalance between the benign and multiple attack classes, specifically the benign class is predominant in all datasets, comprising 80.32\% (CIC-IDS2017) 75.96\% (LycoS-IDS2017) 82.98\% (CSE-CIC-IDS2018) and 73.04\% (LycoS-Unicas-IDS2018).

\subsubsection{CIC-IDS2017}
CIC-IDS2017 (CIC17) is an intrusion detection dataset from the Canadian Institute for Cybersecurity~\cite{sharafaldin2018toward}. The dataset includes 7 days of network traffic acquired from an infrastructure consisting of 14 machines, where several cyber attacks were carried out. Benign background traffic was generated using a system called B-Profile, which extracted abstract behavior from a group of real users through a combination of machine learning and statistical analysis techniques. In addition, features suitable to machine learning are available in CSV files, one for each day of acquisition. These features were extracted using the CICFlowMeter software through distinct phases, including the identification of bidirectional flows, feature extraction, and labeling. Overall the dataset consists of 84 features and 15 classes of which 14 refer to cyber-attacks and one represents benign traffic.

For our study, we excluded the following features: (i) \texttt{Flow ID}, \texttt{Source IP}, \texttt{Source Port}, \texttt{Destination IP}, and \texttt{Protocol} due to their role as flow identifiers; (ii) \texttt{Timestamp} as it could bias the model during training; and (iii) \texttt{Fwd Header Length} since it was redundant.

\subsubsection{CSE-CIC-IDS2018}
The Canadian Institute for Cybersecurity released a second dataset in collaboration with the Communication Security Establishment (CSE) known as CSE-CIC-IDS2018 (CIC18). This dataset utilized the same B-Profiles methodology for generating benign traffic. Notably, the second dataset expanded the generation process to a larger scale by migrating the entire network setup to the Amazon Web Services (AWS) cloud computing platform and deploying a total of 500 machines, a substantial increase with respect to the 14 utilized in the first iteration.

Similar to CIC-IDS2017, CSE-CIC-IDS-2018 is distributed in two formats: raw network packets (PCAPs) and machine learning-friendly CSV files that encapsulate bidirectional flows derived from the PCAPs through the application of CICFlowMeter. It is noteworthy that the CSV files in this instance lack four features related to flow identification. For consistency with CIC-IDS2017, we remove the \texttt{Timestamp} and \texttt{Protocol} features.
 
\subsubsection{LycoS-IDS2017}
In response to identified anomalies within the CIC-IDS2017 dataset, researchers undertook a meticulous correction process, resulting in the creation of the LycoS-IDS2017 (LycoS17) dataset~\cite{rosay2022network}. The comprehensive analysis of the CIC dataset revealed several issues, including feature duplication and miscalculations. Notably, instances of redundancy were observed in features such as \texttt{Average Packet Size} and \texttt{Packet Length Mean}, which shared identical semantic meanings. Furthermore, discrepancies arose from the incorrect calculation of numerous features due to type-related oversights and updating errors concerning the directionality of flow. The protocol detection mechanism in the original dataset exhibited shortcomings, leading to misclassification of ICMP and other protocol packets. Additionally, inconsistencies in TCP termination flows and uncertainties surrounding the labeling process prompted the development of remedial measures. To address these concerns, a tool called LycoStand was developed to accurately extract a set of features based on the predefined feature set of the CIC datasets.

Following the process of feature extraction, a Python-based labeling script was employed, referencing data from the official CIC website that had been refined through meticulous analysis conducted using Wireshark. Noteworthy is the decision to discard an entire afternoon of acquisition data, a precautionary measure taken due to substantial doubts regarding the reliability of labeling.

\subsubsection{LycoS-Unicas-IDS2018}
LycoS-Unicas-IDS2018 (LycoS18) is a novel dataset created by us using LycoStand~\cite{rosay2022network}, a feature extractor of raw network traffic, on the PCAP files of CSE-CIC-IDS2018, and labeling the flows according to the official attack schedule accessible via the dataset website. While certain anomalies have been identified in the annotation of the CIC-IDS2017 dataset, achieving an optimal level of labeling precision remains an intricate challenge. Nevertheless, given the currently available information, the existing labeling methodology represents the most viable approach to address this concern. For the same reason we discarded raw network traffic on days 28/02/2018 and 01/03/2018 containing the infiltration attack, as it has been done for LycoS-IDS2017.
After feature extraction, we obtained 46,950,511 benign samples, so we randomly subsampled this class to 10 million samples both to mitigate the imbalance between classes and for reasons related to computational complexity.
The LycoS-Unicas-IDS2018 dataset is publicly available at \url{https://github.com/MarcoCantone/LycoS-Unicas-IDS2018}.

\begin{table*}
    \small
    \centering
    \begin{tabular}{>{\raggedright}p{0.21\textwidth}>{\raggedleft}p{0.11\textwidth}>{\raggedleft}p{0.18\textwidth}>{\raggedleft}p{0.18\textwidth}>{\raggedleft\arraybackslash}p{0.19\textwidth}}
                                    & CIC-IDS2017   & CSE-CIC-IDS2018   & LycoS-IDS2017     & LycoS-Unicas-IDS2018\\
                                    \toprule
        Benign                      & 2,271,320       & 13,390,249          & 1,395,675           & 10,000,000 \\
        DoS GoldenEye               & 10,293         & 41,508             & 6,765              & 26,861 \\
        DoS Slowloris               & 5,796          & 10,990             & 5,674              & 10,274 \\
        DoS Hulk                    & 230,124        & 461,912            & 158,988            & 1,802,966 \\
        DoS Slowhttptest            & 5,499          & 139,890            & 4,866              & 105,550 \\
        DDoS LOIC-HTTP              & 128,025        & 576,191            & 95,683             & 289,328 \\
        DDoS LOIC-UDP               & 0             & 1,730              & 0                 & 2,382 \\
        DDoS HOIC                   & 0             & 686,012            & 0                 & 1,074,379 \\
        SSH-Patator                 & 5,897          & 187,589            & 2,959              & 92,648 \\
        FTP-Patator                 & 7,935          & 193,354            & 4,003              & 190,300 \\
        PortScan                    & 158,804        & 0                 & 160,106            & 0 \\
        Bot                         & 1,956          & 286,191            & 735               & 96,154 \\
        Infiltration                & 36            & 160,639            & 0                 & 0 \\
        Web Attack - Brute Force    & 1,507          & 611               & 1,360              & 260 \\
        Web Attack - XSS            & 652           & 230               & 661               & 116 \\
        Web Attack - Sql Injection  & 21            & 87                & 12                & 50 \\
        Heartbleed                  & 11            & 0                 & 11                & 0 \\
        \bottomrule

    \end{tabular}
    \caption{Distribution of class instances for all the datasets employed.}
    \label{tab:datasets_composition}
    \normalsize
\end{table*}

\subsection{ML models}
\subsubsection{Linear Discriminant Analysis}
Linear Discriminant Analysis (LDA) is a statistical technique utilized for both dimensionality reduction and classification. As a classifier, it represents one of the simplest methods, functioning by maximizing the separation between multiple classes by projecting the feature space onto a lower-dimensional subspace. The algorithm achieves this by determining the linear combinations of features that best discriminate between classes while minimizing within-class variance. During inference, LDA assigns a class to the input sample based on the learned discriminant functions. We opted for LDA due to its simplicity, facilitating a meaningful comparison with other powerful models and for comprehending the learning difficulty of this task.
\subsubsection{Decision Tree}
 A Decision Tree (DT) is a hierarchical model used in machine learning for tasks like classification or regression. It makes decisions based on the values of features, recursively dividing the data into subsets. The construction involves selecting features at each node to optimize criteria like information gain or variance reduction. The process continues until a stopping criterion is met. Once built, the tree can be used to classify or predict new data by traversing from the root to a leaf node. Decision trees are interpretable but can overfit, so techniques like pruning or ensemble methods are often used for better generalization.
\subsubsection{Random Forest}
Random Forest (RF) is a bagging ensemble of decision trees to improve predictive accuracy and control overfitting. It operates by constructing a multitude of decision trees during training and outputs the average prediction. Each tree is trained on a random subset of the data and may use a random subset of features at each split, introducing diversity and enhancing robustness. Random Forest has been widely adopted in the context of NIDS both for feature selection and classification, also in combination with other models~\cite{farnaaz2016random, li2020building, ustebay2018intrusion, waskle2020intrusion}.
\subsubsection{XGBoost}
XGBoost~\cite{chen2016xgboost} (XGB), short for eXtreme Gradient Boosting, is a boosting ensemble learning algorithm designed for classification and regression tasks. It operates by sequentially building a series of decision trees and combines their predictions to achieve superior accuracy. XGB employs a gradient boosting framework, optimizing model performance by minimizing the gradient of the loss function. It integrates regularization techniques to control model complexity and prevent overfitting and utilizes a strategy to iteratively correct errors made by preceding trees. The algorithm excels in handling large datasets and is renowned for its efficiency, scalability, and ability to capture complex relationships within the data, making it a prominent choice in diverse ML applications. XGB has been employed in many works concerning applications related to network traffic classification~\cite{devan2020efficient, husain2019development, bhati2021improved}.

\subsection{Experimental methodology}
\label{sec:Experimental methodology}

The focus of this work is investigating the generalization capabilities of NIDS based on ML using different datasets for training and evaluating the models. This resembles real-world scenarios wherein these systems must operate. For this purpose, we designed two major categories of experiments: \emph{within-dataset} and \emph{cross-dataset}. Moreover, to understand the motivation behind the discrepancy between within-dataset and cross-dataset results, we conducted additional experiments, utilizing samples from individual attack classes separately and varying the number of features.

\subsubsection{Within-dataset vs. cross-dataset}
We conducted within-dataset experiments to assess the model performance on a designated split of the training dataset. This evaluation provides insights into the models ability to discern patterns from a single source dataset and establishes a reference performance. In cross-dataset experiments, the model is evaluated on a dataset that originates from a network entirely distinct from the one that provided the training dataset. This analysis aims to ascertain the generalization capabilities of the model, determining its effectiveness in applying the learned patterns to novel data sources. The training is conducted using a binary classification approach, i.e. grouping all attacks labels with a single malicious label, allowing us to assess the model capability to distinguish normal traffic from anomalous traffic.

\subsubsection{Single-attack}
\label{sec:Single-attack}
Contrary to grouped binary classification training, where all attack classes are grouped under a single label, we performed also \emph{single-attack} experiments, utilising samples pertaining to a singular class of attack, alongside samples representing benign traffic. This approach provides insights into the models ability to generalize across datasets while considering the nature and modalities of specific attacks and the correlation between attack type and generalization. Since some attacks are not present in both the datasets employed in each cross-dataset experiment, we consider only the following attacks: DoS GoldenEye, DoS Slowloris, DoS Hulk, DoS Slowhttptest, DDoS LOIC-HTTP, SSH-Patator, FTP-Patator, Bot, Infiltration, Web Attack Brute Force, Web Attack - XSS and Web Attack - Sql Injection. In addition, single-attack experiments for the Infliltration class are performed solely with CIC datasets since this attack is absent in the LycoS datasets. Regarding DDoS attacks, we maintain exclusively DDoS LOIC-HTTP samples because, although datasets from 2018 have 3 types of DDoS attacks, CIC17 and LycoS17 comprise only this kind of DDoS. To address high class imbalance, we undersampled the benign class to maintain a 10:1 ratio between benign and attack traffic. This decision was made to trade-off the detrimental effects of heavily imbalanced classes with an adequate representation of the predominant benign class. To decrease the performance variability resulting from undersampling, we performed each single-class experiment three times.

\subsubsection{Feature selection}
In the field of ML, feature selection is often used to mitigate overfitting and enhance generalization. For this reason, we employed the mRMR (minimum Redundancy Maximum Relevance)~\cite{ding2005minimum, peng2005feature} technique. This method involves identifying an optimal subset of features by minimizing redundancy among selected features while maximizing their relevance to the target variable, ensuring the models efficiency and effectiveness in capturing relevant information. It evaluates the mutual information between each feature and the target variable and, at the same time, it considers the redundancy between features. We conducted grouped binary experiments, testing different numbers of features, specifically 1, 2, 3, 4, 5, 10, and 20. Additionally, we performed single-attack experiments using only the two top features.

\subsubsection{Visual feature space analysis}
We conducted an analysis of LycoS datasets structure using visualization techniques to better understand the obtained results.
First, we determine the number of unique values for each of the two best features obtained with mRMR for every attack class. This approach yields a quantitative measure of sample redundancies within the two datasets.
Then, we employ Kernel Density Estimation (KDE) to analyze qualitatively the differences among the attacks distributions between the two datasets in the aforementioned 2D feature subspace. We adopt the implementation of KDE provided by the Seaborn library [66]. Where one of the two features exhibits zero variance or perfect covariance, the employed plotting function fails. Consequently, in these cases, we utilize a scatter plot to effectively depict the distribution.
To further assess diversity in classes distribution between LycoS17 and LycoS18 datasets, we compute the KDE estimation of attacks distributions in a 2D subspace obtained with Principal Component Analysis (PCA) fitted on LycoS17. This technique projects the samples in a subspace where each feature is a linear combination of the original features. Thus, contrary to using only two features, PCA produces a condensed representation mixing the information from all the features.

\subsection{Training strategy}
In the within-dataset experiment, the dataset was randomly split according to an 80:20 ratio in train and test sets. Cross-dataset experiments utilized one entire dataset for training and another for evaluating the model performance.
Before feeding the data to the model, we applied MinMax normalization on each feature. In addition, only for LycoS datasets the categorical feature \texttt{Ip Prot} is encoded using a one-hot encoder, with categories representing less than 0.1\% frequency mapped into a single feature.
A grid search was adopted for selecting the best hyperparameters for each classifier on a 20\% random subset of the training data. The hyperparameter space for each classifier is reported in Table~\ref{tab:GS_hyperparameters}. When the optimal hyperparameter configuration is determined, the classifier is retrained on the entire training set.

\begin{table}
    \centering
    \small
    \begin{tabular}{lll}
        \textbf{Classifier} & \textbf{Hyperparameter} & \textbf{Values} \\
        \toprule
        \multirow{2}*{LDA} & solver & ["svd", "lsqr", "eigen"] \\
         & shrinkage & [None, "auto", 0.1, 0.5, 0.9]\\
        \midrule
        \multirow{7}*{DT} & criterion & ["gini", "entropy"] \\
         & splitter & ["best", "random"] \\
         & max\_depth & [None, 20] \\
         & min\_samples\_split & [2, 4, 8, 16] \\
         & min\_samples\_leaf & [1, 2, 4] \\
         & max\_features & [None, "sqrt", "log2"] \\
         & max\_leaf\_nodes & [None, 10,000, 1,000,000] \\
        \midrule
        \multirow{5}*{RF} & criterion & ["gini", "entropy"] \\
         & max\_depth & [None, 10] \\
         & min\_samples\_split & [4, 16] \\
         & max\_features & [None, "sqrt", "log2"] \\
         & n\_estimators & [10, 50] \\
         \midrule
        \multirow{5}*{XGBoost} & max\_depth& [3, 6, 12]\\
         & n\_estimators &[10, 50] \\
         & learning\_rate &[0.1, 0.3, 1] \\
         & booster &["gbtree", "gblinear"] \\
         & min\_child\_weight &[0.5, 1, 2] \\
         & gamma &[0, 1, 10] \\
        \bottomrule
    \end{tabular}
    \normalsize
    \caption{Hyperparameter search space for each classifier.}
    \label{tab:GS_hyperparameters}
\end{table}

\subsection{Performance evaluation}
For evaluating model performance we computed the Matthews Correlation Coefficient (MCC), the F1-score and the Area Under the Receiver Operating Characteristic (AUROC) curve.
The MCC \cite{chicco2020advantages, yao2020assessing, boughorbel2017optimal} is a correlation coefficient between classifier output and true label, hence it yields a more informative and truthful score in evaluating binary classifications than accuracy, especially in the case of unbalanced datasets. 
The F1-score represents the harmonic mean between precision and recall. Precision measures the accuracy of positive predictions, while recall measures the ability of the model to correctly identify all positive instances in the dataset. The harmonic mean combines these two measures into a single value, making it a useful indicator of model performance.
The Receiver Operating Characteristic (ROC) curve shows the prediction ability of a binary classifier by representing the sensitivity over the false-positive rate varying the threshold value. The area under the ROC curve is a widely used metric that offers a simple way to summarize the overall performance of a model.
We decided not to report the accuracy value since, given the datasets inherent imbalance, reporting accuracy might be misleading, especially as a classifier outputting only the most frequent value can yield a seemingly high performance. For instance, 11 out of 14 attacks present in CIC17 have a benign-attack ratio greater than 200, resulting in a 99.5\% accuracy for a model predicting always the Benign class.

\section{Results and discussion}
\label{sec:Results and discussion}
\subsection{Within-dataset vs. cross-dataset}
Table~\ref{tab:res_binary} reports the results obtained for the 4 within-dataset and 4 cross-dataset experiments for each classifier employed. Within-dataset experiments show significantly superior performance compared to cross-dataset ones, with an average MCC among the diverse classifier and train-test set combinations of 94.63\% and 29.35\%, respectively. It follows that neither of the classifiers was able to transfer the learned patterns when evaluated on a new dataset originating from a different network than the one used for training. The cross-dataset combination that obtained the worst result was the LycoS17 on LycoS18 with an average MCC of 10.83\%, while the best result was obtained by CIC18 on CIC17 with an average value of 40.24\%.
This is likely due to a too homogeneous class representation that is very simple to learn in within-dataset experiments but not sufficient to generalize on an external dataset.
In support of this hypothesis, 9 out of 16 within-dataset experiments achieved near-perfect classification performance with an MCC of over 99.6\%. This is unusual in many ML problems and can be a consequence of low variability in the training datasets and sample redundancies.

\begin{table}
    \centering
    \footnotesize
    \begin{tabular}{@{}llllll@{}}
    \textbf{Train set} & \textbf{Test set} & \textbf{Classifier} & \textbf{MCC} & \textbf{F1} & \textbf{AUROC} \\
    \toprule
    \multirow{8}*{CIC17} & \multirow{4}*{CIC17} & LDA & 64.86\%	& 69.02\% & 96.45\% \\
     & & DT & 99.72\% & 99.78\% & 99.97\% \\
     & & RF & 99.74\% & 99.79\% & 99.98\% \\
     & & XGB & 99.74\% & 99.79\% & 100.00\% \\
    \cmidrule{2-6}	
     & \multirow{4}*{CIC18} & LDA & 32.62\% & 39.12\% & 81.70\% \\
     & & DT & 30.82\% & 31.20\% & 59.14\% \\
     & & RF & 25.96\% & 25.76\% & 66.93\% \\
     & & XGB & 35.68\% & 26.43\% & 80.44\% \\
    \cmidrule{1-6}	
    \multirow{8}*{CIC18} & \multirow{4}*{CIC18} & LDA & 76.52\% & 79.82\% & 94.13\% \\
     & & DT & 96.33\% & 96.89\% & 98.85\% \\
     & & RF & 96.45\% & 96.99\% & 98.87\% \\
     & & XGB & 96.48\% & 97.02\% & 99.10\% \\
    \cmidrule{2-6}	
     & \multirow{4}*{CIC17} & LDA & 39.29\% & 48.41\% & 85.06\% \\
     & & DT & 35.50\% & 33.86\% & 57.80\% \\
     & & RF & 44.44\% & 51.21\% & 75.77\% \\
     & & XGB & 41.75\% & 44.58\% & 79.15\% \\
    \midrule
    \multirow{8}*{LycoS17} & \multirow{4}*{LycoS17} & LDA & 94.14\% & 95.56\% & 99.54\% \\
     & & DT & 99.83\% & 99.87\% & 99.92\% \\
     & & RF & 99.89\% & 99.92\% & 100.00\% \\
     & & XGB & 99.87\% & 99.90\% & 100.00\% \\
    \cmidrule{2-6}
     & \multirow{4}*{LycoS18} & LDA & -15.32\% & 10.24\% & 57.24\% \\
     & & DT & 8.55\% & 17.04\% & 53.11\% \\
     & & RF & 9.29\% & 13.97\% & 65.46\% \\
     & & XGB & 10.17\% & 14.42\% & 61.54\% \\
    \cmidrule{1-6}
    \multirow{8}*{LycoS18} & \multirow{4}*{LycoS18} & LDA & 91.32\% & 93.44\% & 99.78\% \\
     & & DT & 99.77\% & 99.83\% & 99.99\% \\
     & & RF & 99.76\% & 99.83\% & 99.99\% \\
     & & XGB & 99.67\% & 99.76\% & 99.96\% \\
    \cmidrule{2-6}
     & \multirow{4}*{LycoS17} & LDA & 60.41\% & 70.46\% & 91.66\% \\
     & & DT & 1.24\% & 2.56\% & 50.15\% \\
     & & RF & 38.33\% & 37.29\% & 77.49\% \\
     & & XGB & 40.29\% & 35.54\% & 81.37\% \\
    \bottomrule
    \end{tabular}
    \normalsize
    \caption{Performance obtained for each classifier for all the within-dataset and cross-dataset configurations tested.}
    \label{tab:res_binary}
\end{table}

When trained on CIC18, neither of the four classifiers achieved a within-dataset MCC greater than 97\%, while for all the other within-dataset combinations many models easily reached and surpassed a value of 99\%.  In the cross-dataset scenario, the average MCC during training on CIC18 was 40.24\%, while training on the CIC17 dataset yielded a lower average of 31.27\%. Consequently, pretraining on CIC18 yielded suboptimal within-dataset results but demonstrated superior performance in cross-dataset evaluation. This observed difference may be attributed to the larger dimensionality of the CIC18 dataset, which potentially provides a more comprehensive representation of network traffic, encompassing 13,691,268 instances compared to the 1,837,498 instances of the CIC17 dataset.

Among all classifiers, LDA is the one that obtained the worst results in the within-dataset experiments whereas in cross-dataset experiments, instead, its performances are aligned with other classifiers and achieved the best result among all the cross-dataset experiments when trained on LycoS18 and evaluated on LycoS17 with a 60.41\% MCC. This can be due to the lower model complexity that reduces overfitting and thus benefits generalization.

\subsection{Single-attack}
\begin{figure*}
    \centering    
    \subfloat{
        \includegraphics[width=.33\linewidth]{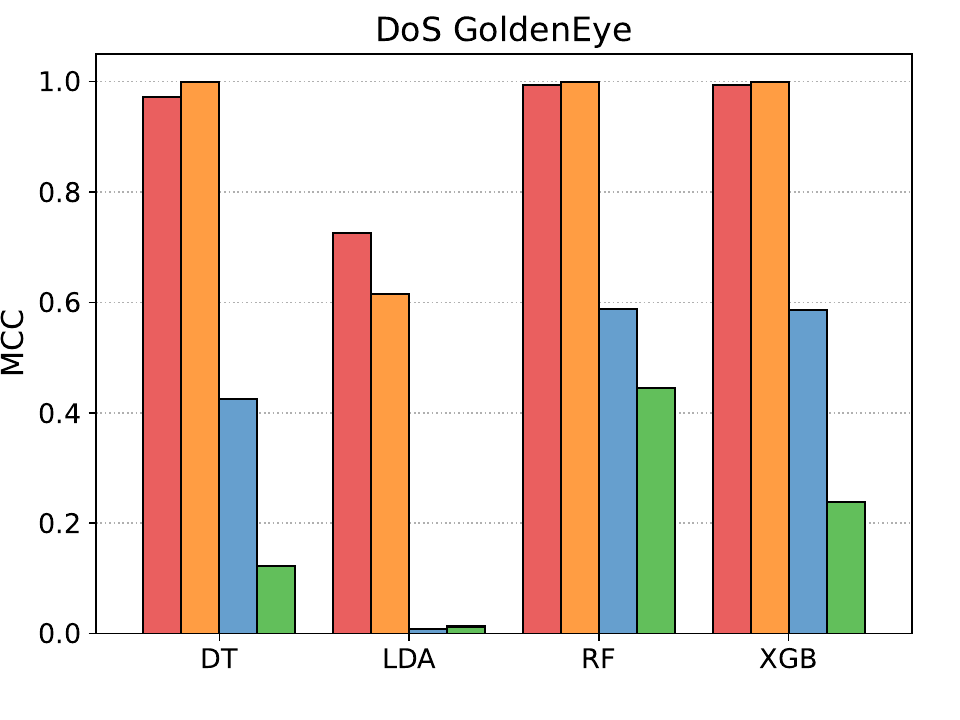}
    }
    \subfloat{
        \includegraphics[width=.33\linewidth]{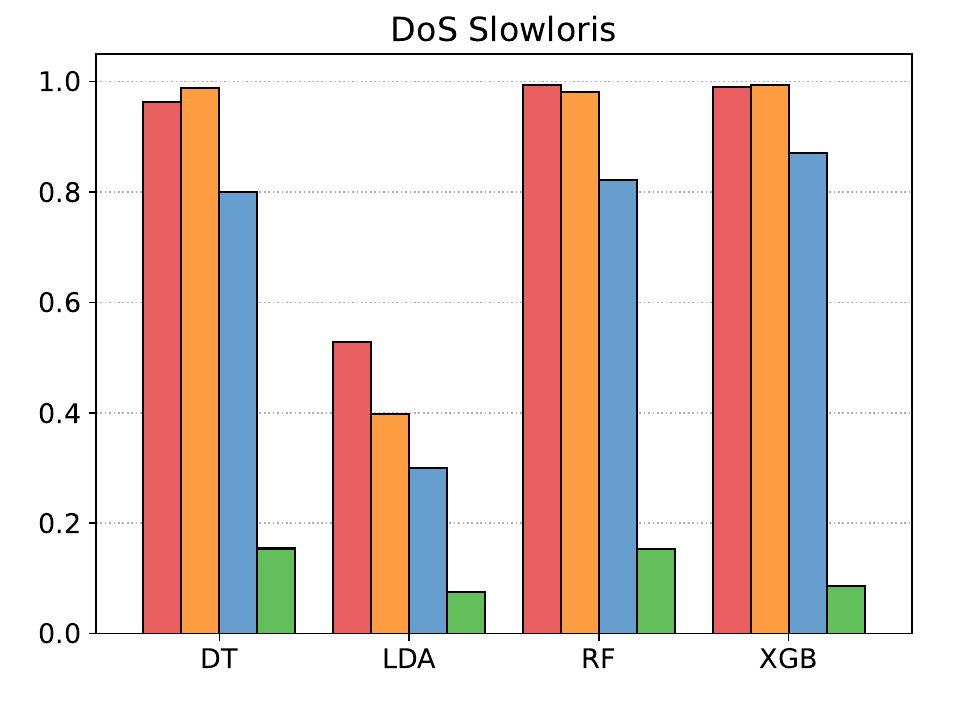}
    }
    \subfloat{
        \includegraphics[width=.33\linewidth]{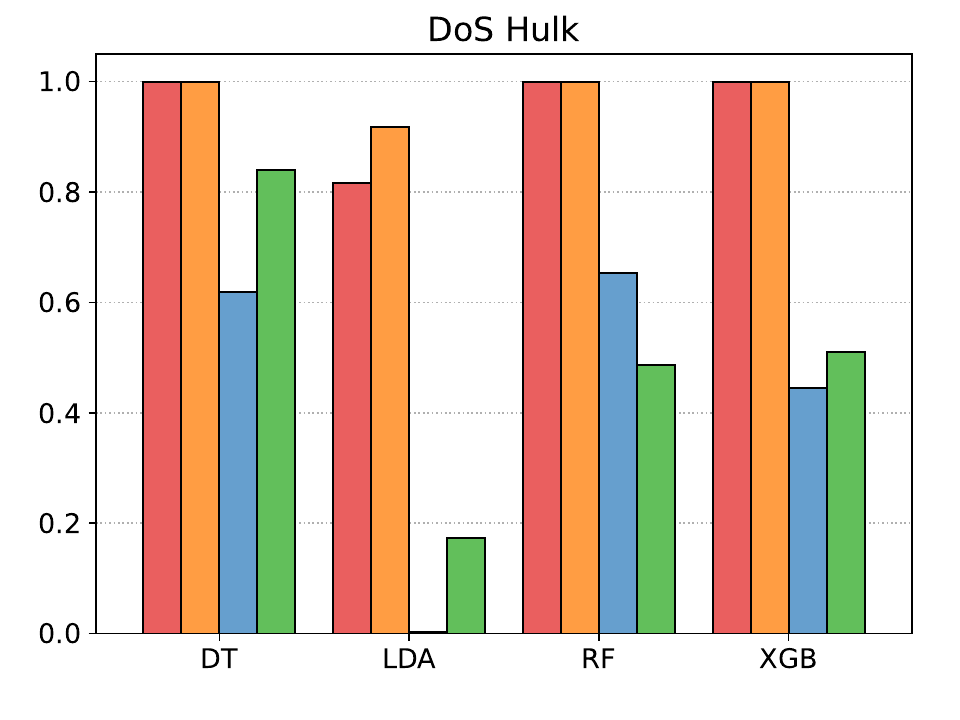}
    }\\
        \subfloat{
        \includegraphics[width=.33\linewidth]{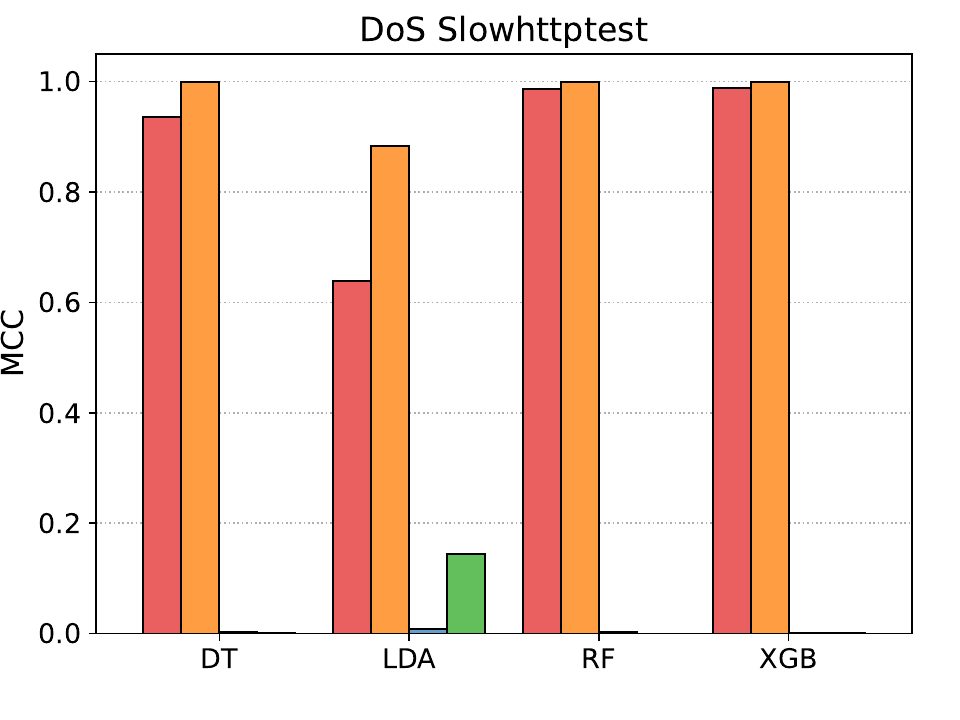}
    }
    \subfloat{
        \includegraphics[width=.33\linewidth]{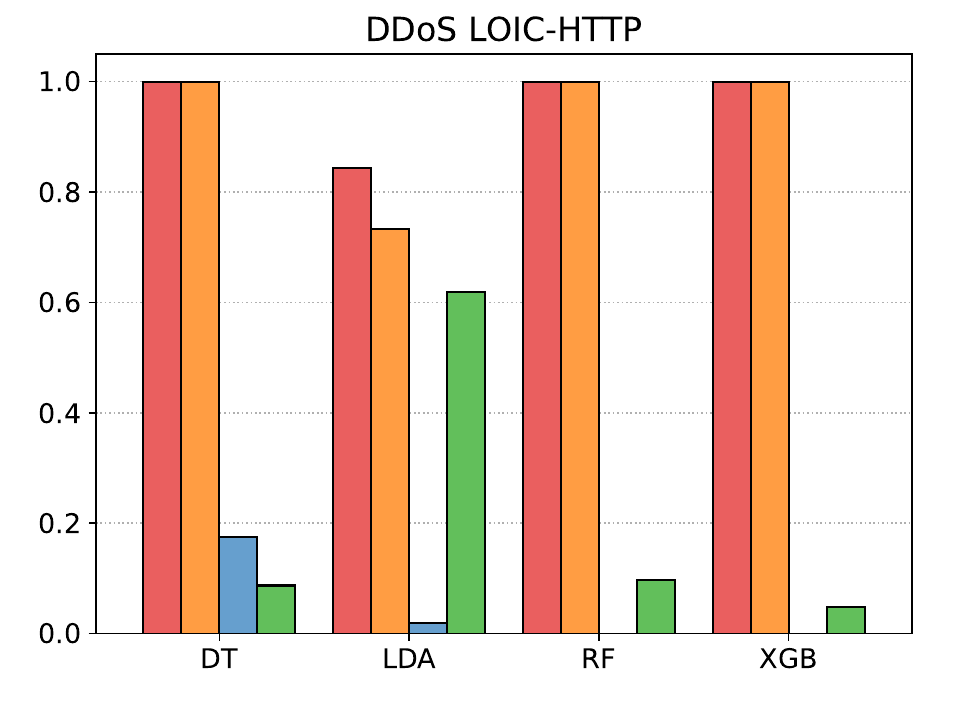}
    }
    \subfloat{
        \includegraphics[width=.33\linewidth]{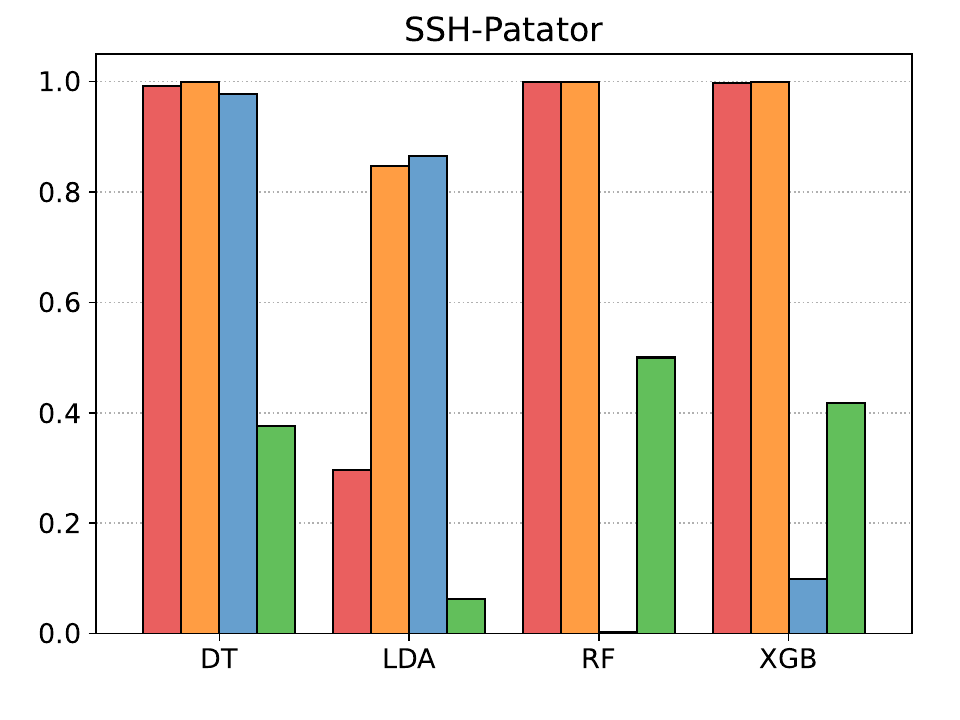}.
    }\\
    \subfloat{
        \includegraphics[width=.33\linewidth]{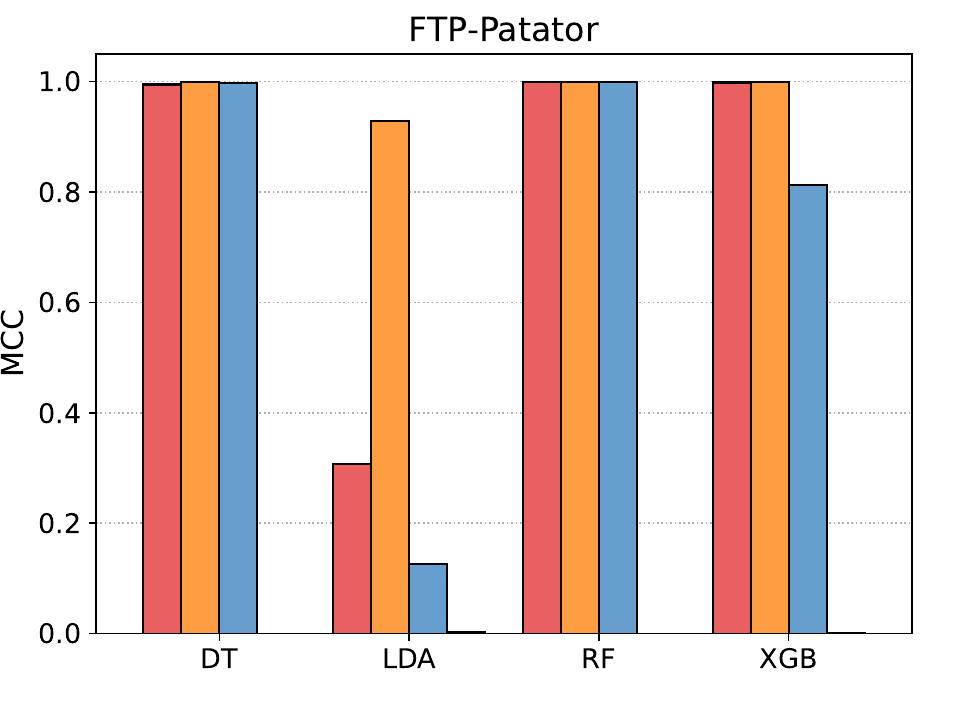}
    }
    \subfloat{
        \includegraphics[width=.33\linewidth]{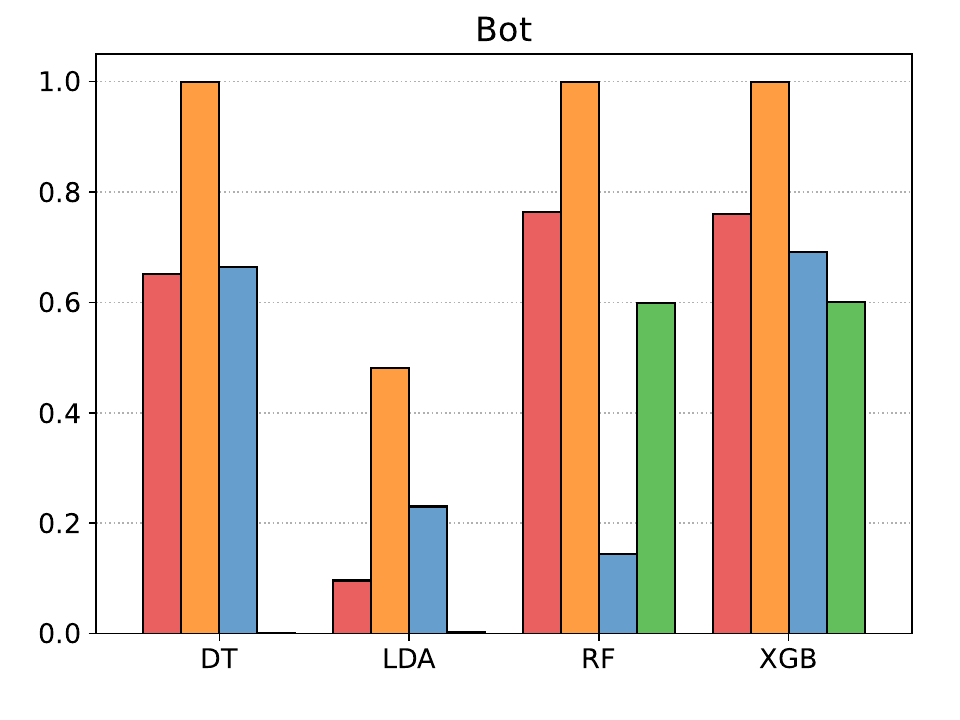}
    }
    \subfloat{
        \includegraphics[width=.33\linewidth]{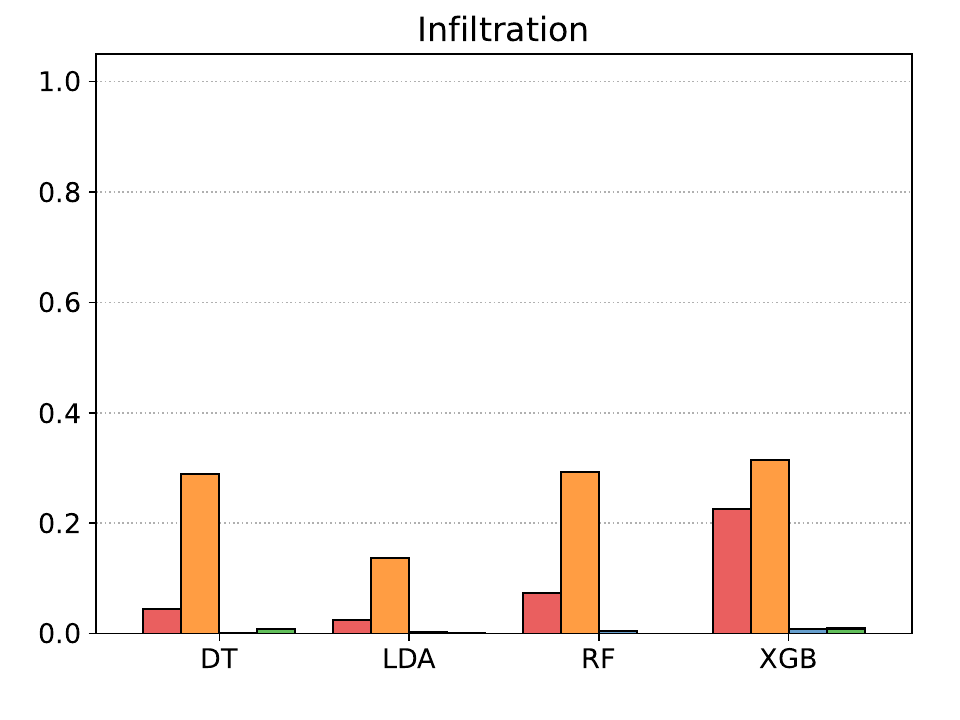}
    }\\
    \subfloat{
        \includegraphics[width=.33\linewidth]{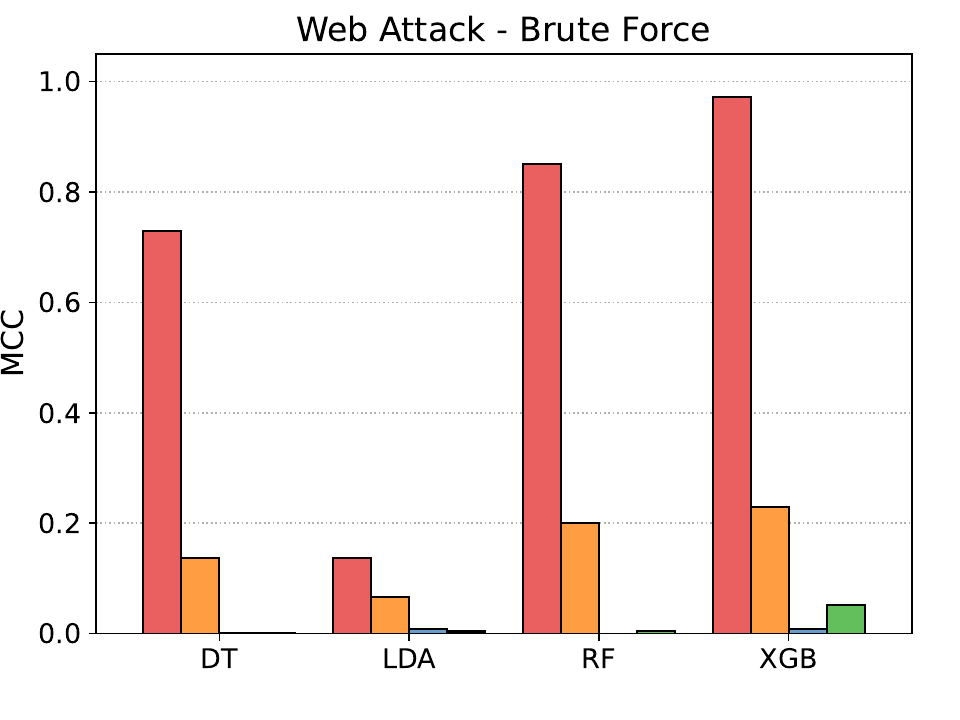}
    }
    \subfloat{
        \includegraphics[width=.33\linewidth]{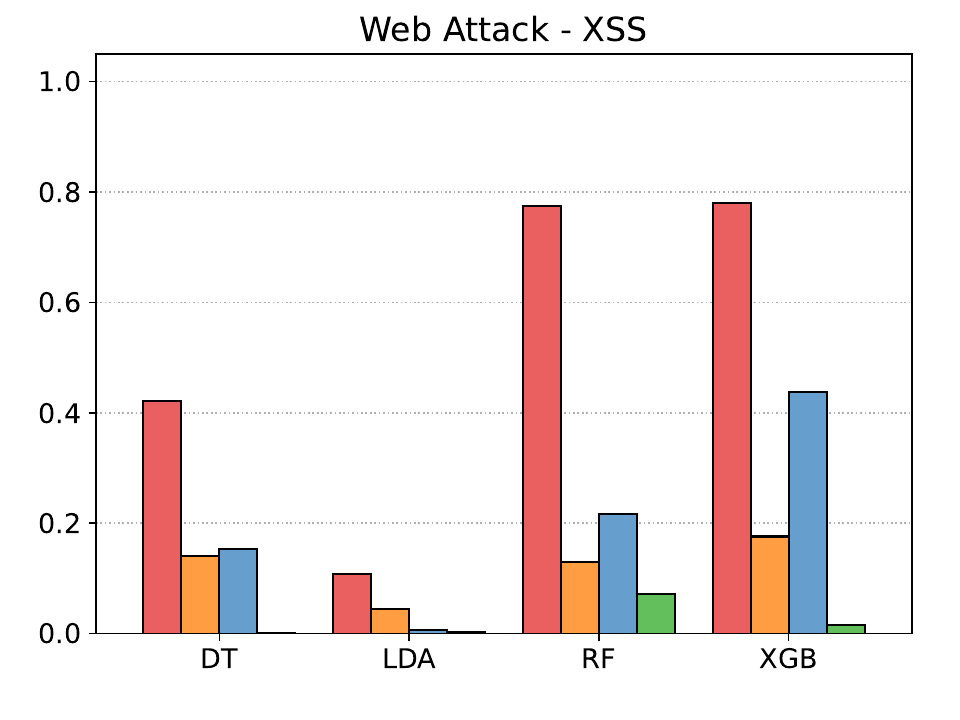}
    }
    \subfloat{
        \includegraphics[width=.33\linewidth]{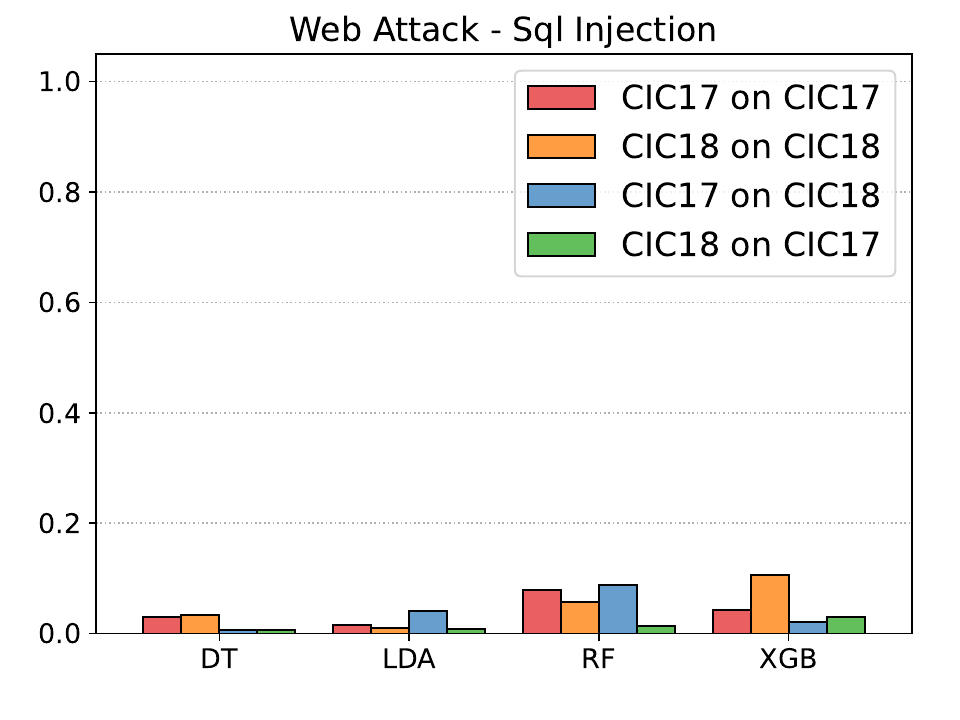}
    }
    \caption{MCC value obtained for each classifier and train-test combination with CIC datasets in single-attack experiments.}
    \label{fig:singleClass_cic}
\end{figure*}

\begin{figure*}
    \centering    
    \subfloat{
        \includegraphics[width=.33\linewidth]{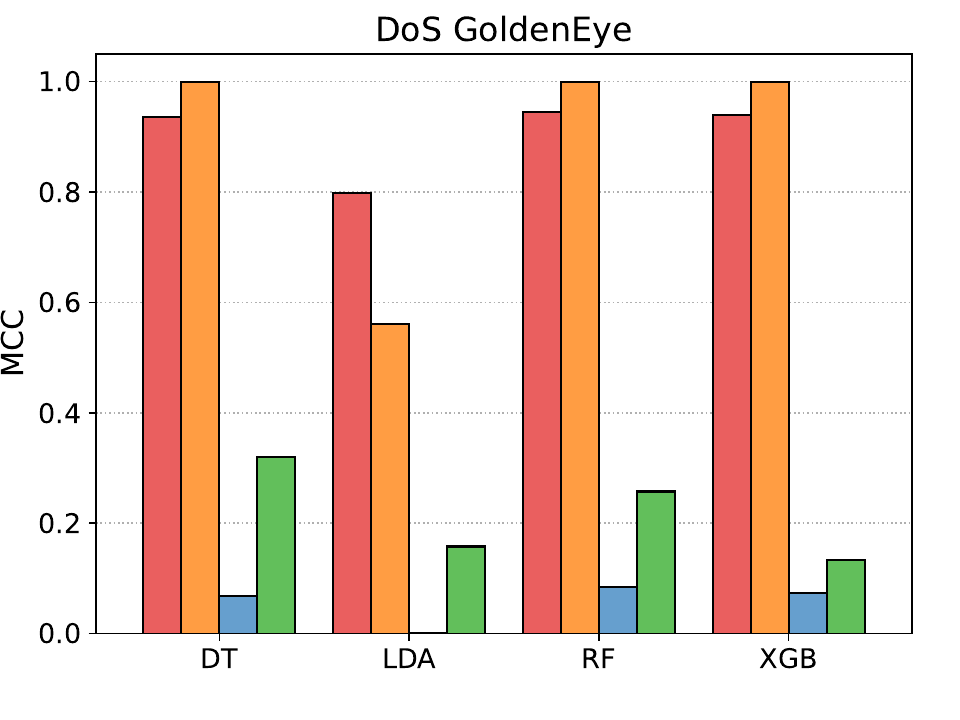}
    }
    \subfloat{
        \includegraphics[width=.33\linewidth]{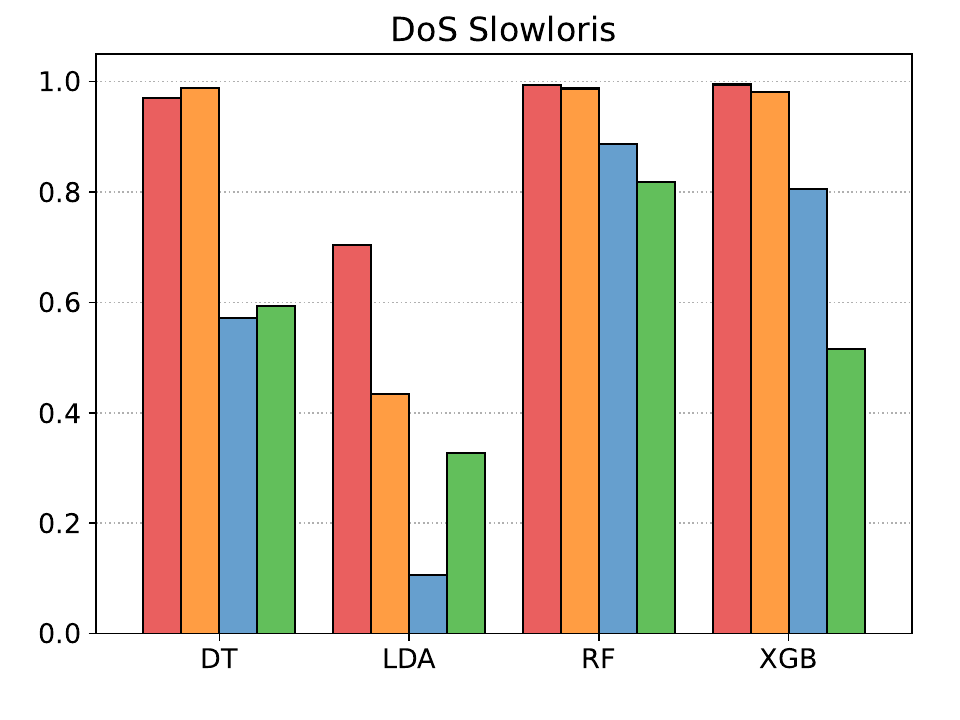}
    }
    \subfloat{
        \includegraphics[width=.33\linewidth]{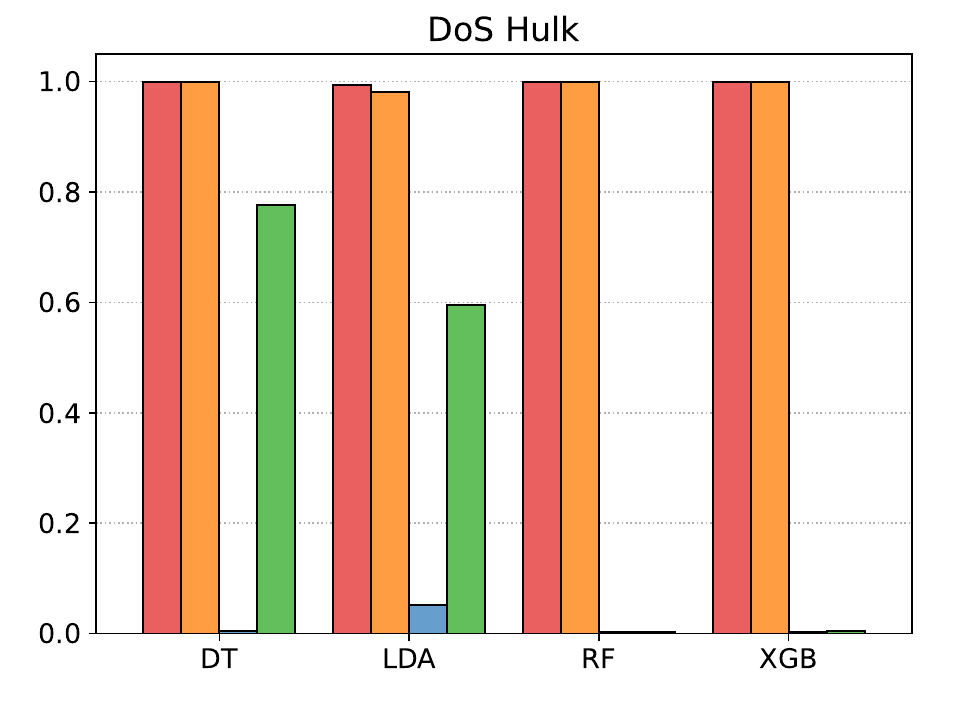}
    }\\
        \subfloat{
        \includegraphics[width=.33\linewidth]{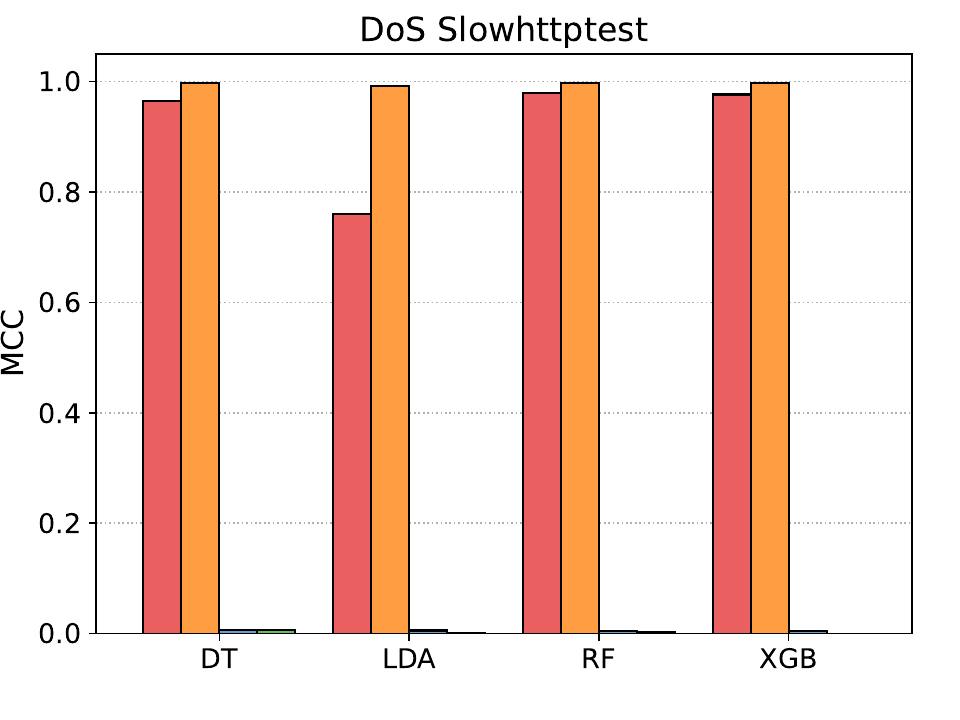}
    }
    \subfloat{
        \includegraphics[width=.33\linewidth]{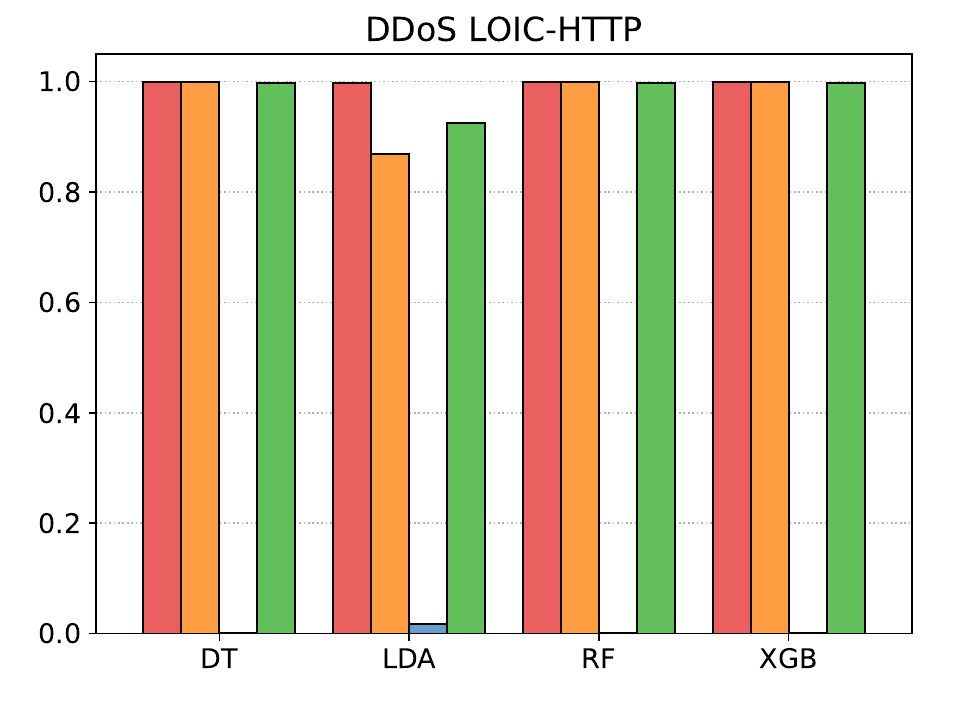}
    }
    \subfloat{
        \includegraphics[width=.33\linewidth]{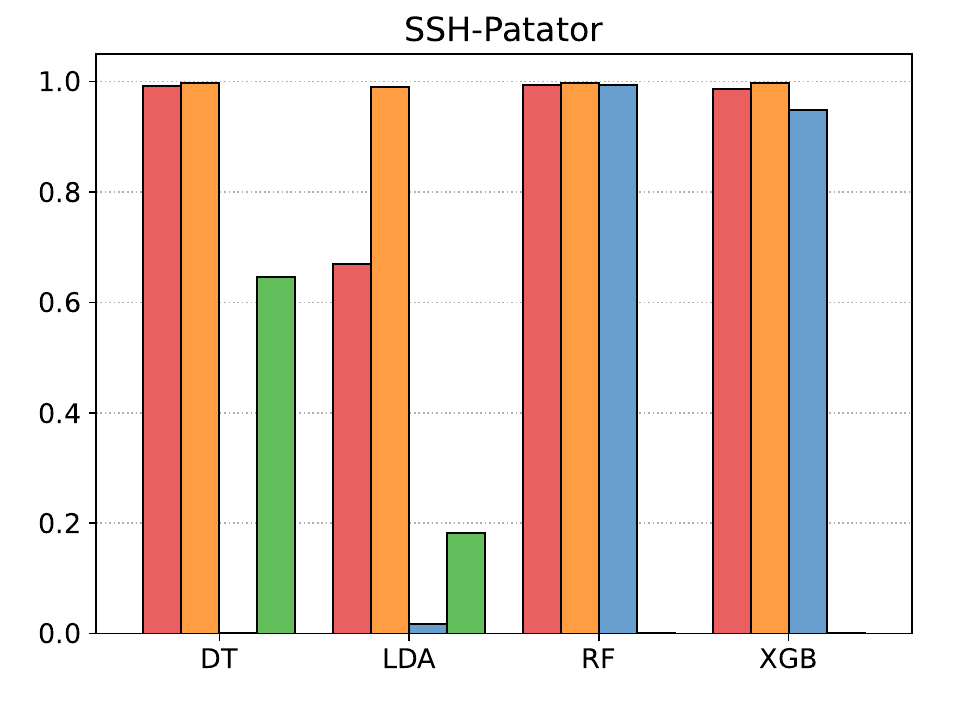}
    }\\
    \subfloat{
        \includegraphics[width=.33\linewidth]{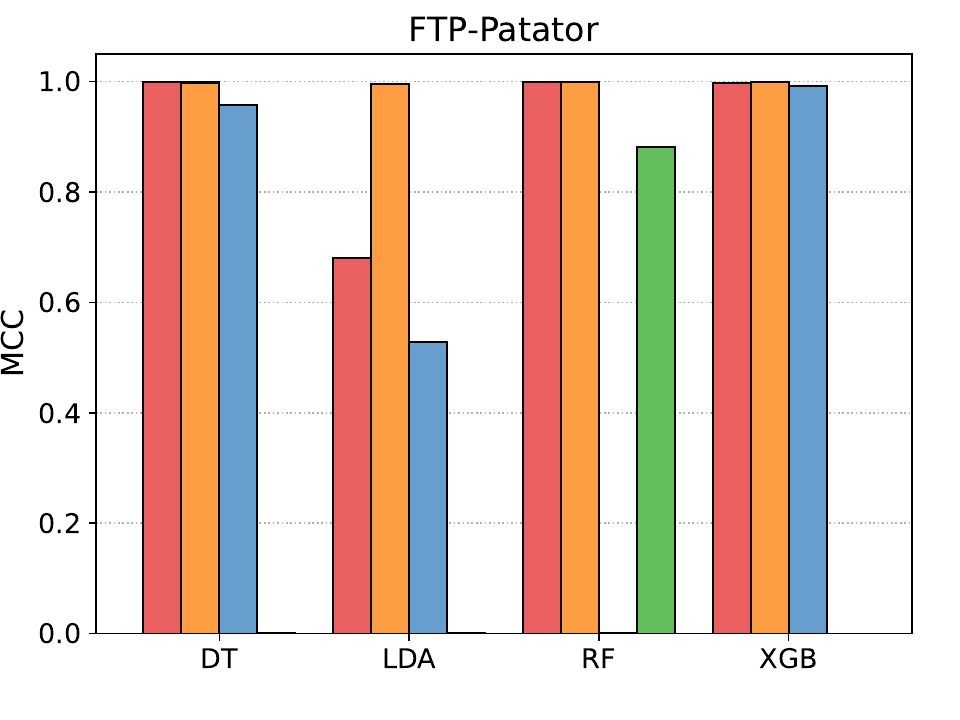}
    }
    \subfloat{
        \includegraphics[width=.33\linewidth]{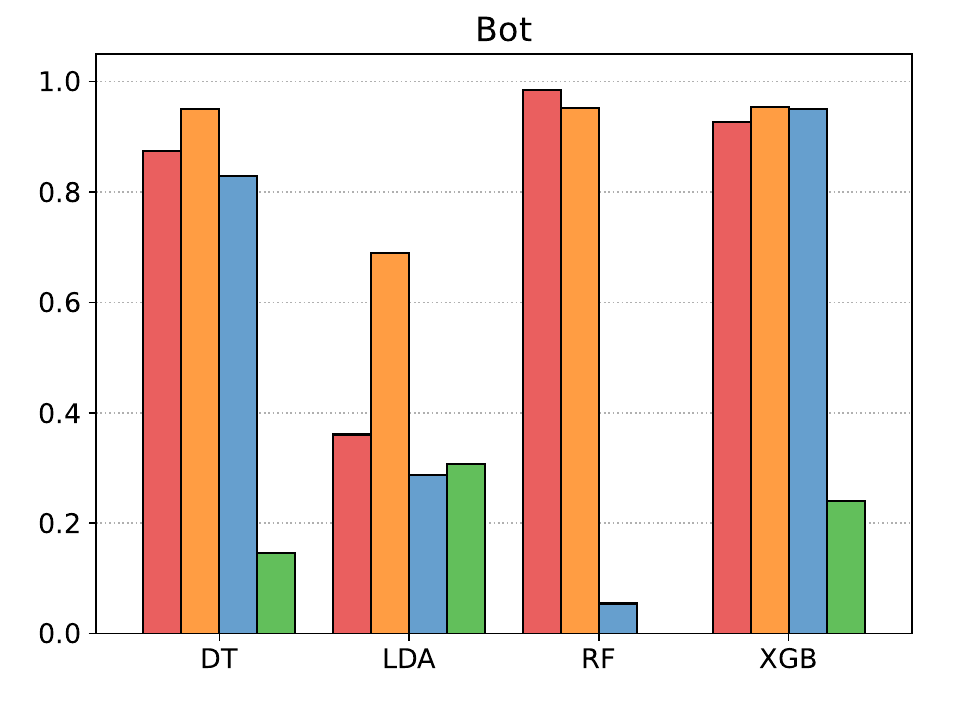}
    }
    \subfloat{
        \includegraphics[width=.33\linewidth]{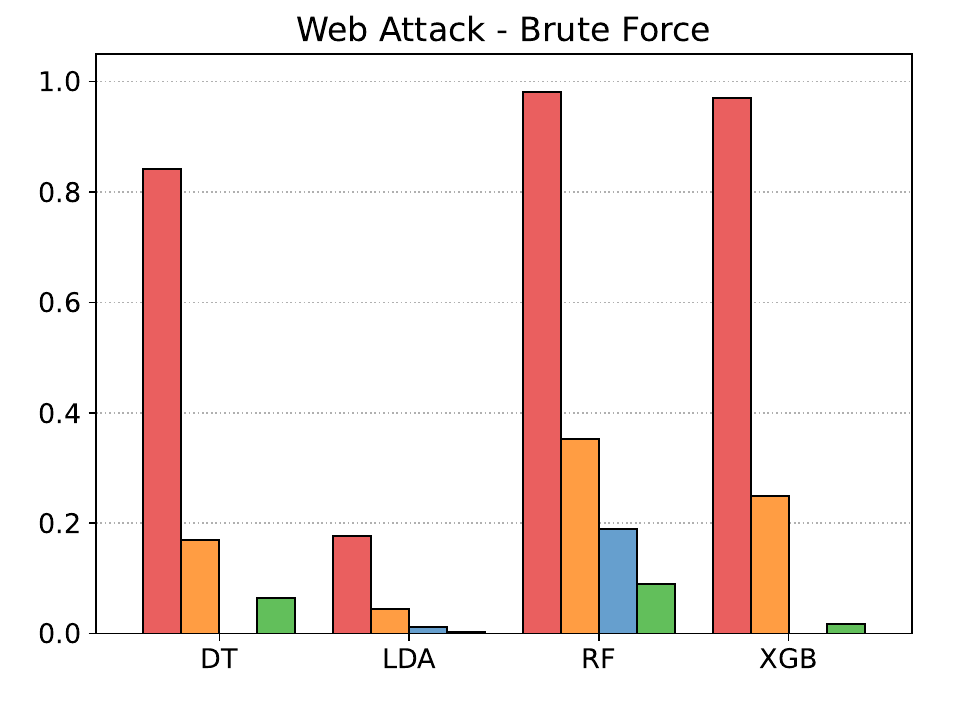}
    }\\
    \subfloat{
        \includegraphics[width=.33\linewidth]{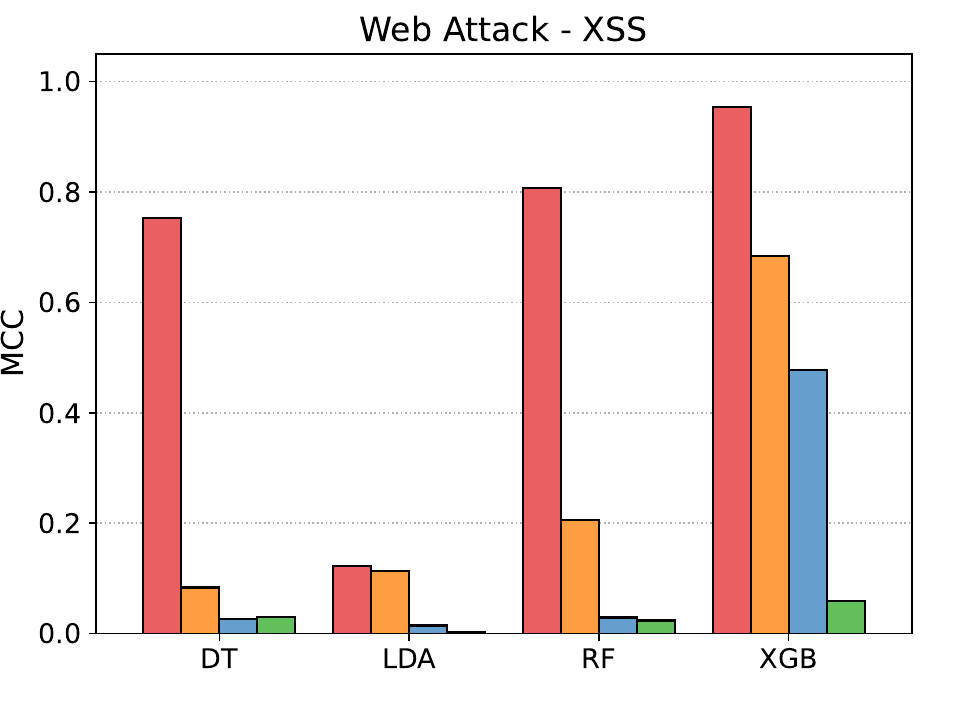}
    }
    \subfloat{
        \includegraphics[width=.33\linewidth]{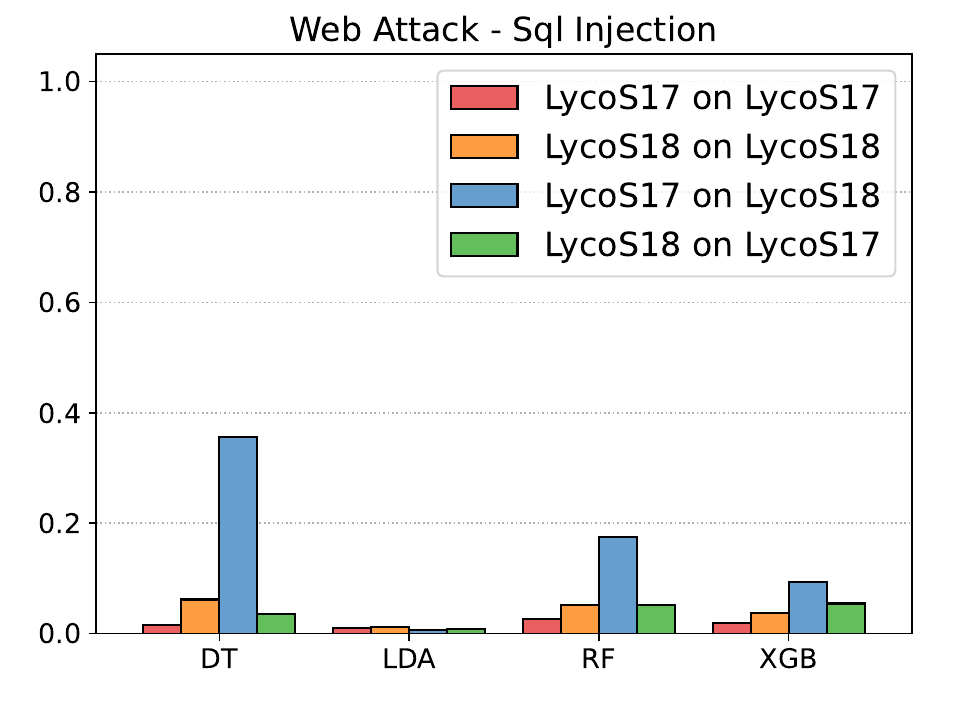}
    }
    \subfloat{
        \includegraphics[width=.33\linewidth]{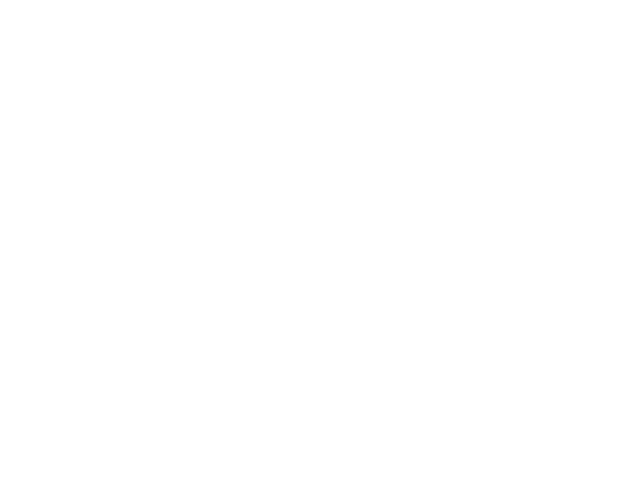}
    }
    \caption{MCC value obtained for each classifier and train-test combination with LycoS datasets in single-attack experiments.}
    \label{fig:singleClass_lycos}
\end{figure*}

Figure~\ref{fig:singleClass_cic} and Figure~\ref{fig:singleClass_lycos} present the results obtained by single-attack experiments on CIC and LycoS datasets, respectively. For each experiment we report the highest MCC value across the three runs (see Section~\ref{sec:Single-attack}).
It can be observed that there was no classifier able to generalize the patterns of a specific attack across all combinations of train and test sets.

There are 7 attacks for which the classifiers exhibited high performance in within-dataset experiments, namely DoS GoldenEye, DoS Slowloris, DoS Hulk, DoS Slowhttptest, DDoS LOIC-HTTP, SSH-Patator and FTP-Patator.
Other 4 attacks generally showed weak performance in within-dataset evaluation, namely all 3 web attacks and Infiltration. This is due to the very low number of samples that made it difficult the learning process.
Regarding the Bot attack, the results are strongly dependant on the specific dataset used for training. When the models are trained and evaluated on CIC18, the performances are near 100\% MCC, whereas if CIC17 is chosen as training set the MCC value dropped slightly above 60\%. When considering the LycoS datasets, this discrepancy is not present and the within-dataset results when training on LycoS17 and LycoS18 are similar. This suggests that the approach of~\cite{rosay2022network}, that we adopted to generate LycoS18, has probably corrected the issues present in the CIC datasets.

Cross-dataset experiments showed some generalization capabilities depending on the attack type and also on the specific train-test combination used for cross-dataset evaluation.
For the following attacks, DoS GoldenEye, DoS Slowloris, DoS Hulk, DDoS LOIC-HTTP, SSH-Patator, FTP-Patator and Bot, 40 out of 112 combinations of classifier and train-test datasets showed MCC superior to 50\%, indicating partial transfer of the learned patterns into the new dataset.
In the case of DoS Slowhttptest, Infiltration, Web Attack - Brute Force, Web Attack - XSS and Web Attack - Sql Injection, none of the classifiers in any of the train-test combinations were able to achieve MCC greater than 50\%, and in many cases, the performances are almost zero.
For the attack FTP-Patator, the models trained on CIC17 showed very high MCC both in within-dataset and cross-dataset evaluation but the models trained on CIC18, although performed well in within-dataset configuration, obtained MCC less than 1\% when evaluated on the external test set.
The models trained with DDoS LOIC-HTTP attack samples showed near perfect generalization when trained on LycoS18 and tested on LycoS17. However, in the other 3 cross-dataset configurations their performances approached MCC 0\%, with the only exception of LDA in the configuration CIC18 on CIC17 which achieved a 61.89\% MCC. This phenomenon may be attributed to the possibility that, in one of the two datasets, the attack samples do not adequately represent the true distribution of that class. Instead, they might represent a subclass of the class distribution of the other dataset, leading to cross-dataset generalization occurring predominantly in one direction.

\subsection{Feature selection}
\begin{figure*}
    \centering
    \subfloat{
        \includegraphics[trim= 0cm 0cm 0cm 0cm, width=.5\linewidth]{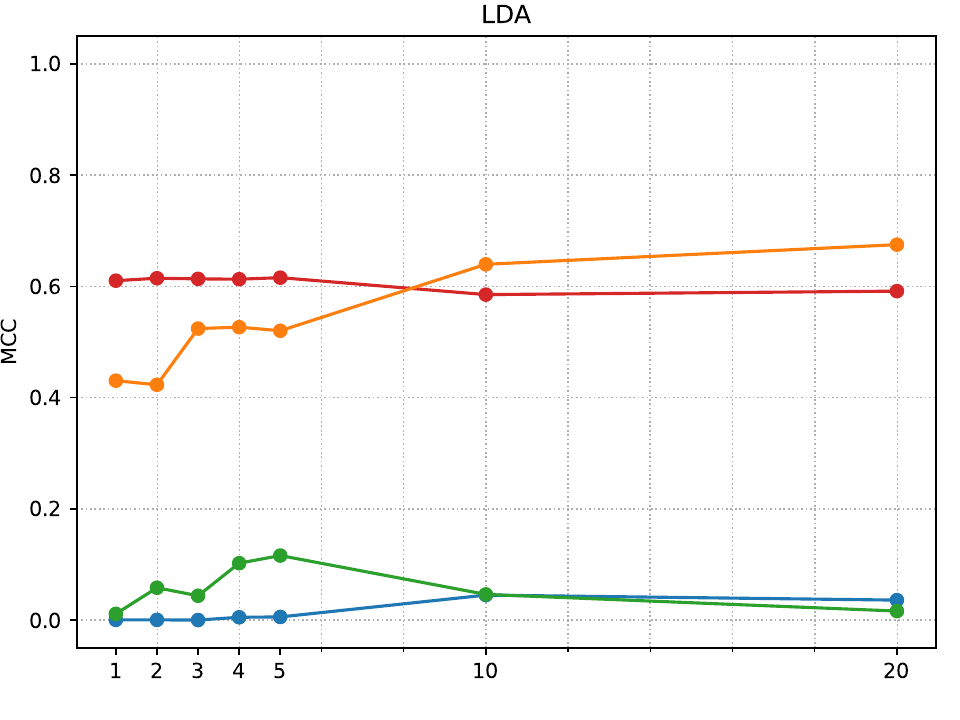}
        \label{subfig:lda_cic}
    }
    \subfloat{
        \includegraphics[trim= 0cm 0cm 0cm 0cm, width=.5\linewidth]{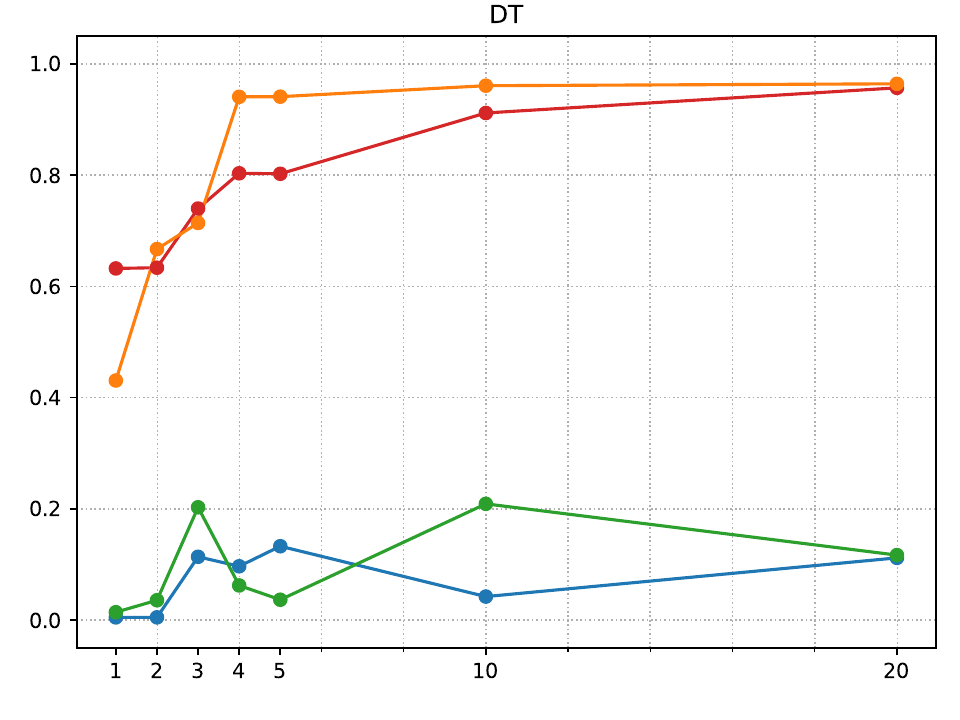}
        \label{subfig:dt_cic}
    }\\
    \subfloat{
        \includegraphics[trim= 0cm 0cm 0cm 0cm, width=.5\linewidth]{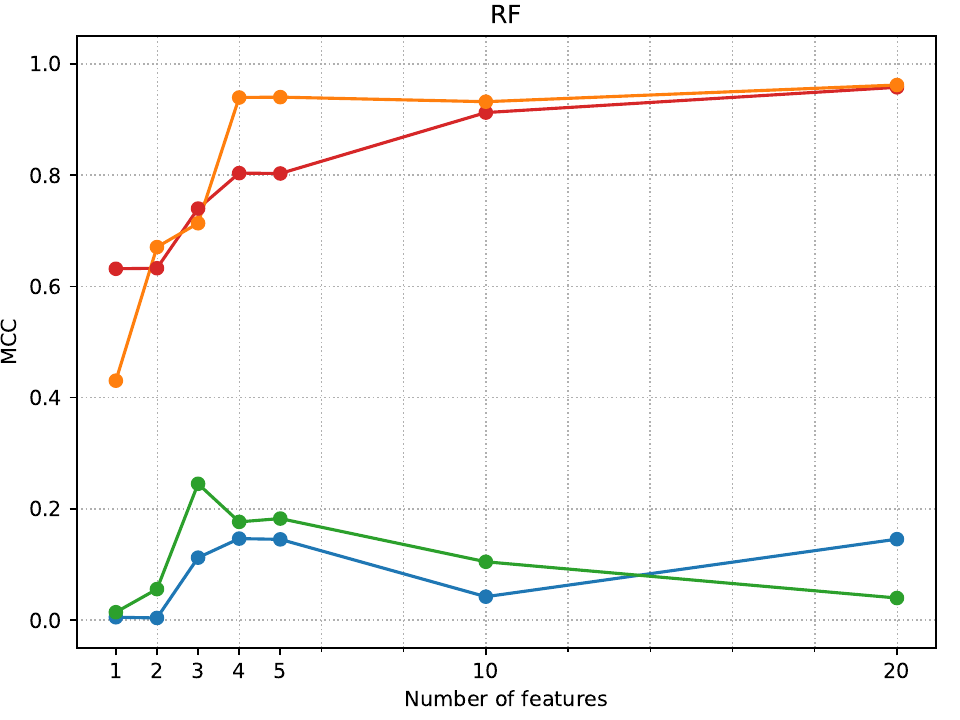}
        \label{subfig:rf_cic}
    }
    \subfloat{
        \includegraphics[trim= 0cm 0cm 0cm 0cm, width=.5\linewidth]{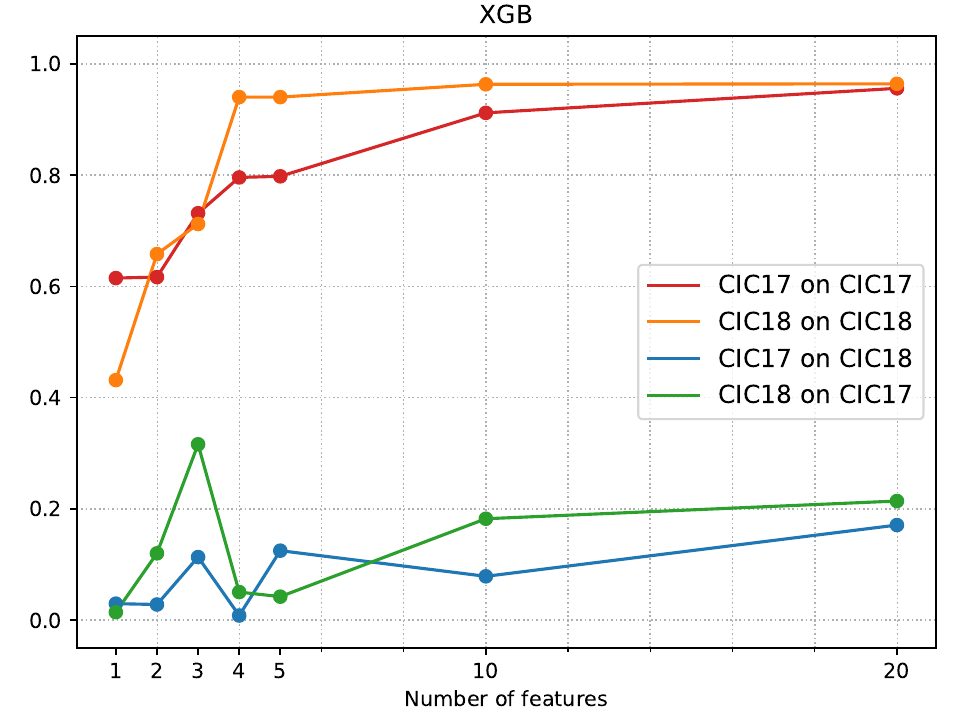}
        \label{subfig:xgb_cic}
    }
    \caption{MCC for each classifier varying the number of features for the 4 train-test set combination of CIC datasets.}
    \label{fig:allClass_featureSelection_cic}
\end{figure*}

\begin{figure*}
    \centering    
    \subfloat{
        \includegraphics[trim= 0cm 0cm 0cm 0cm, width=.5\linewidth]{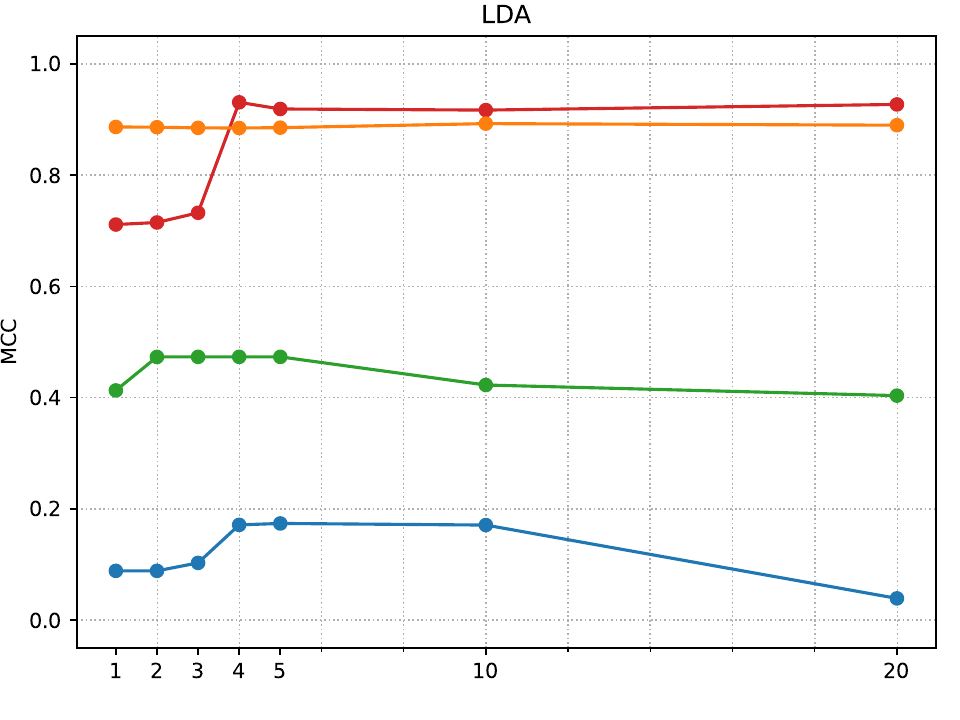}
        \label{subfig:lda_lycos}
    }
    \subfloat{
        \includegraphics[trim= 0cm 0cm 0cm 0cm, width=.5\linewidth]{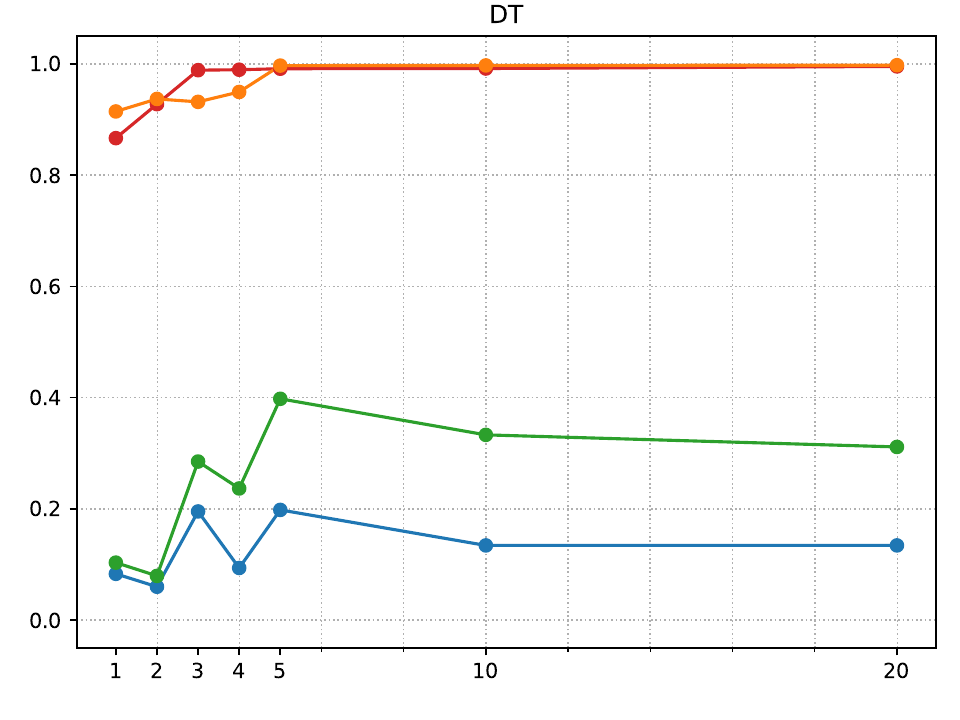}
        \label{subfig:dt_lycos}
    }\\
    \subfloat{
        \includegraphics[trim= 0cm 0cm 0cm 0cm, width=.5\linewidth]{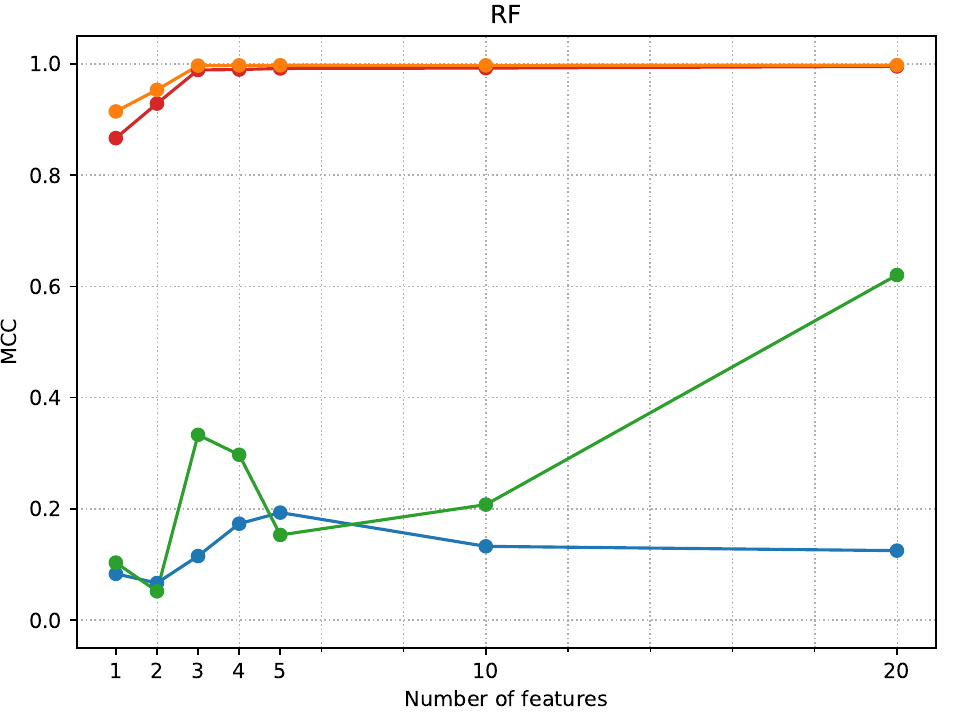}
        \label{subfig:rf_lycos}
    }
    \subfloat{
        \includegraphics[trim= 0cm 0cm 0cm 0cm, width=.5\linewidth]{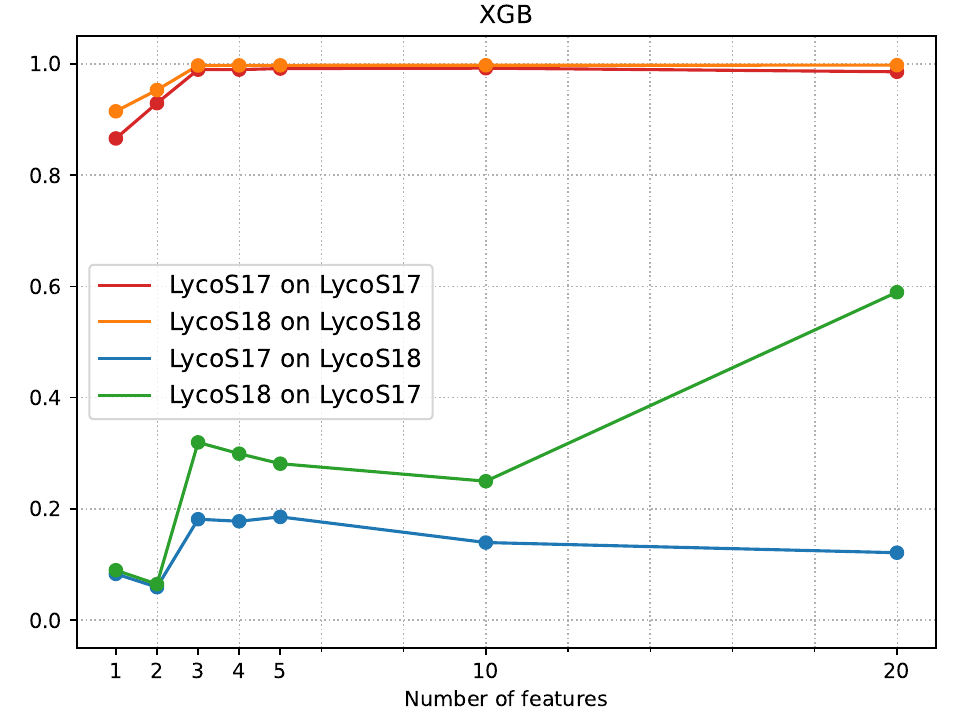}
        \label{subfig:xgb_lycos}
    }
    \caption{MCC for each classifier varying the number of features for the 4 train-test set combination of LycoS datasets.}
    \label{fig:allClass_featureSelection_lycos}
\end{figure*}

In Figure~\ref{fig:allClass_featureSelection_cic} and Figure~\ref{fig:allClass_featureSelection_lycos} are reported the results obtained by varying the number of features used for training with CIC and LycoS datasets, respectively. The experiments are performed in a binary manner using a malicious label that represents all the different attacks. The features employed in each experiment are reported in Table~\ref{tab:best_features_binary}. Since mRMR is an incremental method, the optimal subset of $n$ features is given by the first $n$ features in one of the reported lists. Compared to the experiments employing the original feature set, also in this case within-dataset experiments achieved significantly better performance with respect to cross-dataset experiments with an average MCC of 83.86\% and 14.91\%, respectively.

As expected, performance in within-dataset experiments increases with the number of features, whereas in cross-dataset experiments we observe a different behavior. In 11 out of 16 experiments, the best result was achieved with a low number of features, within the range $[3, 5]$. This value is considerably low compared to the total original number of features, that is 77 both for CIC and LycoS datasets. This can be due to the fact that using a low number of features reduces overfitting, encouraging the models to learn more general patterns.

34 out of 64 experiments achieved MCC greater than 85\% employing 4 or less features. Using only one feature, the average MCC among the within-dataset experiments is 69.7\%, which corresponds to an accuracy of 91.22\%. This indicates a strong correlation of some features with the label in the specific dataset. Unfortunately, these features are not sufficient to detect the attack since the performance in cross-dataset experiments drops sharply. 
We believe that the attacks are excessively homogeneous, allowing the attack samples of a single dataset to be represented using a minimal number of features.

\begin{table}
\centering
\footnotesize
\begin{tabular}{p{0.12\linewidth}p{0.78\linewidth}}
\textbf{Dataset} & \textbf{Best 20 features} \\
\toprule
CIC17 & Bwd Packet Length Std, ACK Flag Count, Fwd Packet Length Mean, Bwd Packet Length Max, Bwd Packet Length Mean, Avg Bwd Segment Size, Fwd IAT Std, Min Packet Length, Packet Length Variance, Packet Length Std, Idle Max, Max Packet Length, Bwd Packet Length Min, Idle Mean, Average Packet Size, Fwd IAT Max, Packet Length Mean, Flow IAT Max, Idle Min, Flow IAT Std \\
\midrule
CIC18 & min\_seg\_size\_forward, Packet Length Mean, Bwd Packets/s, Init\_Win\_bytes\_forward, ACK Flag Count, Fwd Packet Length Mean, Bwd Packet Length Min, Destination Port, Avg Fwd Segment Size, Min Packet Length, Init\_Win\_bytes\_backward, Average Packet Size, Flow Packets/s, Fwd Packet Length Min, Total Fwd Packets, Bwd IAT Total, Fwd Packet Length Max, Bwd Packet Length Mean, Fwd PSH Flags, Avg Bwd Segment Size \\
\midrule
LycoS17 & pkt\_len\_std, fwd\_non\_empty\_pkt\_cnt, fwd\_pkt\_hdr\_len\_min, flag\_rst, down\_up\_ratio, bwd\_pkt\_len\_std, pkt\_len\_mean, ip\_prot\_6, pkt\_len\_var, bwd\_pkt\_len\_mean, ip\_prot\_17, bwd\_pkt\_len\_max, pkt\_len\_max, bwd\_bulk\_rate\_mean, pkt\_len\_min, dst\_port, bwd\_pkt\_hdr\_len\_min, flag\_SYN, fwd\_pkt\_len\_min, fwd\_tcp\_init\_win\_bytes \\
\midrule
LycoS18 & fwd\_tcp\_init\_win\_bytes, fwd\_iat\_tot, pkt\_per\_s, flag\_fin, down\_up\_ratio, fwd\_pkt\_hdr\_len\_min, bwd\_pkt\_hdr\_len\_min, bwd\_pkt\_per\_s, fwd\_pkt\_per\_s, ip\_prot\_6, ip\_prot\_17, flag\_SYN, fwd\_pkt\_cnt, dst\_port, flow\_duration, pkt\_len\_min, bwd\_iat\_tot, fwd\_pkt\_len\_min, fwd\_iat\_std, flag\_rst \\
\bottomrule
\end{tabular}
\normalsize
\caption{List of best features calculated with mRMR for binary experiments.}
\label{tab:best_features_binary}
\end{table}

\begin{figure*}
    \centering    
    \subfloat{
        \includegraphics[width=.33\linewidth]{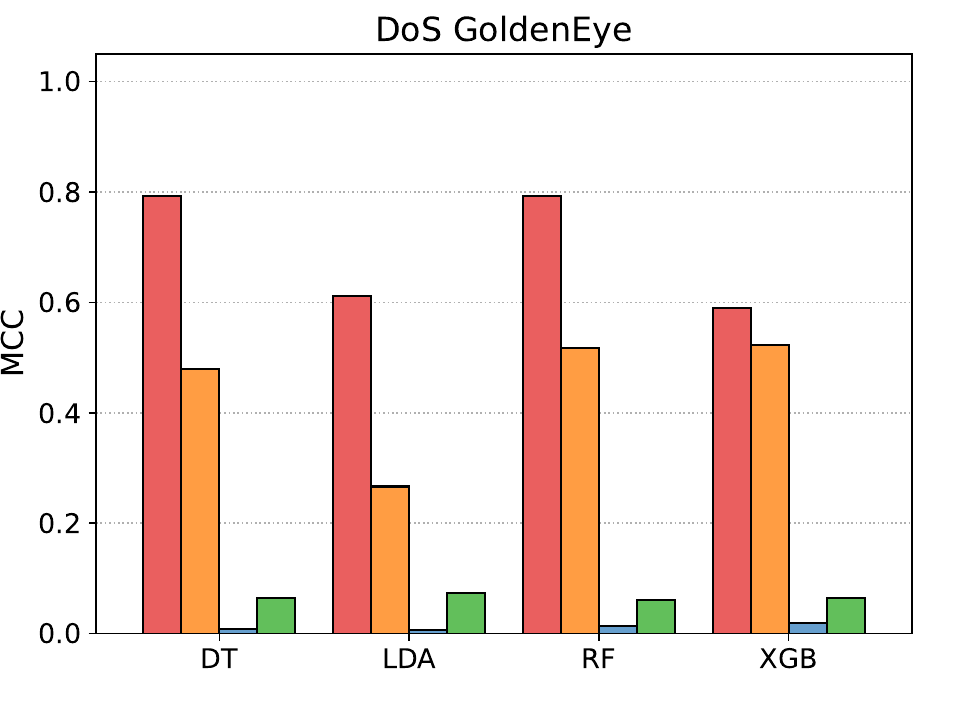}
    }
    \subfloat{
        \includegraphics[width=.33\linewidth]{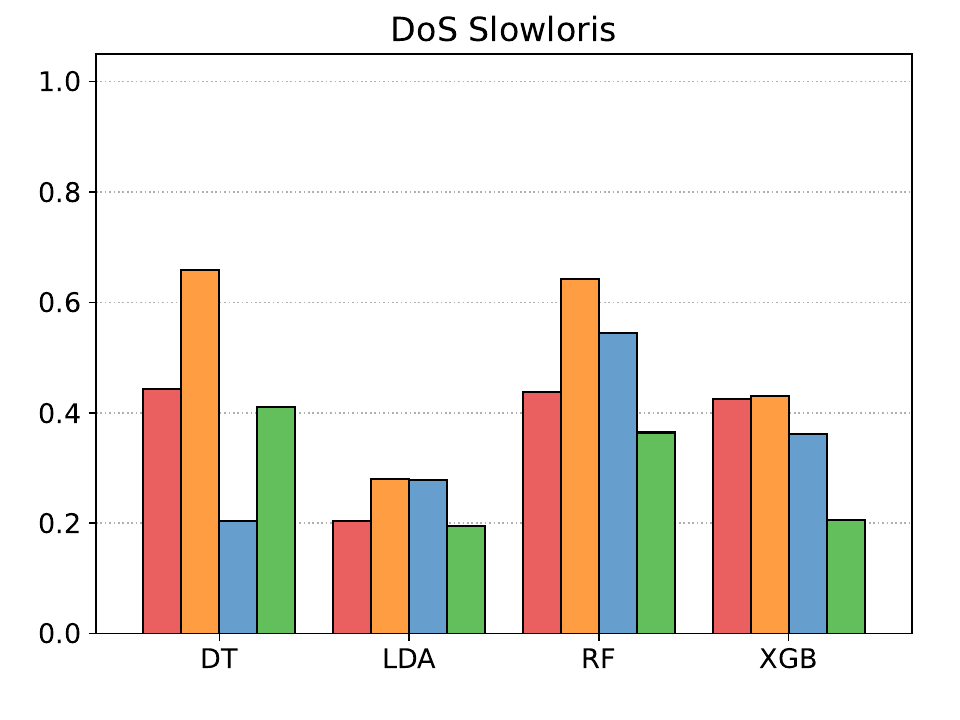}
    }
    \subfloat{
        \includegraphics[width=.33\linewidth]{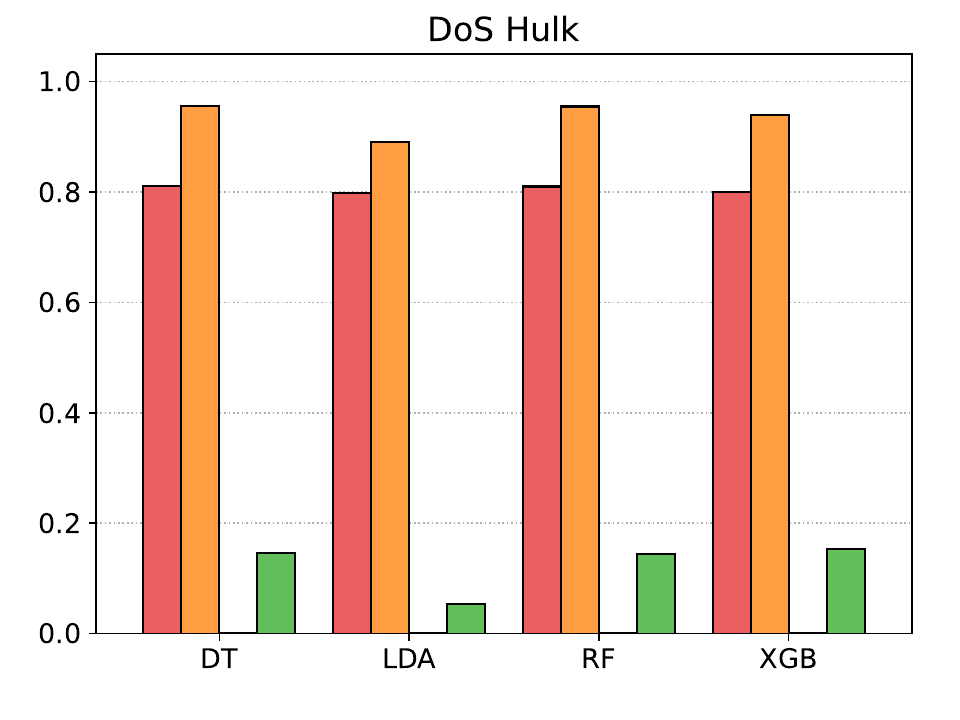}
    }\\
        \subfloat{
        \includegraphics[width=.33\linewidth]{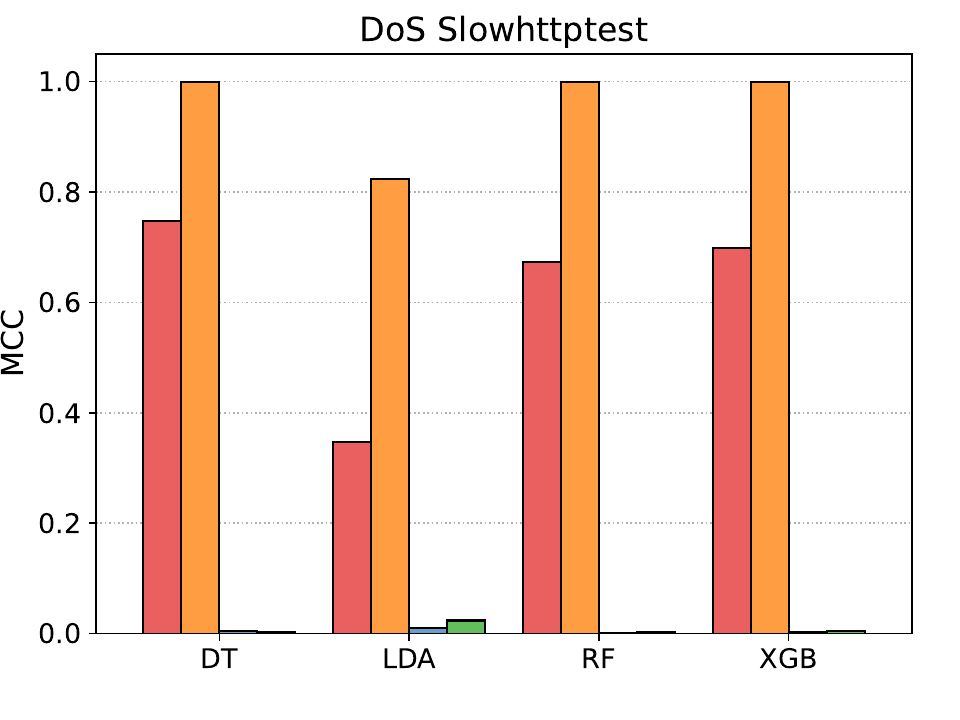}
    }
    \subfloat{
        \includegraphics[width=.33\linewidth]{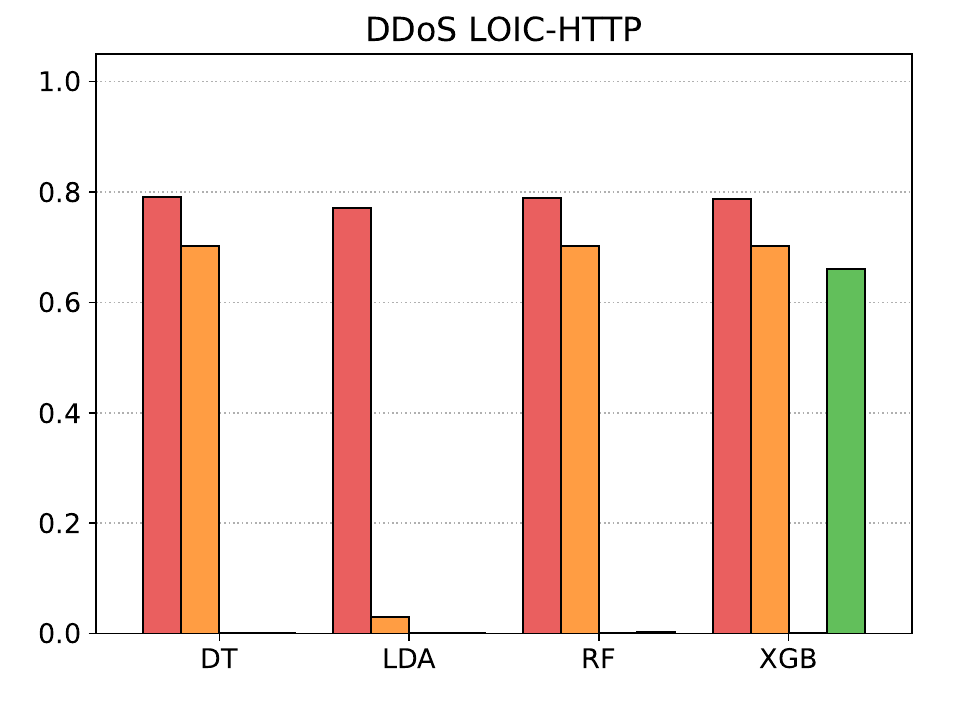}
    }
    \subfloat{
        \includegraphics[width=.33\linewidth]{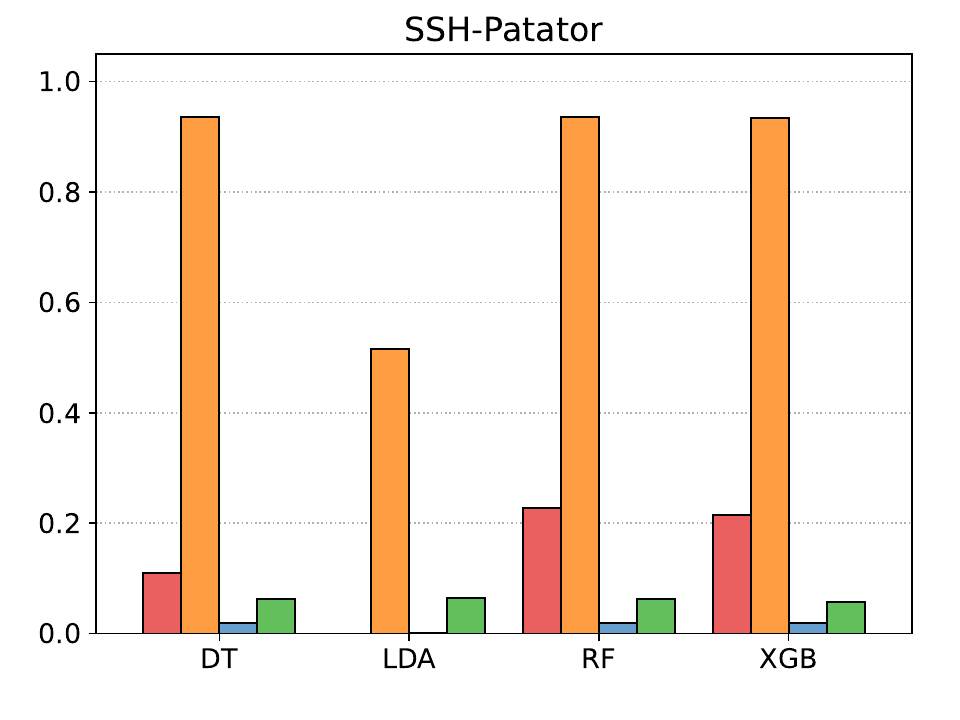}
    }\\
    \subfloat{
        \includegraphics[width=.33\linewidth]{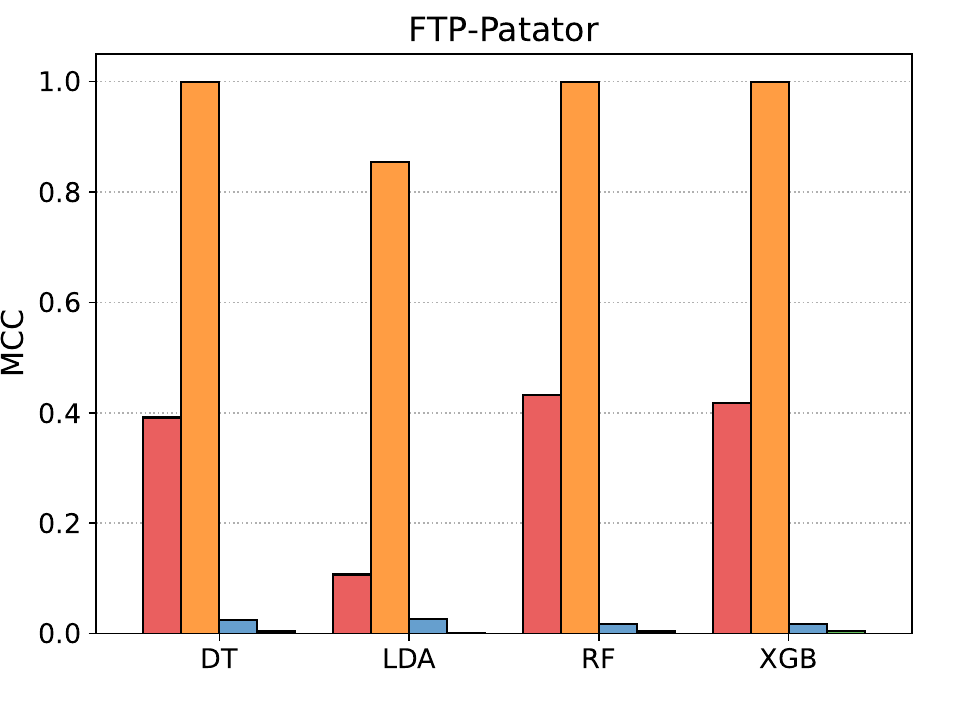}
    }
    \subfloat{
        \includegraphics[width=.33\linewidth]{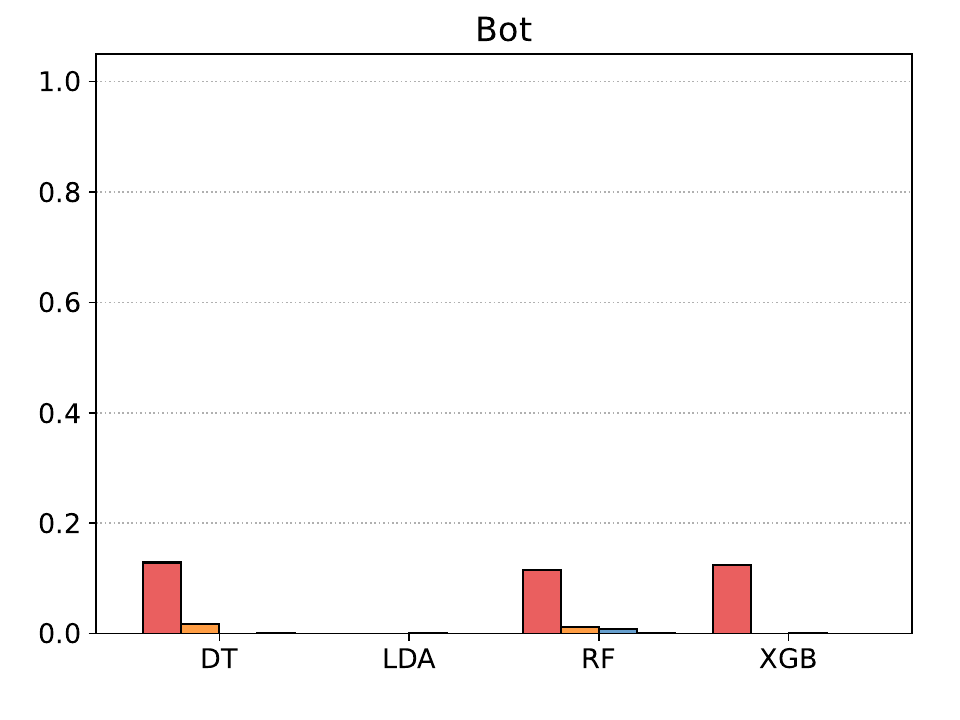}
    }
    \subfloat{
        \includegraphics[width=.33\linewidth]{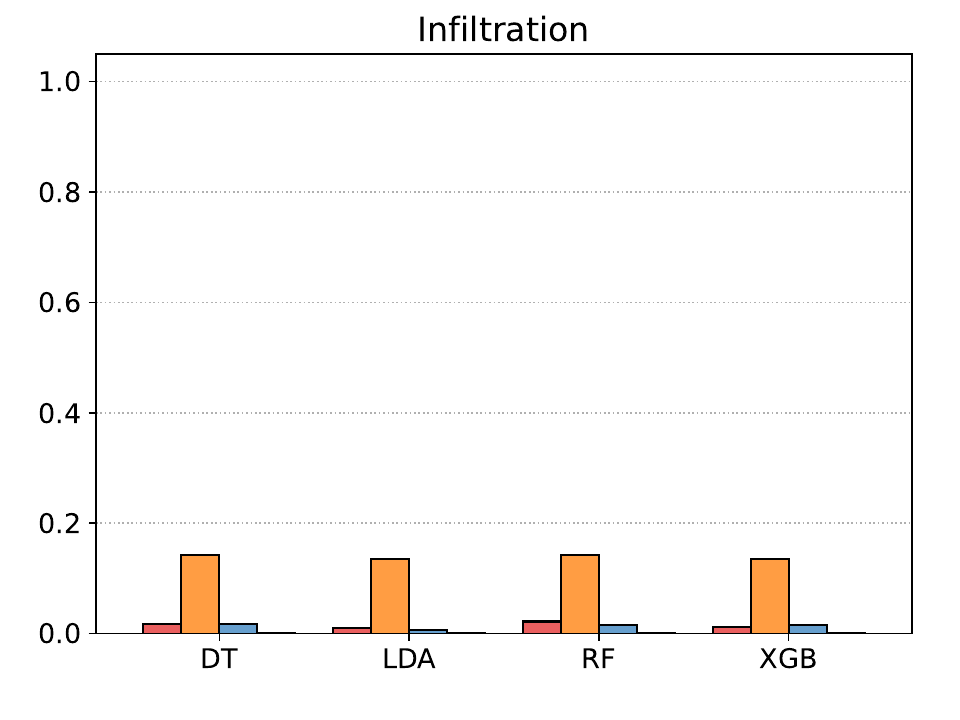}
    }\\
    \subfloat{
        \includegraphics[width=.33\linewidth]{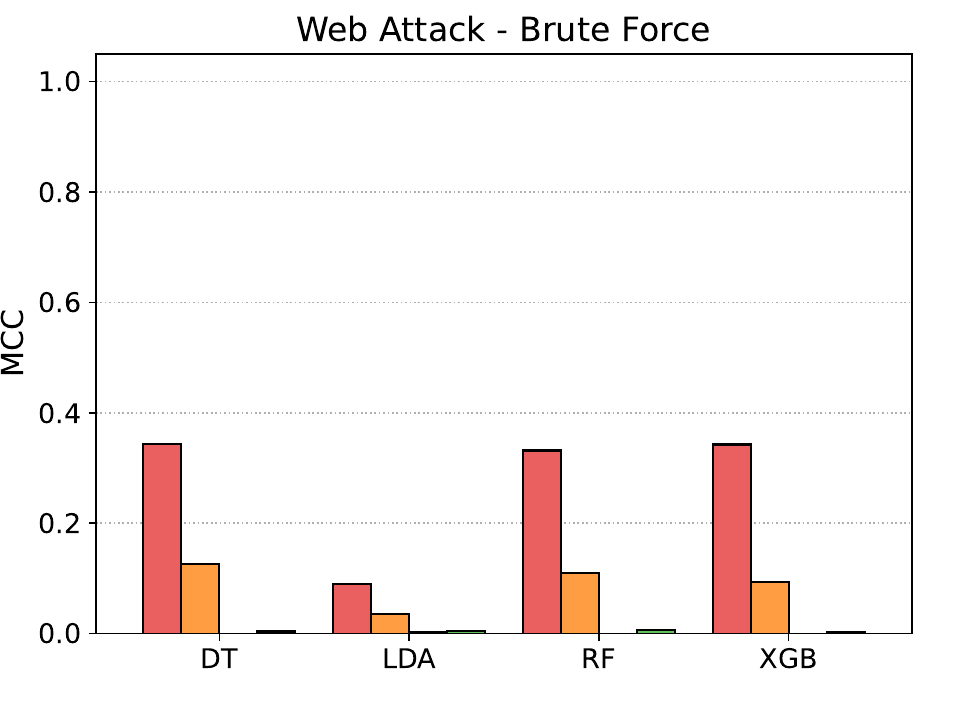}
    }
    \subfloat{
        \includegraphics[width=.33\linewidth]{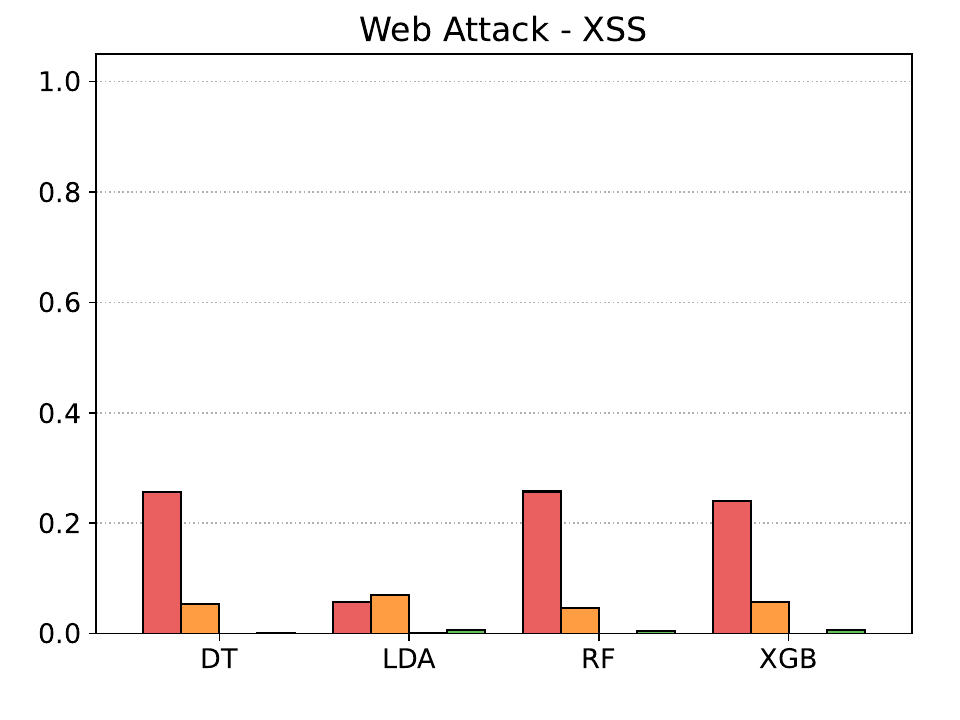}
    }
    \subfloat{
        \includegraphics[width=.33\linewidth]{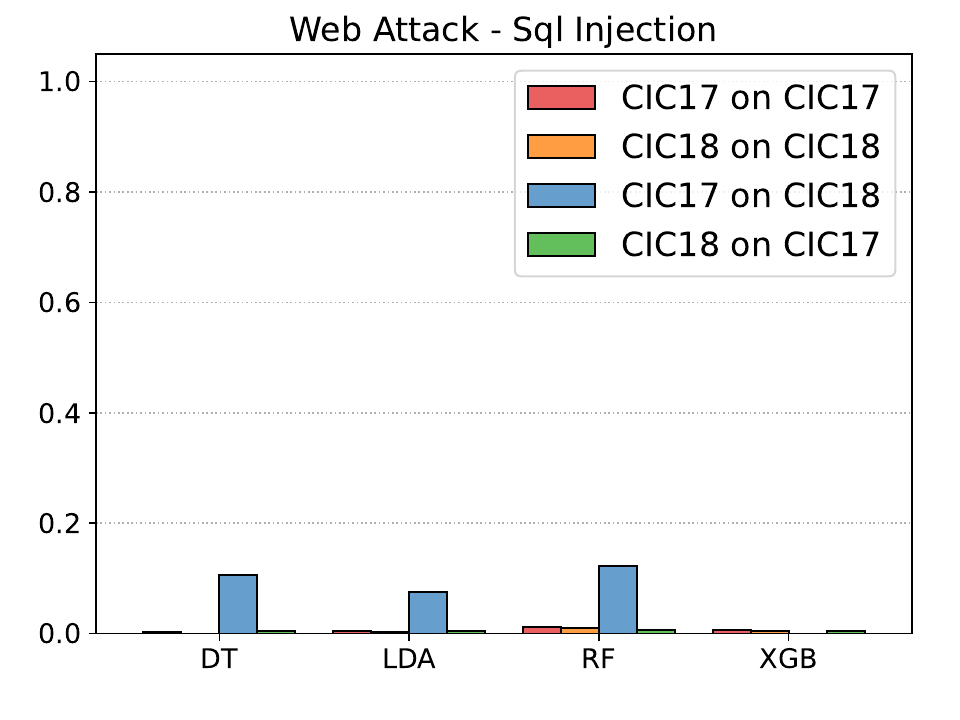}
    }
    \caption{MCC value obtained for each classifier and train-test combination with CIC datasets in single-attack experiments using the best two features selected with mRMR.}
    \label{fig:singleClass_cic_fs}
\end{figure*}

\begin{figure*}
    \centering    
    \subfloat{
        \includegraphics[width=.33\linewidth]{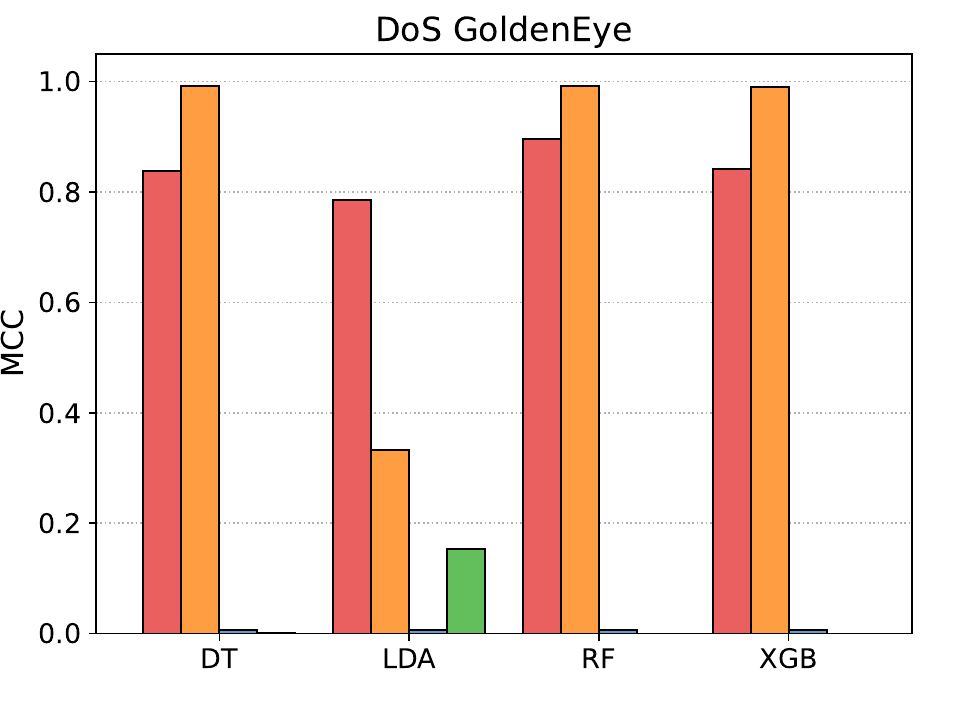}
    }
    \subfloat{
        \includegraphics[width=.33\linewidth]{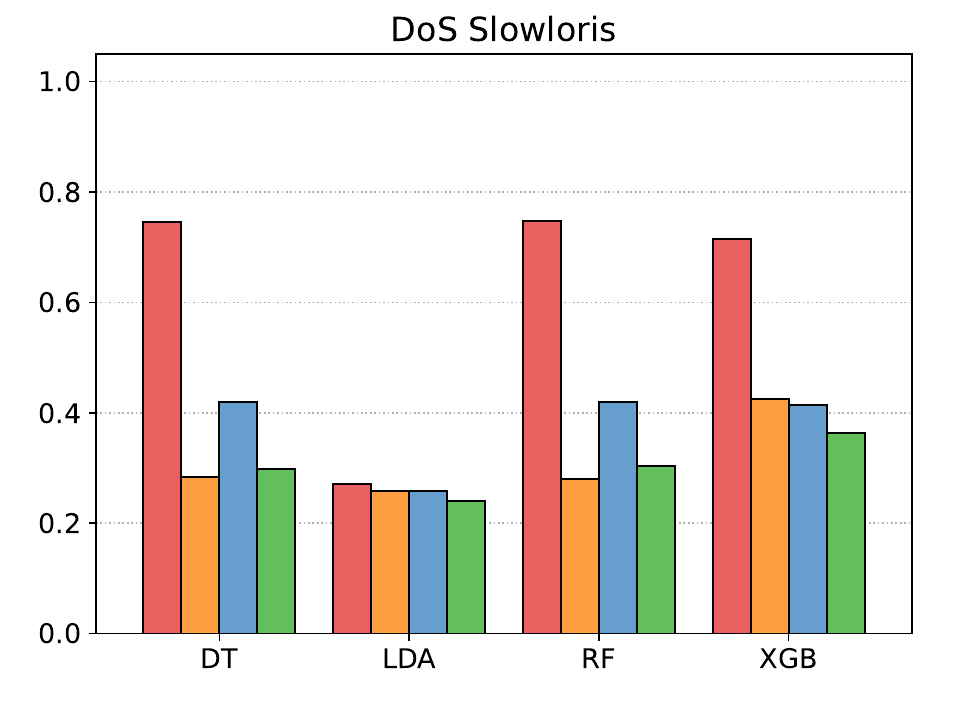}
    }
    \subfloat{
        \includegraphics[width=.33\linewidth]{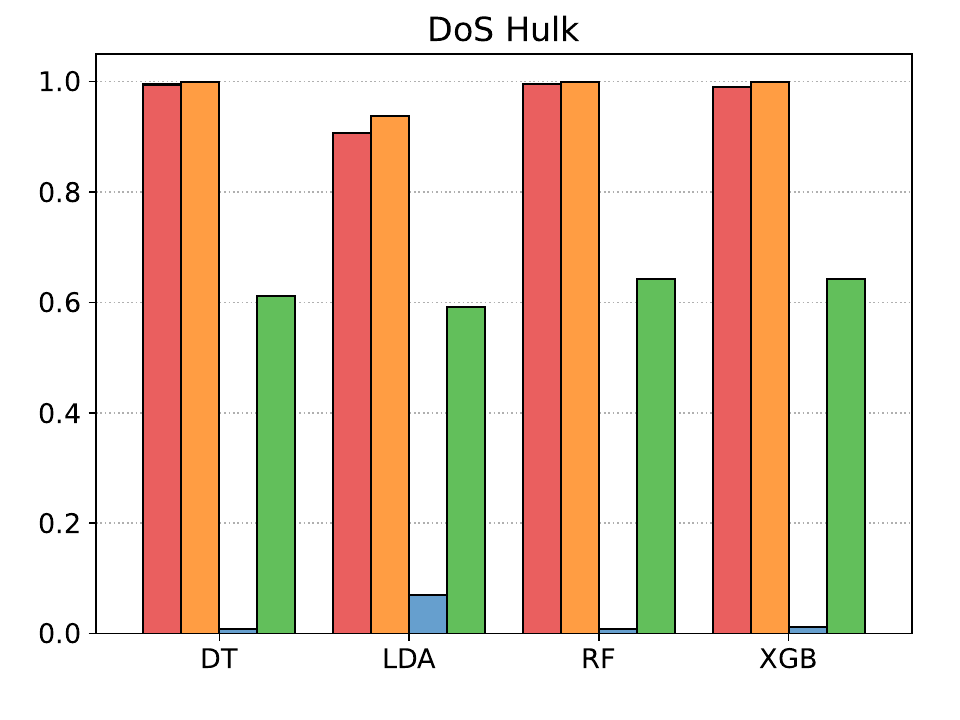}
    }\\
        \subfloat{
        \includegraphics[width=.33\linewidth]{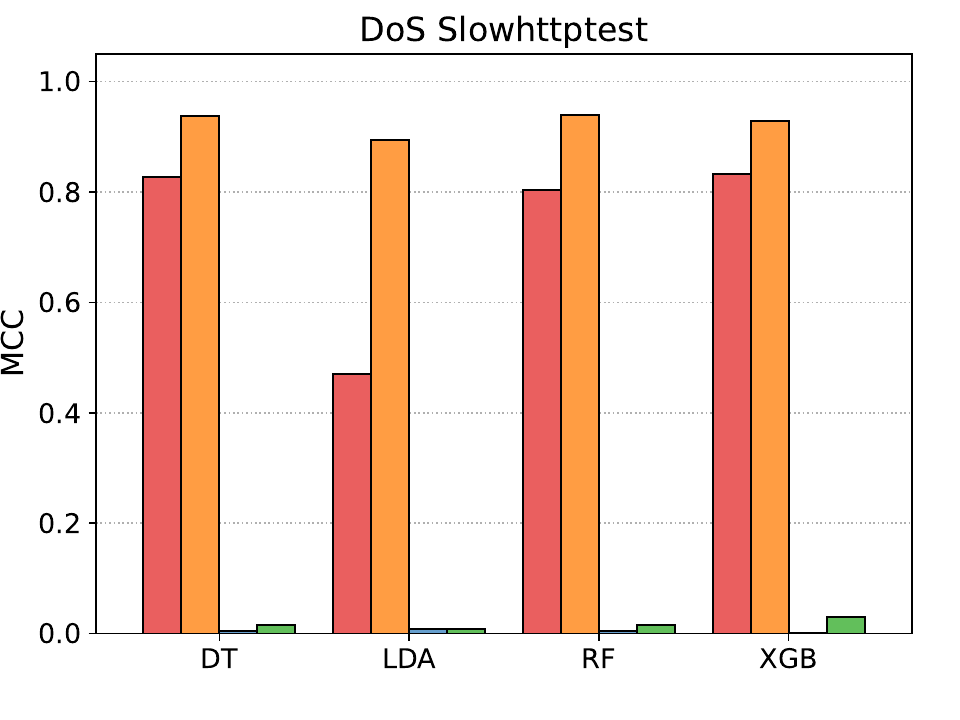}
    }
    \subfloat{
        \includegraphics[width=.33\linewidth]{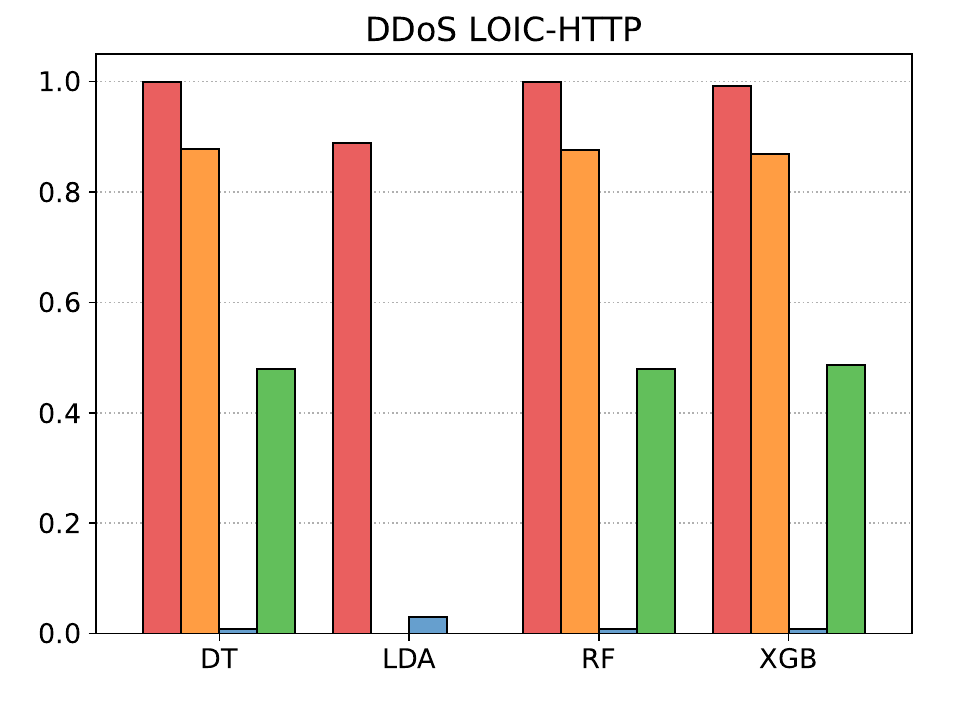}
    }
    \subfloat{
        \includegraphics[width=.33\linewidth]{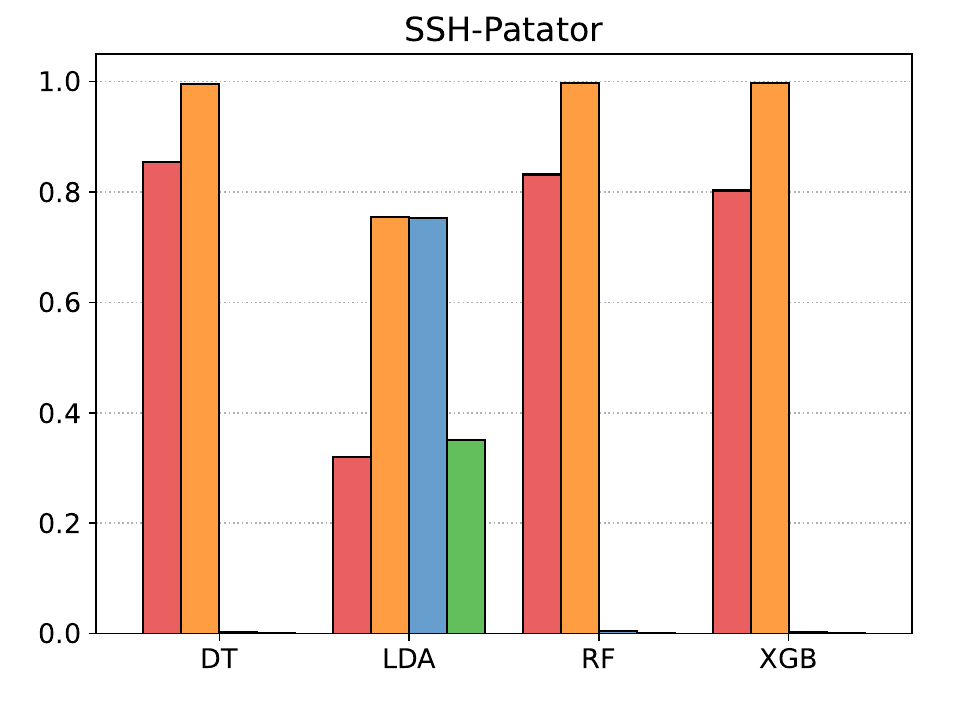}
    }\\
    \subfloat{
        \includegraphics[width=.33\linewidth]{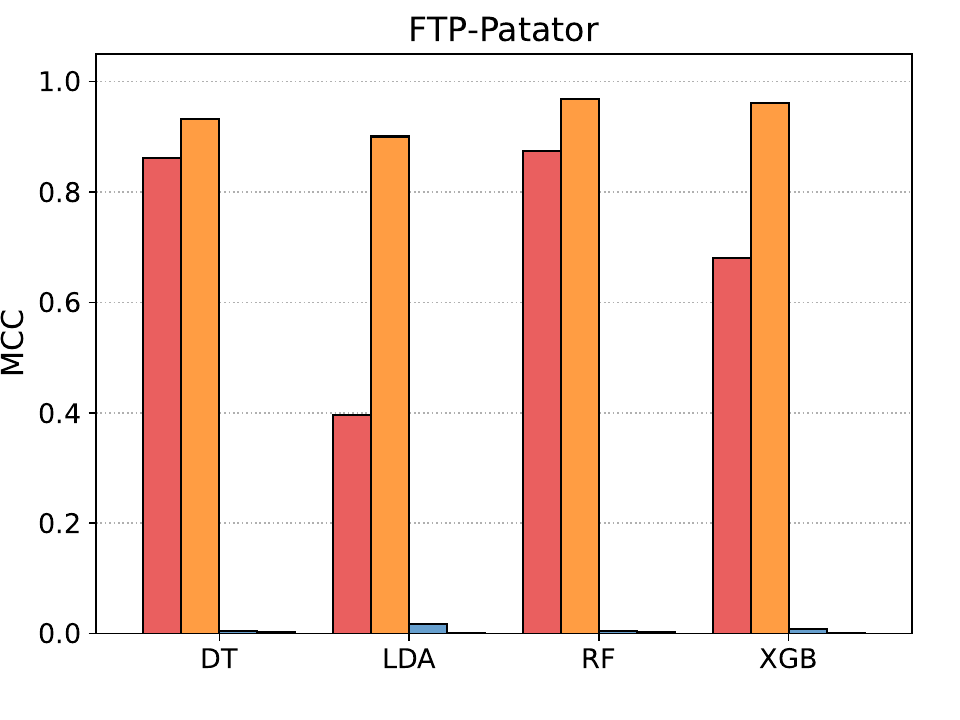}
    }
    \subfloat{
        \includegraphics[width=.33\linewidth]{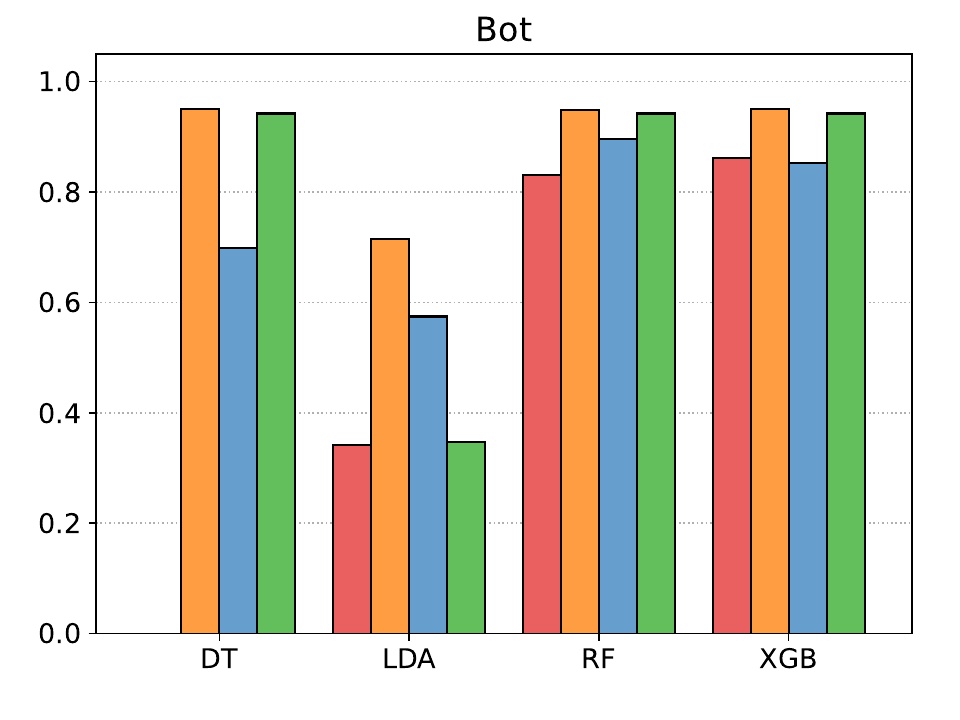}
    }
    \subfloat{
        \includegraphics[width=.33\linewidth]{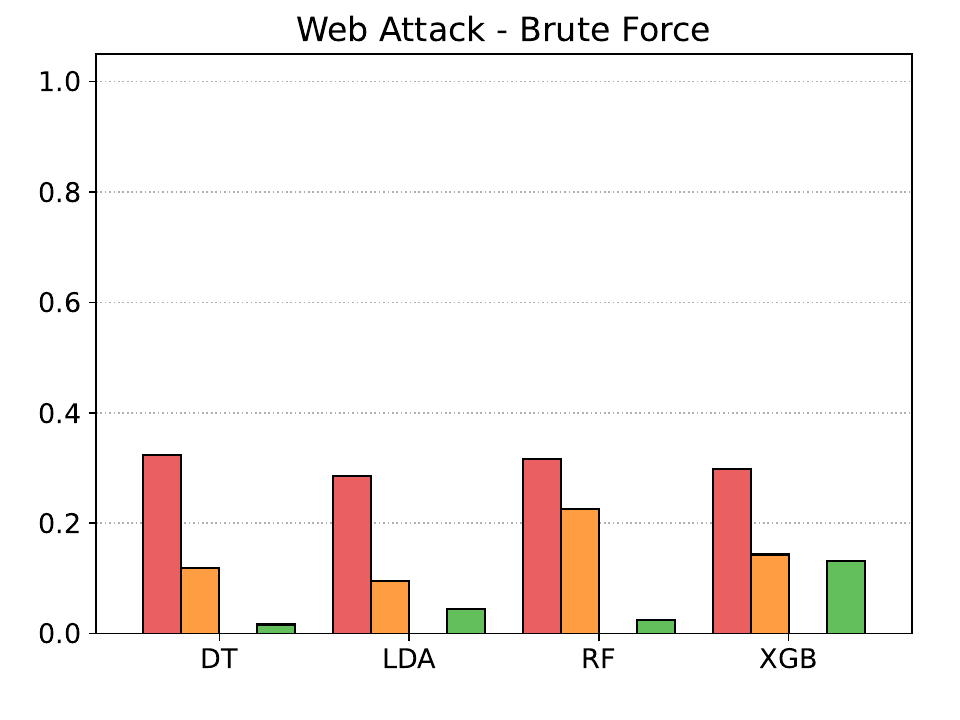}
    }\\
    \subfloat{
        \includegraphics[width=.33\linewidth]{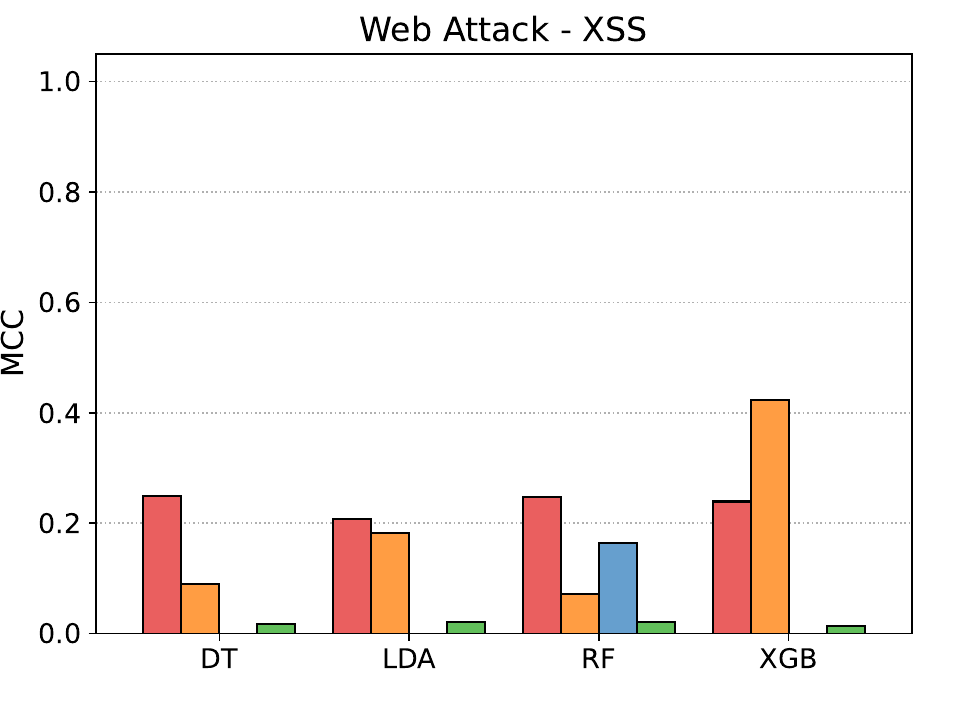}
    }
    \subfloat{
        \includegraphics[width=.33\linewidth]{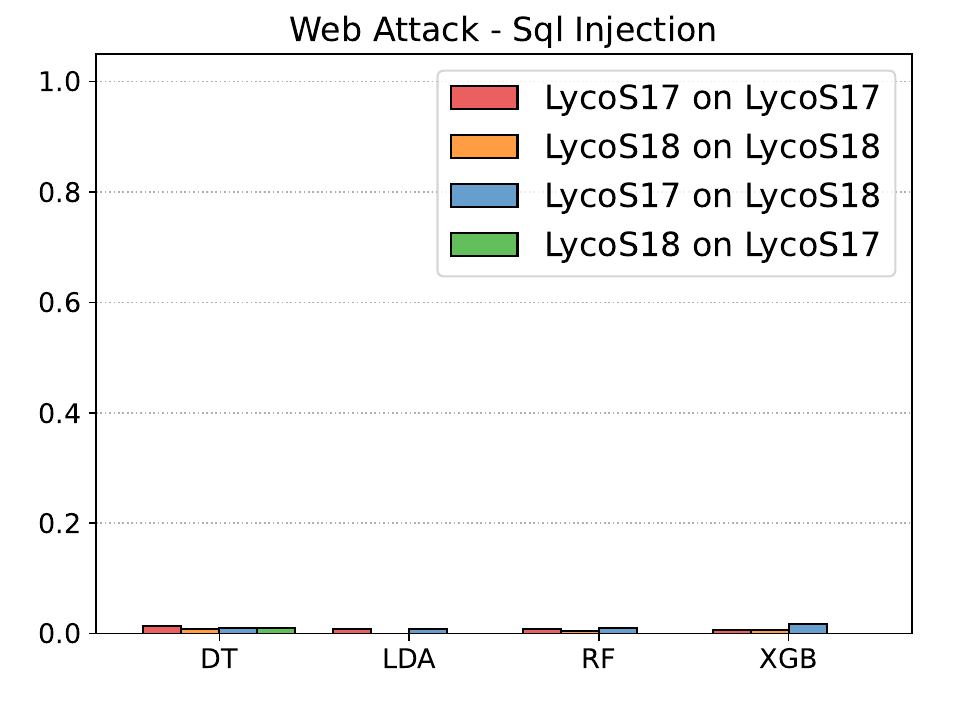}
    }
    \subfloat{
        \includegraphics[width=.33\linewidth]{blank.png}
    }
    \caption{MCC value obtained for each classifier and train-test combination with LycoS datasets in single-attack experiments using the best two features selected with mRMR.}
    \label{fig:singleClass_lycos_fs}
\end{figure*}

Figures~\ref{fig:singleClass_cic_fs} and~\ref{fig:singleClass_lycos_fs} show the results obtained by single-attack experiments using only the best two features listed in Table~\ref{tab:best_two_features} for each combination of attack and training set.
When examining the DoS Slowloris attack in the context of the CIC datasets, we note a degradation in within-dataset performance, coupled with an increase in cross-dataset performance.
The Bot attack is not effectively classified in any of the configurations with the CIC datasets, whereas when using the LycoS datasets it is the second attack, alongside DoS Slowloris, to achieve similar within-dataset and cross-dataset performance. This discrepancy is likely attributed to errors in the labeling of Bot samples in the CIC datasets that have been corrected in the LycoS datasets.

\begin{table*}[t]
\centering
\footnotesize
\begin{tabular}{p{0.18\linewidth}p{0.18\linewidth}p{0.2\linewidth}p{0.18\linewidth}p{0.18\linewidth}}
\textbf{Attack} & \textbf{CIC17} & \textbf{CIC18} & \textbf{LycoS17} & \textbf{LycoS18} \\
\toprule
\multirow{2}*{DoS GoldenEye} & Bwd Packet Length Std & min\_seg\_size\_forward & bwd\_pkt\_len\_std & fwd\_pkt\_hdr\_len\_min \\
 & Flow IAT Mean & Flow Bytes/s & pkt\_len\_std & fwd\_tcp\_init\_win\_bytes \\
\cmidrule{1-5}
\multirow{2}*{DoS Slowloris} & Bwd IAT Mean & Bwd IAT Mean & fwd\_pkt\_hdr\_len\_min & bwd\_iat\_mean \\
 & Init\_Win\_bytes\_forward & Down/Up Ratio & bwd\_iat\_mean & fwd\_pkt\_hdr\_len\_min \\
\cmidrule{1-5}
\multirow{2}*{DoS Hulk} & Bwd Packet Length Std & min\_seg\_size\_forward & pkt\_len\_mean & fwd\_pkt\_hdr\_len\_min \\
 & ACK Flag Count & Packet Length Mean & flow\_duration & bwd\_iat\_min \\
\cmidrule{1-5}
\multirow{2}*{DoS Slowhttptest} & Active Mean & Bwd Packets/s & active\_mean & pkt\_per\_s \\
 & Bwd IAT Std & min\_seg\_size\_forward & bwd\_iat\_std & bwd\_pkt\_per\_s \\
\cmidrule{1-5}
\multirow{2}*{DDoS LOIC-HTTP} & Bwd Packet Length Std & Avg Fwd Segment Size & pkt\_len\_mean & flag\_fin \\
 & Bwd Packet Length Mean & Fwd Header Length & bwd\_iat\_tot & iat\_std \\
\cmidrule{1-5}
\multirow{2}*{SSH-Patator} & URG Flag Count & min\_seg\_size\_forward & fwd\_flag\_psh & fwd\_flag\_psh \\
 & Flow IAT Mean & Average Packet Size & down\_up\_ratio & iat\_mean \\
\cmidrule{1-5}
\multirow{2}*{FTP-Patator} & Fwd PSH Flags & Bwd Packets/s & down\_up\_ratio & pkt\_per\_s \\
 & Flow IAT Max & min\_seg\_size\_forward & fwd\_iat\_min & bwd\_pkt\_per\_s \\
\cmidrule{1-5}
\multirow{2}*{Bot} & PSH Flag Count & RST Flag Count & dst\_port & dst\_port \\
 & Fwd IAT Mean & Idle Min & fwd\_non\_empty\_pkt\_cnt & flag\_fin \\
\cmidrule{1-5}
\multirow{2}*{Web Attack - Brute Force} & Init\_Win\_bytes\_backward & Fwd Packet Length Mean & bwd\_pkt\_hdr\_len\_min & fwd\_flag\_psh \\
 & ACK Flag Count & Destination Port & down\_up\_ratio & idle\_max \\
\cmidrule{1-5}
\multirow{2}*{Web Attack - XSS} & Init\_Win\_bytes\_backward & Total Length of Fwd Packets & bwd\_pkt\_hdr\_len\_min & fwd\_flag\_psh \\
 & ACK Flag Count & Fwd Header Length & down\_up\_ratio & flag\_psh \\
\cmidrule{1-5}
\multirow{2}*{Web Attack - Sql Injection} & Bwd Packet Length Std & Packet Length Variance & bwd\_pkt\_len\_std & flag\_fin \\
 & min\_seg\_size\_forward & Down/Up Ratio & fwd\_pkt\_len\_mean & bwd\_flag\_psh \\
\cmidrule{1-5}
\multirow{2}*{Infiltration} & Flow Duration & Bwd Packets/s & & \\
 & ACK Flag Count & Fwd URG Flags & & \\
\cmidrule{1-5}

\end{tabular}
\caption{Best two features selected with mRMR for each attack and dataset.}
\label{tab:best_two_features}
\end{table*}

\subsection{Visual feature space analysis}
\label{sec:Visual feature space analysis}
Experiments using only two features confirm the trend already shown using the entire feature set, with high within-dataset and low cross-dataset performances. This provides a sound foundation for the visual analysis of the datasets in the 2D feature subspace.

Tables~\ref{tab:unique_instancies_mrmr17} and \ref{tab:unique_instancies_mrmr18} report the counts of unique values for the two top features derived with mRMR for each attack type. Table~\ref{tab:unique_instancies_mrmr17} refers to features calculated using LycoS17 as training set while Table~\ref{tab:unique_instancies_mrmr18} refers to features calculated using Lycos18 as training set. Comparing this information with those reported in Table~\ref{tab:datasets_composition} clearly shows strong sample redundancy for most classes in both datasets. 
Several attacks are represented by a single repeated instance. This is aggravated, but not due to the use of only two features. For instance, DoS Slowhttptest attack has only 56 different instances in the LycoS18 dataset using the entire feature set and a total number of samples of 105,550. The aforementioned conditions may negatively affect ML applications where the data should ideally exhibit sufficient variability.

\begin{table}
    \centering
    \footnotesize
    \begin{tabular}{@{}p{0.38\linewidth}>{\raggedleft}p{0.1\linewidth}>{\raggedleft}p{0.1\linewidth}>{\raggedleft}p{0.1\linewidth}>{\raggedleft\arraybackslash}p{0.12\linewidth}@{}}
\multirow{2}*{ } & \multicolumn{2}{c}{\textbf{LycoS17}} & \multicolumn{2}{c}{\textbf{LycoS18}} \\
 & $\phi_1$ & $\phi_2$ & $\phi_1$ & $\phi_2$ \\
\toprule
DoS GoldenEye & 98 & 4,915 & 7 & 2,381 \\
DoS Slowloris & 3 & 2,891 & 2 & 6,415 \\
DoS Hulk & 2,899 & 82,383 & 1,908 & 378,266 \\
DoS Slowhttptest & 3,215 & 1,142 & 1 & 1 \\
DDoS LOIC-HTTP & 11 & 90,365 & 1 & 251,453 \\
\textbf{SSH-Patator} & 15 & 56 & 2 & 33 \\
FTP-Patator & 5 & 230 & 1 & 1 \\
\textbf{Bot} & 1 & 22 & 1 & 35 \\
\textbf{Web Attack-Brute Force} & 2 & 13 & 2 & 11 \\
Web Attack - XSS & 2 & 15 & 2 & 10 \\
Web Attack - Sql Injection & 9 & 10 & 37 & 38 \\
\bottomrule
    \end{tabular}
    \normalsize
    \caption{Number of unique values for each of the two best features, named $\phi_1$ and $\phi_2$, calculated with mRMR using LycoS17 as training set (fourth column of Table~\ref{tab:best_two_features}). In bold are highlighted the attacks for which in both datasets the number of unique instances is less than 5\% of the total.}
    \label{tab:unique_instancies_mrmr17}
\end{table}

\begin{table}
    \centering
    \footnotesize
    \begin{tabular}{@{}p{0.38\linewidth}>{\raggedleft}p{0.1\linewidth}>{\raggedleft}p{0.1\linewidth}>{\raggedleft}p{0.1\linewidth}>{\raggedleft\arraybackslash}p{0.12\linewidth}@{}}
\multirow{2}*{ } & \multicolumn{2}{c}{\textbf{LycoS17}} & \multicolumn{2}{c}{\textbf{LycoS18}} \\
 & $\phi_1$ & $\phi_2$ & $\phi_1$ & $\phi_2$ \\
\toprule
\textbf{DoS GoldenEye} & 2 & 1 & 1 & 2 \\
DoS Slowloris & 2,891 & 3 & 6,415 & 2 \\
\textbf{DoS Hulk} & 2 & 829 & 1 & 48,024 \\
DoS Slowhttptest & 1,492 & 1,076 & 56 & 56 \\
DDoS LOIC-HTTP & 2 & 95,683 & 1 & 289,327 \\
SSH-Patator & 15 & 2,956 & 2 & 77,713 \\
FTP-Patator & 3,988 & 3,973 & 54 & 54 \\
\textbf{Bot} & 1 & 1 & 1 & 2 \\
\textbf{Web Attack-Brute Force} & 10 & 1 & 7 & 1 \\
Web Attack - XSS & 7 & 7 & 6 & 7 \\
Web Attack - Sql Injection & 1 & 1 & 1 & 4 \\
\bottomrule
    \end{tabular}
    \normalsize
    \caption{Number of unique values for each of the two best features, named $\phi_1$ and $\phi_2$, calculated with mRMR using LycoS18 as training set (fifth column of Table~\ref{tab:best_two_features}). In bold are highlighted the attacks for which in both datasets the number of unique instances is less than 5\% of the total.}
    \label{tab:unique_instancies_mrmr18}
\end{table}

\begin{figure*}
    \centering    
    \subfloat{
        \includegraphics[width=.33\linewidth]{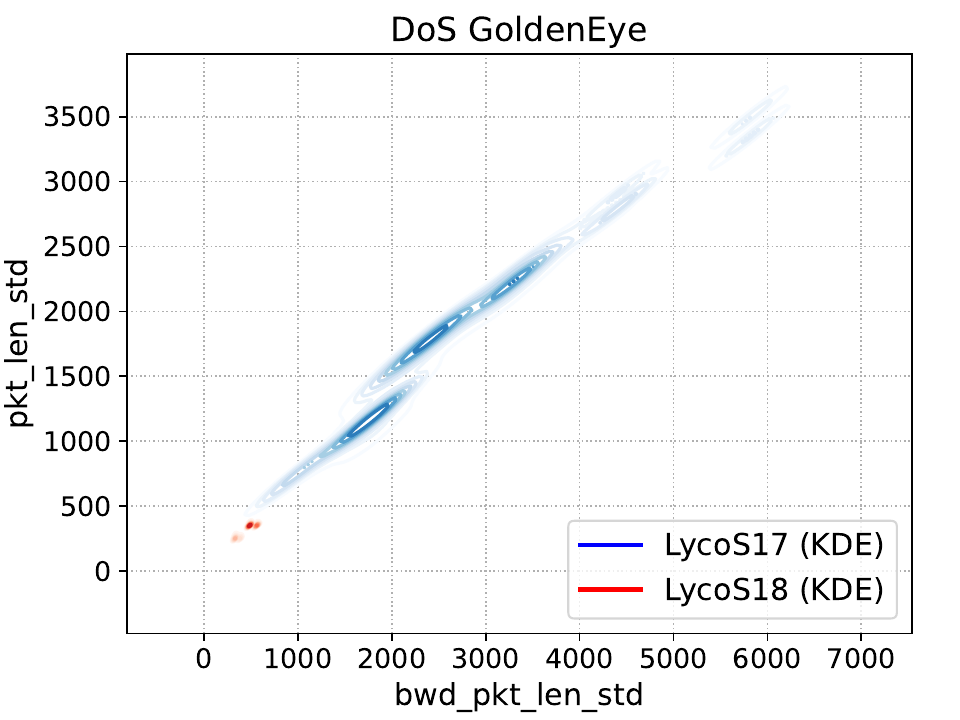}
    }
    \subfloat{
        \includegraphics[width=.33\linewidth]{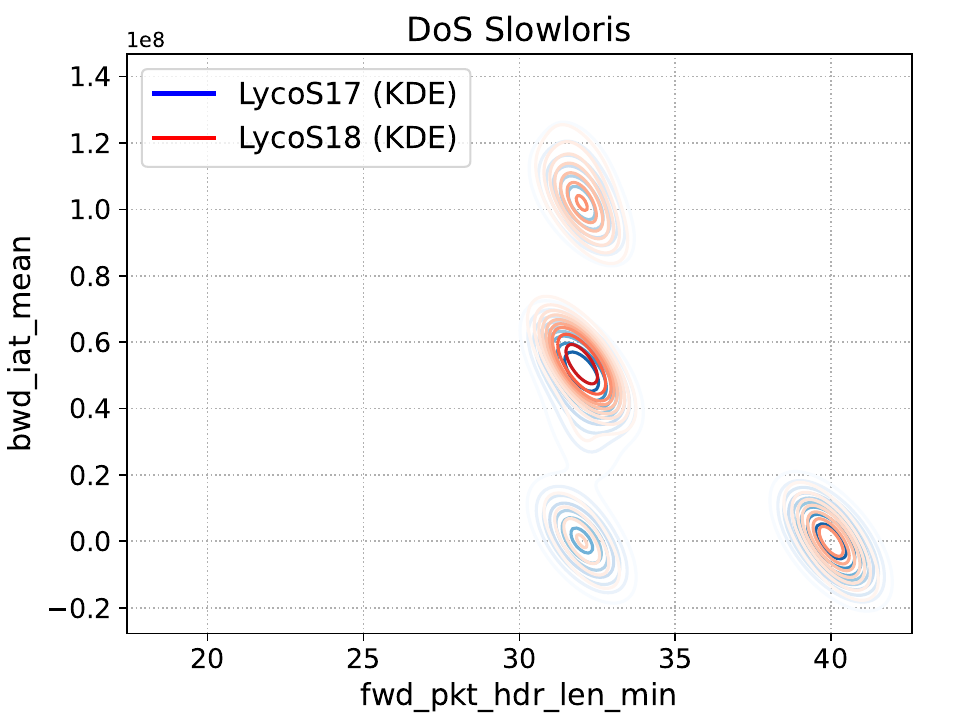}
    }
    \subfloat{
        \includegraphics[width=.33\linewidth]{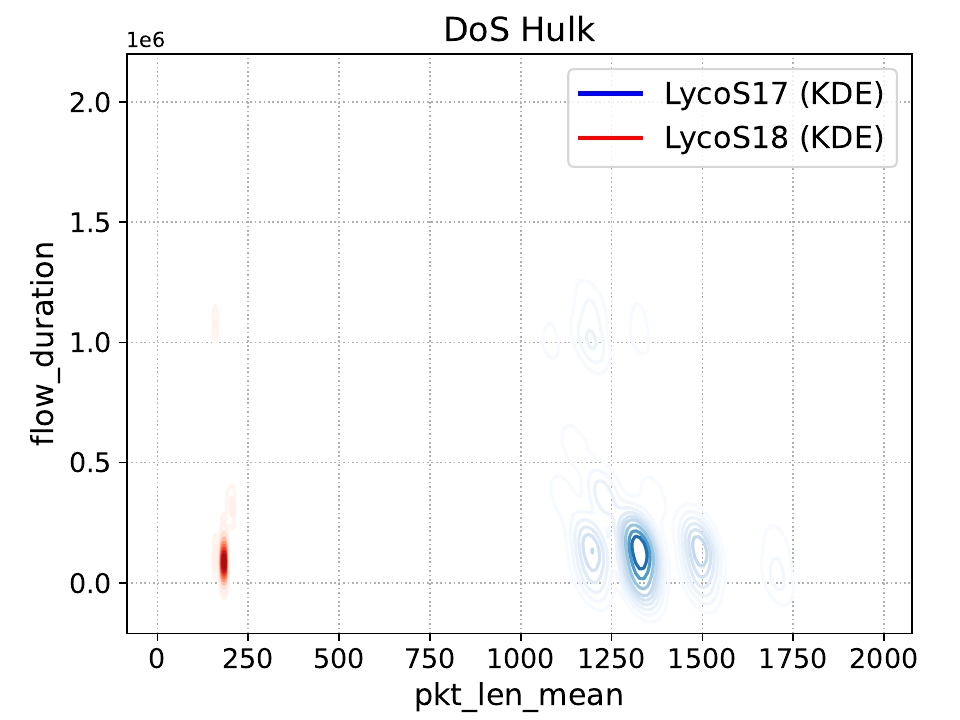}
    }\\
        \subfloat{
        \includegraphics[width=.33\linewidth]{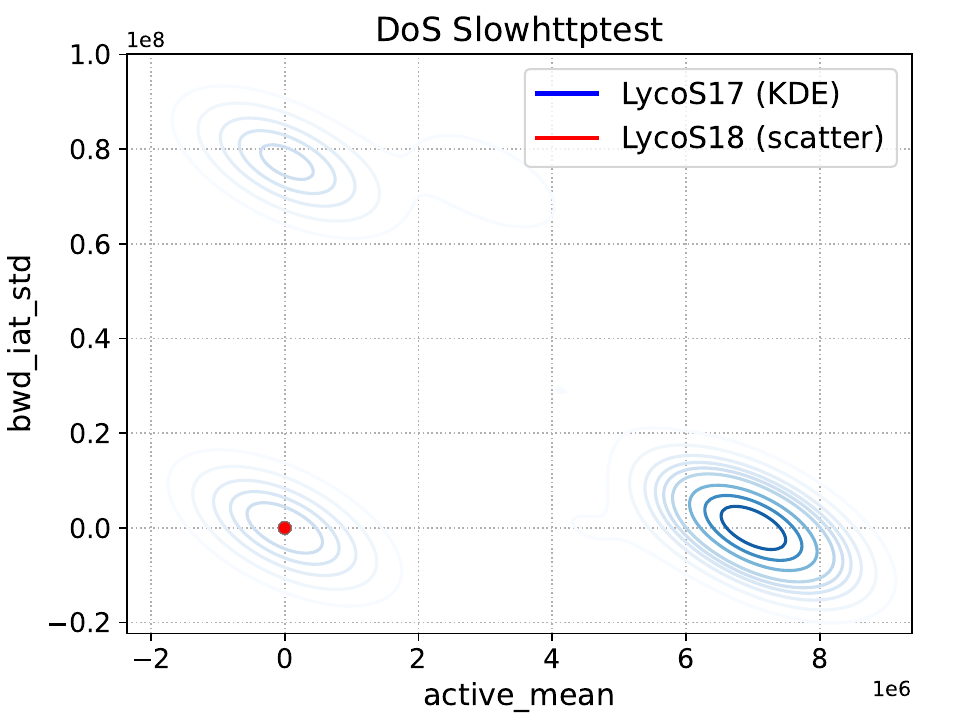}
    }
    \subfloat{
        \includegraphics[width=.33\linewidth]{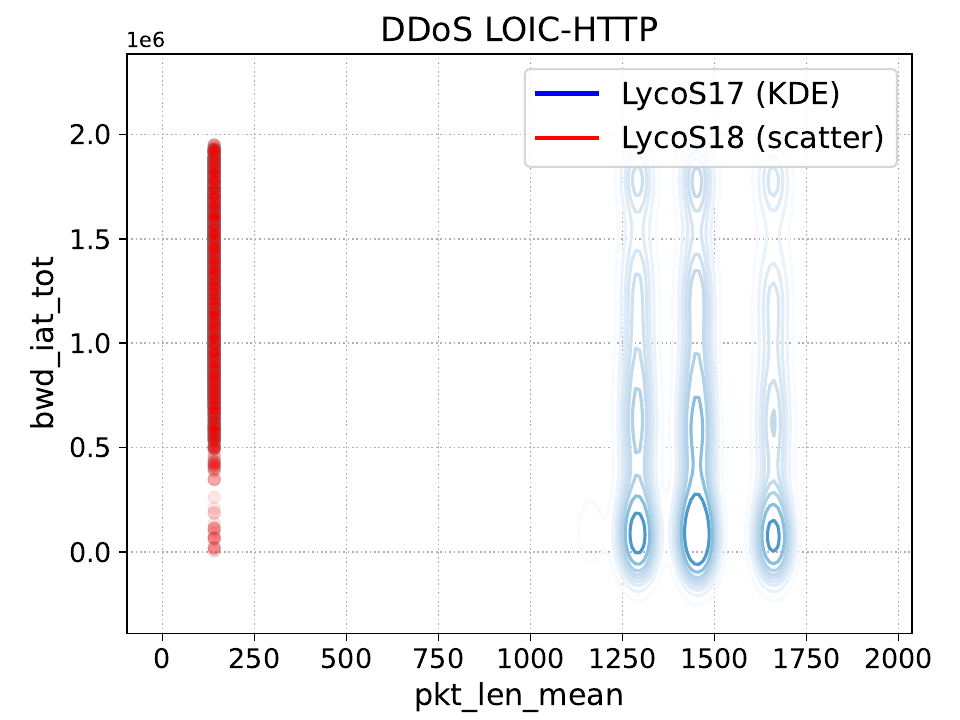}
    }
    \subfloat{
        \includegraphics[width=.33\linewidth]{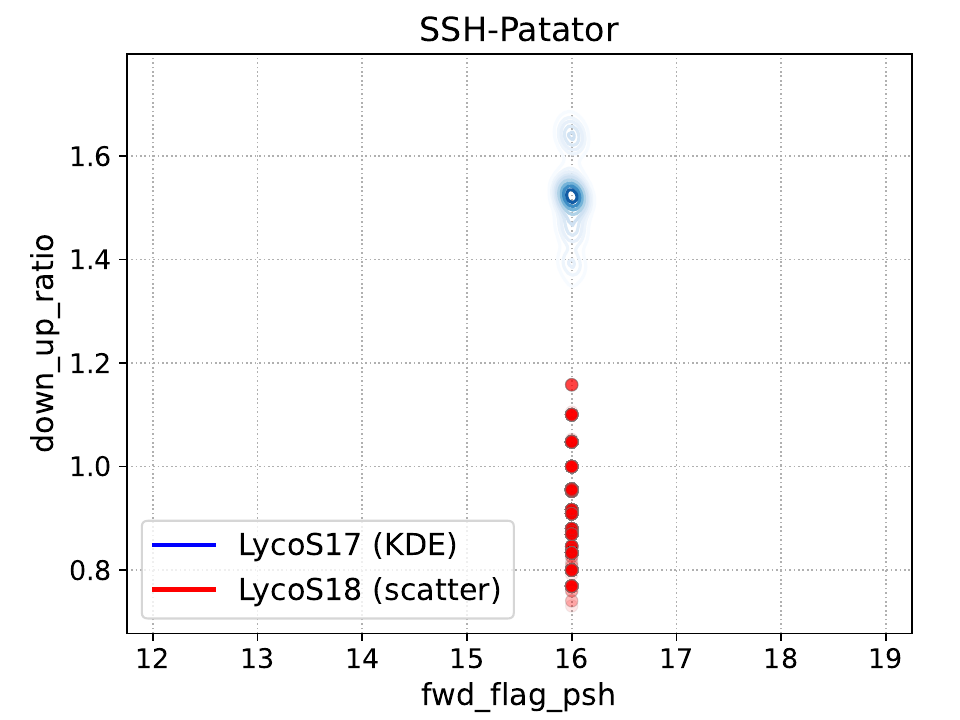}
    }\\
    \subfloat{
        \includegraphics[width=.33\linewidth]{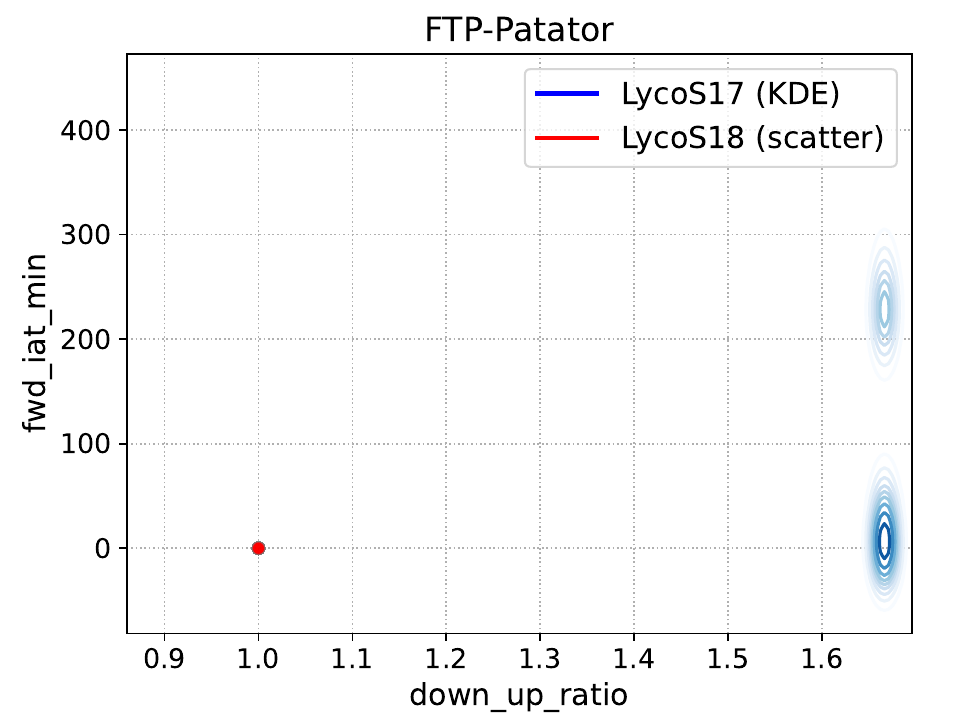}
    }
    \subfloat{
        \includegraphics[width=.33\linewidth]{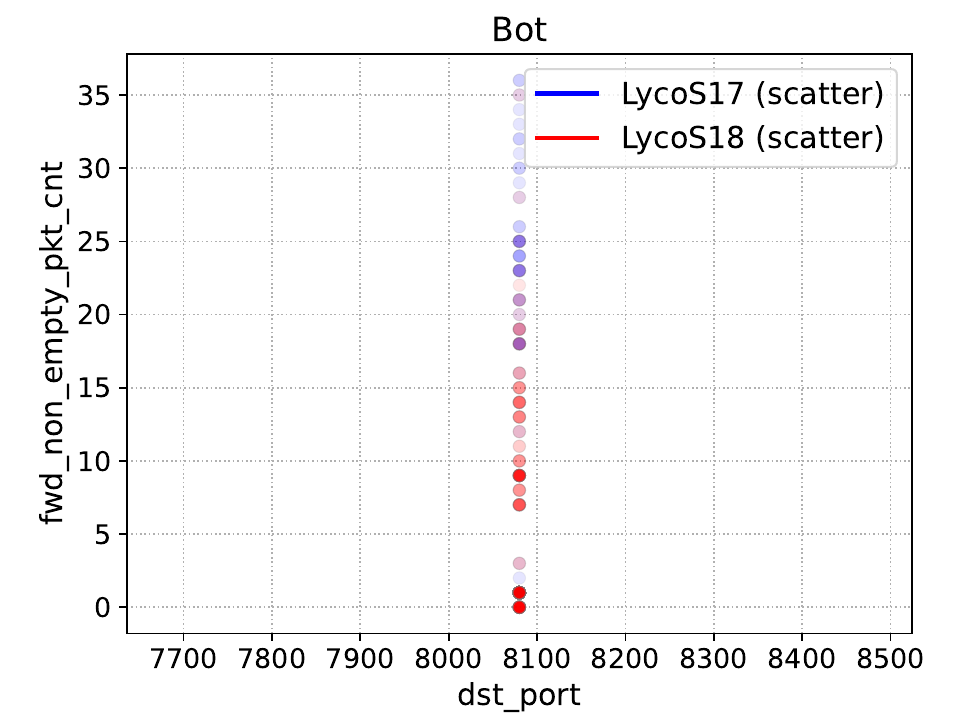}
    }
    \subfloat{
        \includegraphics[width=.33\linewidth]{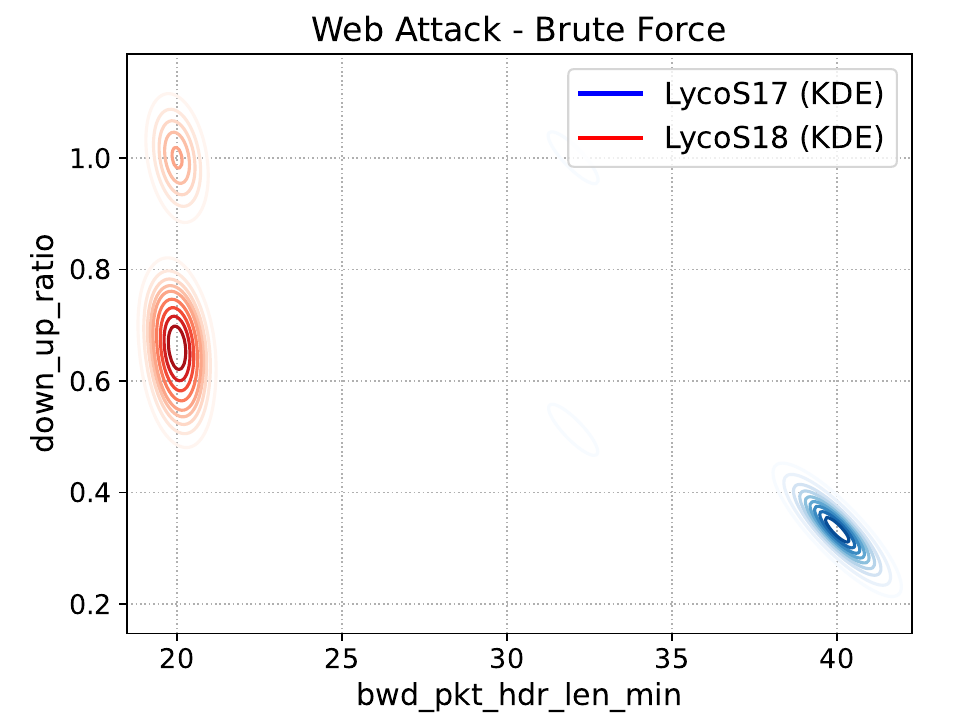}
    }\\
    \subfloat{
        \includegraphics[width=.33\linewidth]{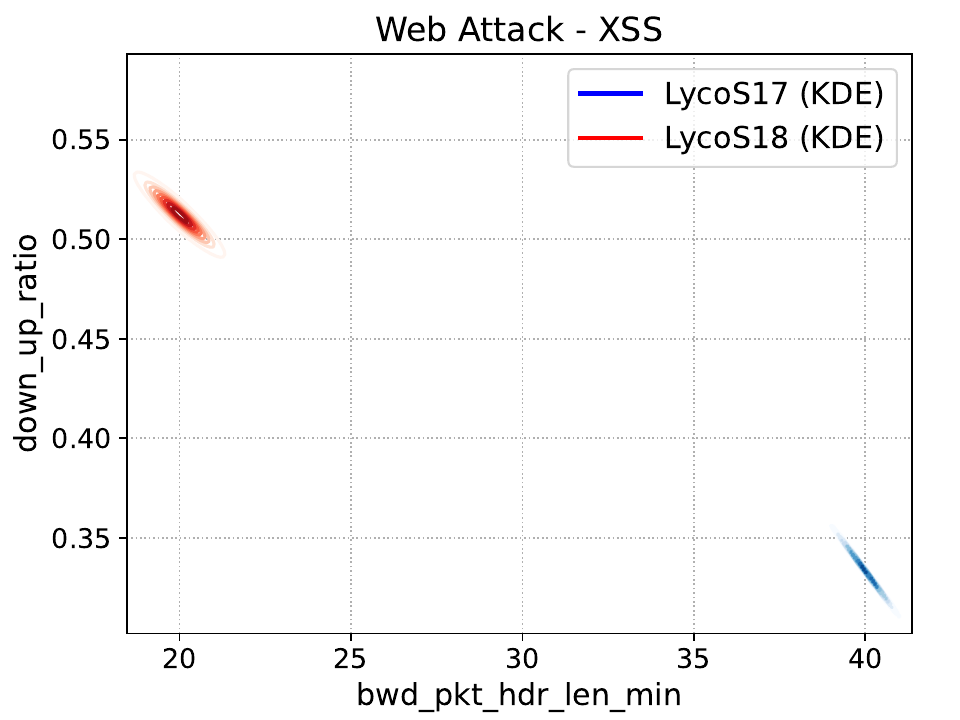}
    }
    \subfloat{
        \includegraphics[width=.33\linewidth]{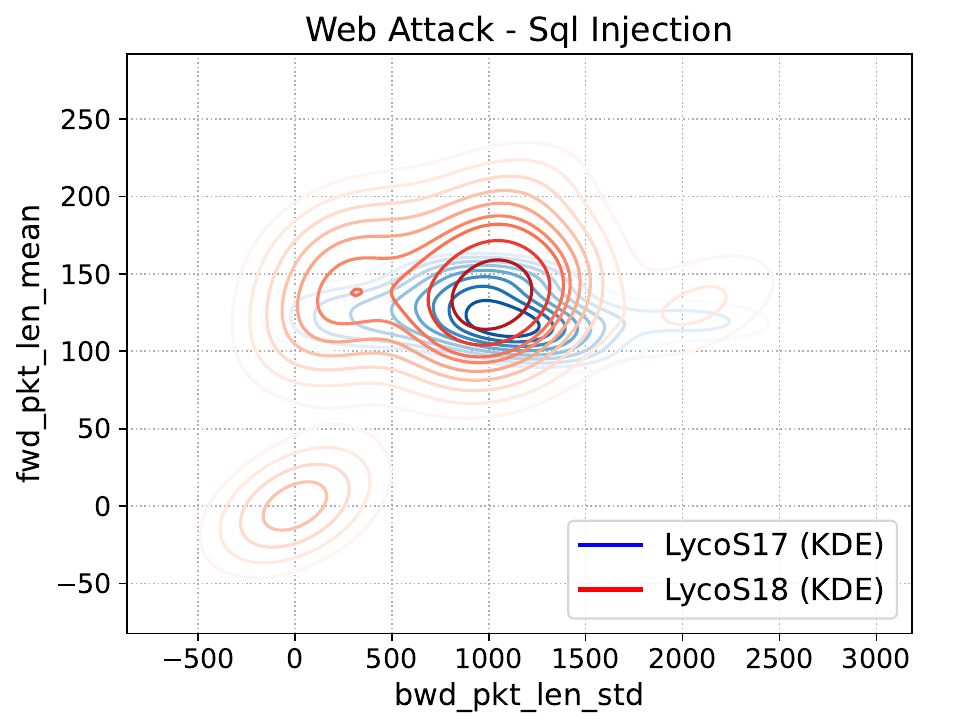}
    }
    \subfloat{
        \includegraphics[width=.33\linewidth]{blank.png}
    }
    \caption{Attacks distribution of the LycoS17 and LycoS18 datasets in the subspace of the best two features obtained with mRMR using Lycos2017 as training set. The legend, in parentheses, shows the type of representation employed to visualize the distribution.}
    \label{fig:kde_mrmr}
\end{figure*}

Figure~\ref{fig:kde_mrmr} depicts the distributions computed with KDE for each attack, utilizing the top two features derived from the LycoS17 dataset. 
All 11 attacks, with the exception of DoS Slowloris, show substantially different distributions for the two datasets. In some cases, such as DoS Hulk, DoS Slowhttptest and FTP-Patator, the distribution in the LycoS17 dataset is spread over multiple clusters while the counterpart is represented by a single point or a distribution with a significantly lower variance.
It is also interesting to note that the DoS Slowhttptest attack in the LycoS18 dataset, despite having only one possible instance, is actually a subclass of the same attack in the LycoS17 dataset. Other cases, such as that of DoS GoldenEye, show similar shape distribution, but with a large difference in scale.
As to DDoS LOIC-HTTP attack, the pkt\_len\_mean feature was selected as the most informative, and it is indeed an important parameter in the execution of this attack. It can be seen how in the LycoS18 dataset this parameter always takes exactly the value 140.571429, while in the Lycos17 dataset it takes different values distributed over 3 clusters. This indicates that a different configuration was employed when generating the DDoS LOIC-HTTP attack for the two datasets. Furthermore, by arbitrarily fixing a parameter during the attack execution, it is not possible to generate traffic samples that adequately represent that specific attack, as they depict only one potential instance. The only attack that has a similar representation in both datasets is DoS Slowloris. Hence the average reduction in performance between within-dataset and cross-dataset experiments is only 38.93\%, whereas for the other attacks the average reduction is 84.36\%. 

\begin{figure*}
    \centering    
    \subfloat{
        \includegraphics[width=.33\linewidth]{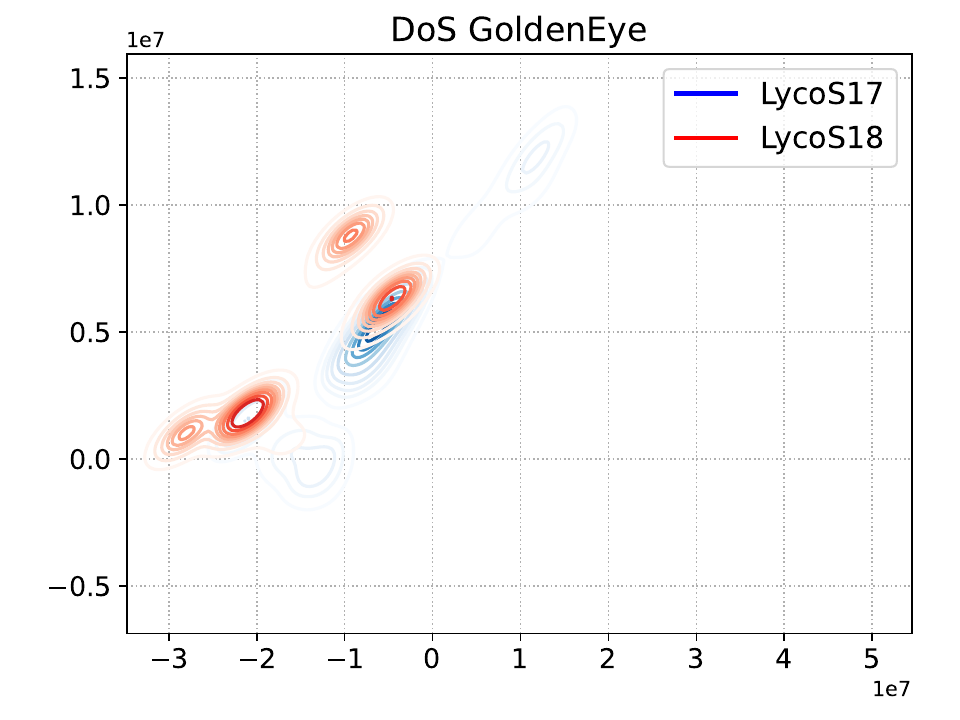}
    }
    \subfloat{
        \includegraphics[width=.33\linewidth]{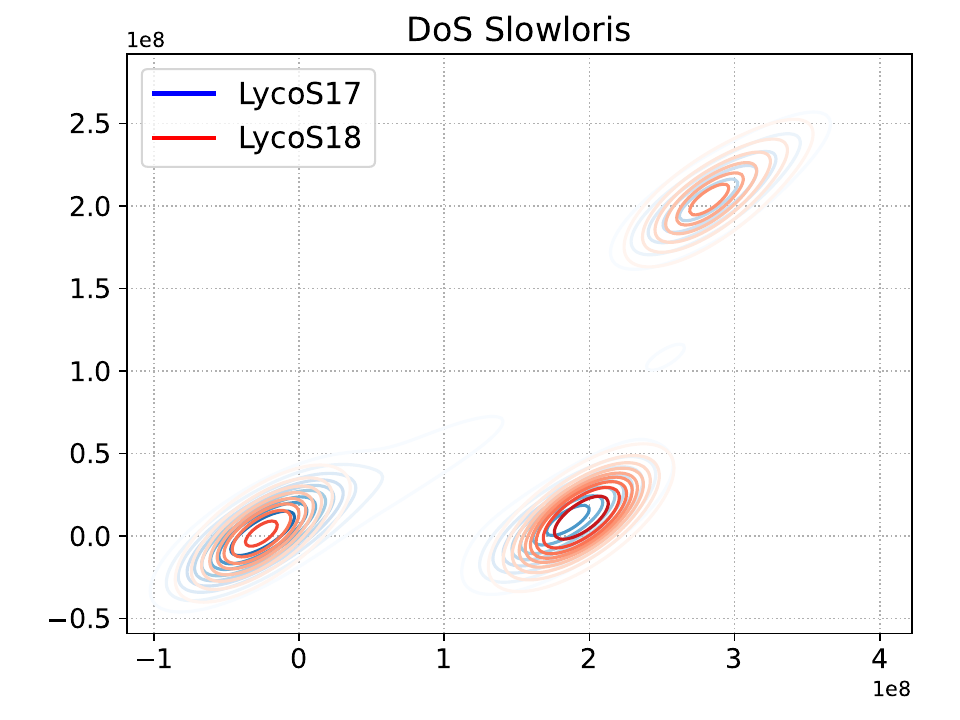}
    }
    \subfloat{
        \includegraphics[width=.33\linewidth]{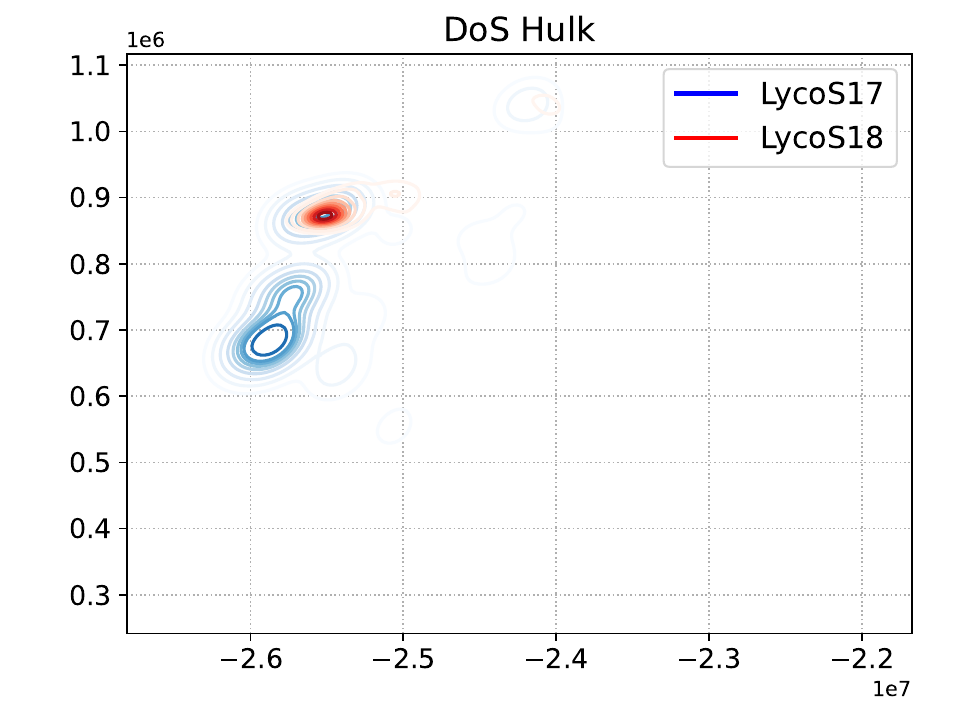}
    }\\
        \subfloat{
        \includegraphics[width=.33\linewidth]{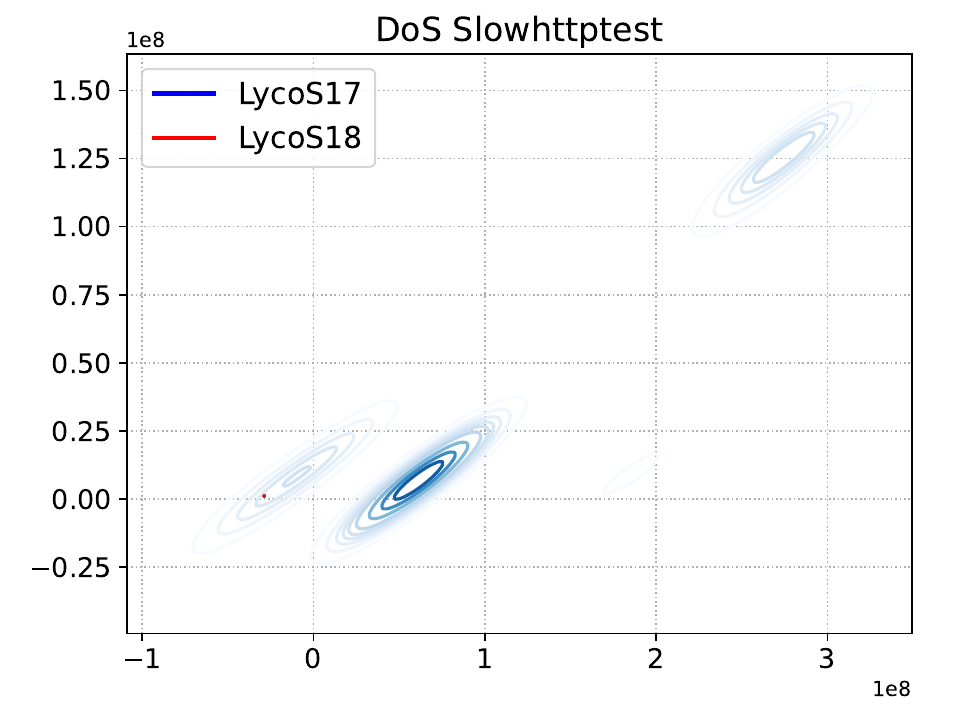}
    }
    \subfloat{
        \includegraphics[width=.33\linewidth]{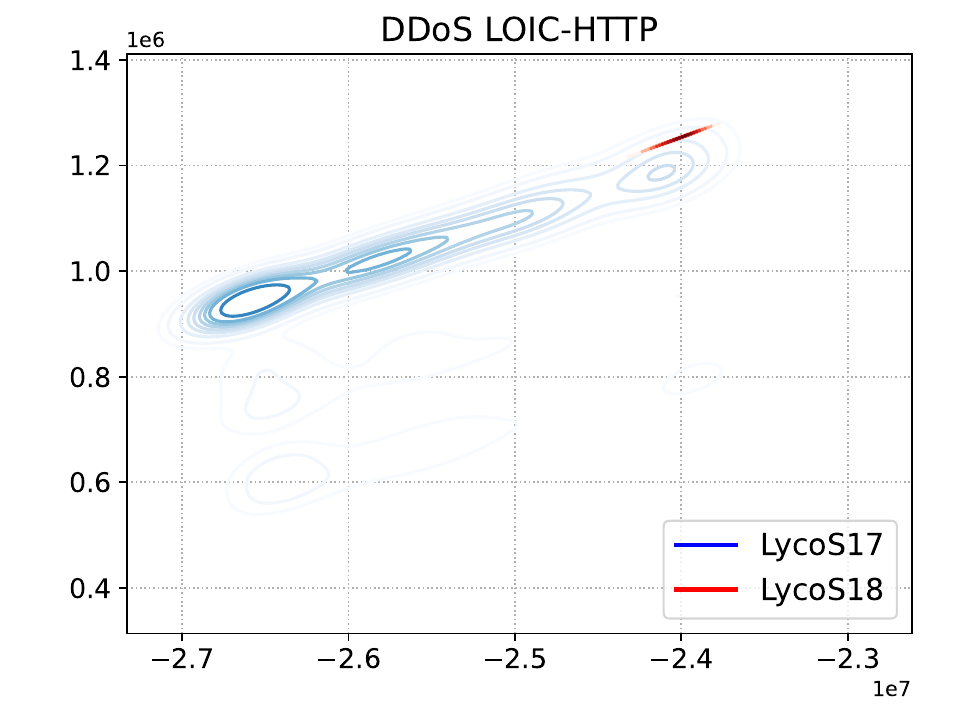}
    }
    \subfloat{
        \includegraphics[width=.33\linewidth]{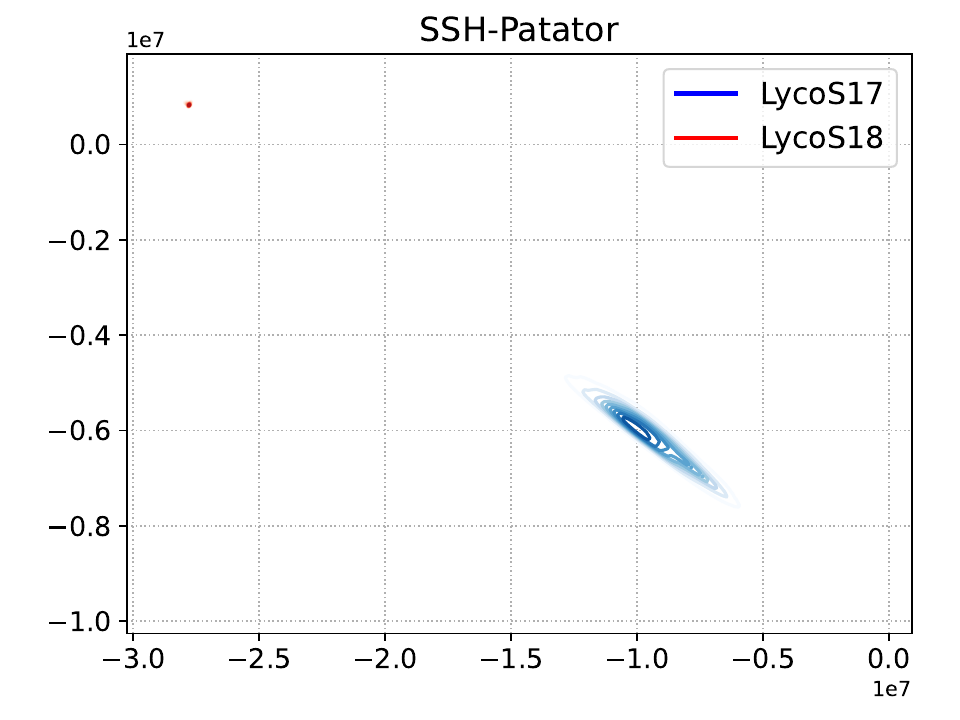}
    }\\
    \subfloat{
        \includegraphics[width=.33\linewidth]{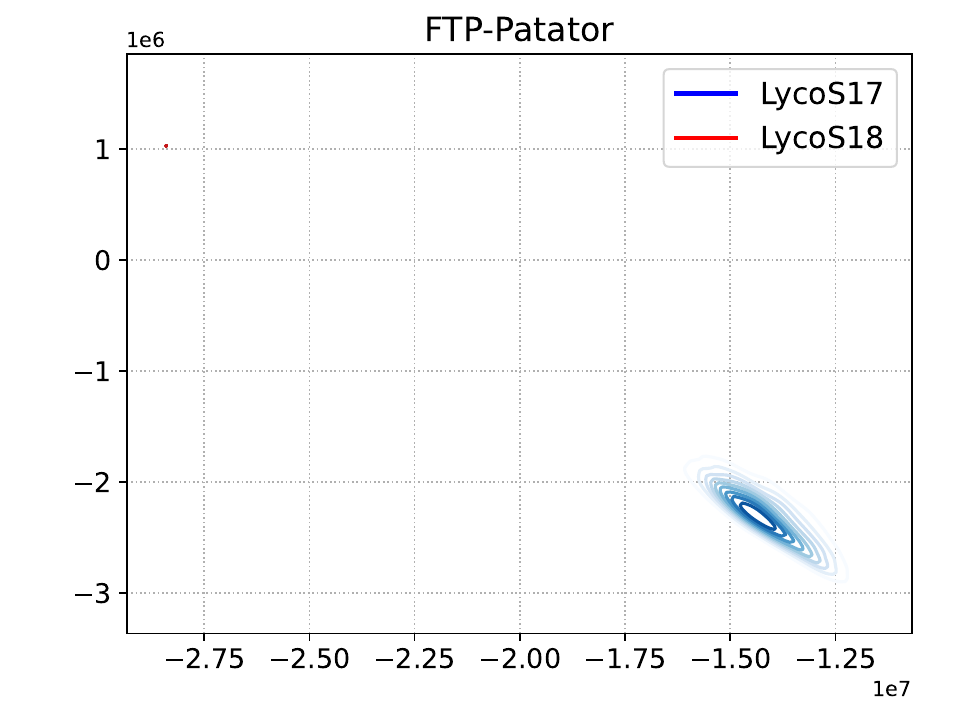}
    }
    \subfloat{
        \includegraphics[width=.33\linewidth]{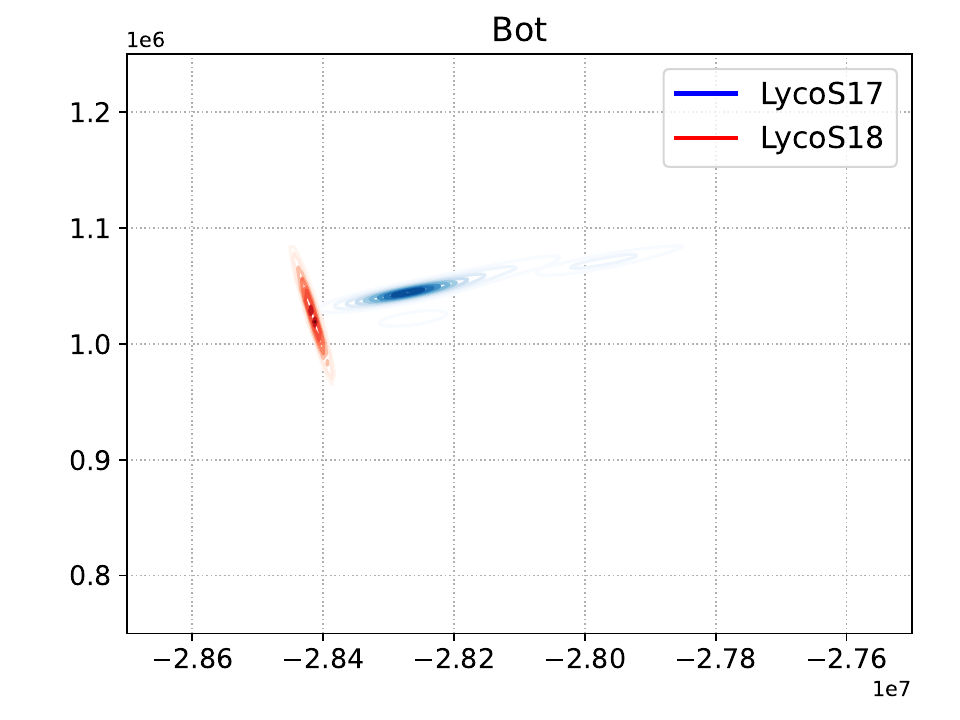}
    }
    \subfloat{
        \includegraphics[width=.33\linewidth]{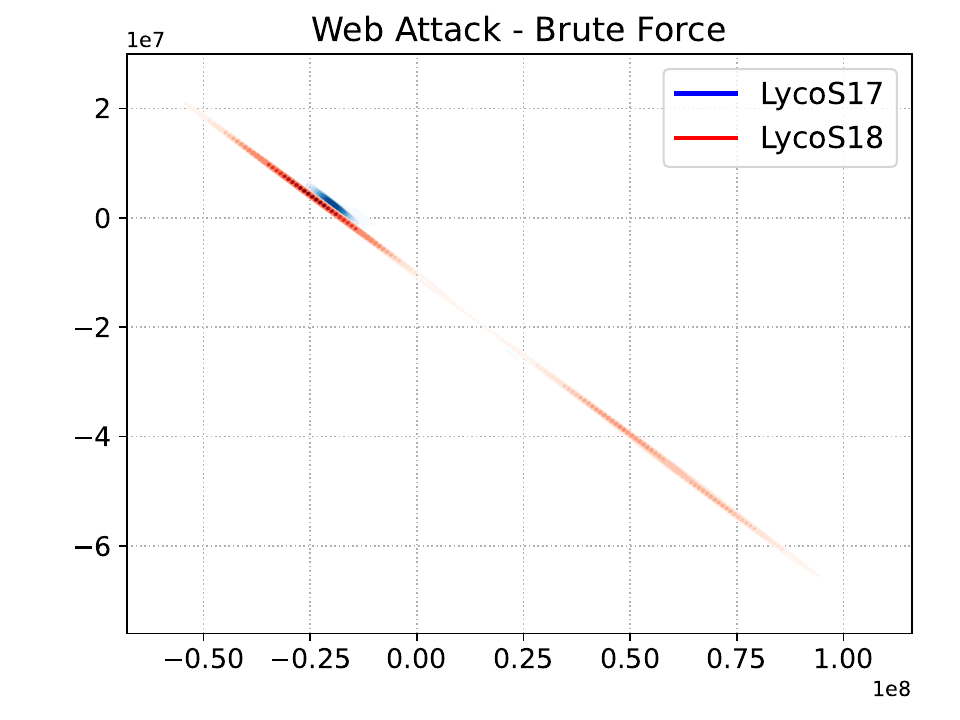}
    }\\
    \subfloat{
        \includegraphics[width=.33\linewidth]{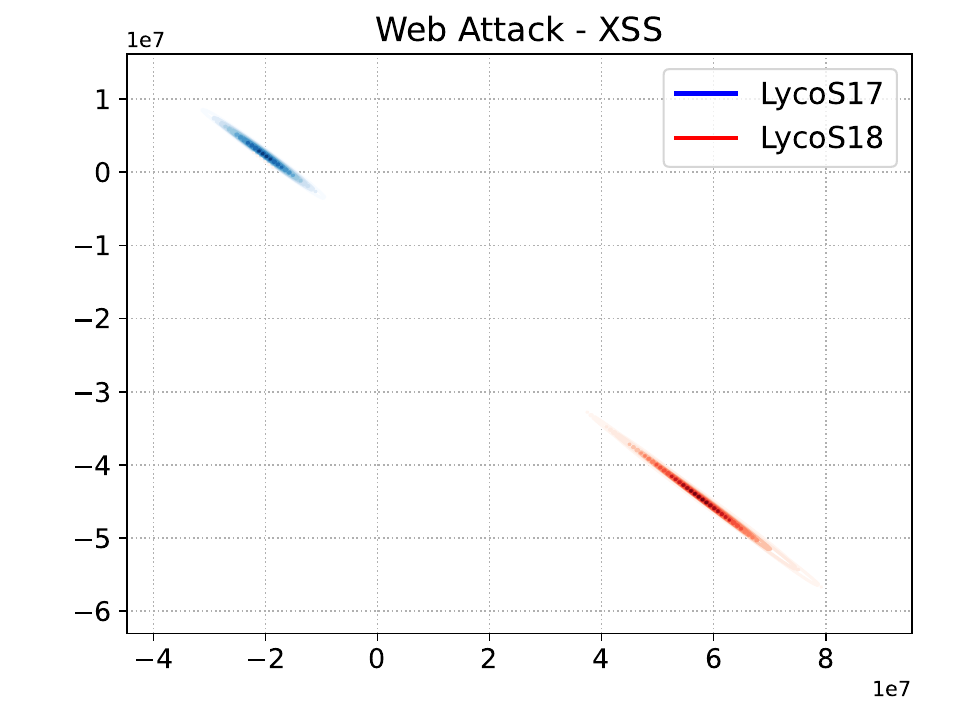}
    }
    \subfloat{
        \includegraphics[width=.33\linewidth]{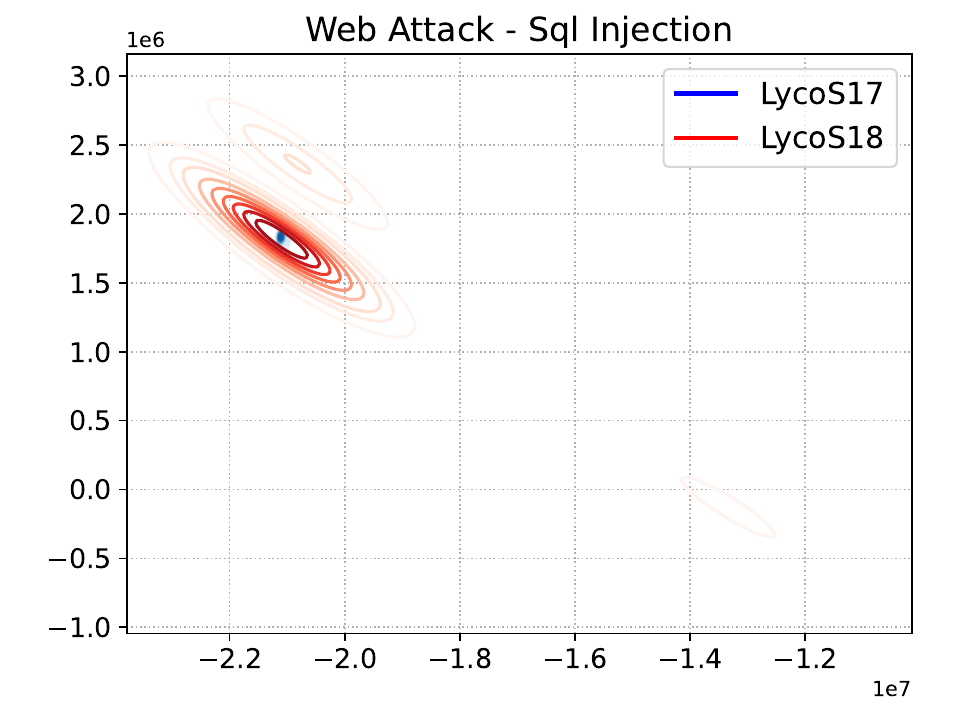}
    }
    \subfloat{
        \includegraphics[width=.33\linewidth]{blank.png}
    }
    \caption{Attacks distribution of the LycoS17 and LycoS18 datasets in the feature subspace obtained through PCA fitted with attack and benign samples from LycoS17. The legend, in parentheses, shows the type of representation employed to visualize the distribution.}
    \label{fig:kde_pca}
\end{figure*}

To further confirm that the distribution of the classes between LycoS17 and LycoS18 datasets is substantially different, in Figure~\ref{fig:kde_pca} we plot the KDE estimation of the attack distributions in a 2D subspace obtained with PCA fitted on LycoS17.
Although the distributions in this different subspace are obviously different from those shown in Figure~\ref{fig:kde_mrmr}, analogous conclusions can be inferred. Specifically, it can be seen that for some attacks, namely DoS Slowhttptest, SSH-Patator, and FTP-Patator, the distribution in the LycoS17 dataset consists of one or more clusters with a higher variance with respect to distribution in the LycoS18 dataset that is concentrated around a single point.
Differently from Figure~\ref{fig:kde_mrmr}, the distributions of DoS Hulk attack between the two datasets exhibit partial overlap. Hence, the mRMR method, seeing only the information of the training set, has selected an optimal 2D subspace that exacerbates the difference between the two distributions.
Also in this 2D subspace, all the attack classes are differently distributed among the two datasets except in the case of DoS Slowloris.

\section{Conclusions}
\label{sec:Conclusions}
This work focuses on analyzing the generalization capabilities of ML-based NIDS using different datasets for training and evaluating the models. We found that training a ML model on a single publicly available dataset is not sufficient to create a robust NIDS capable of working in different datasets and real-world applications.
The main motivations behind this conclusion are related to dataset composition. The classes in the four datasets employed do not adequately represent the real distribution of attacks, limiting them to representing the particular instance of the attack that was executed when the dataset was created and thus dependent on specific tools and parameters used.
Additionally,  we found that in some cases only a few features are needed to correctly identify an attack. This indicates a dataset-specific overcorrelation with the label that prevents the model from learning the true attack pattern.
Other possible motivations are related to the difficulties in correctly labeling an attack and other factors that influence the distribution of samples in the feature space, such as the topology of the networks and the hardware used for communication.

In the literature, many works utilize federated learning to address the generalization problem of ML-based NIDS, demonstrating promising results. However, unlike our methodology, these models were exposed to data from all the diverse sources employed in the testing phase, potentially biasing the evaluation of their generalization capability. Future research could explore the potential of federated learning using a sufficient number of diverse datasets, where each local model is trained on a single dataset, and the global model is tested on a new dataset. This approach may enhance the heterogeneity of the training set, yielding enhanced generalization potential.

Training models capable of identifying anomalous traffic in different network contexts presents considerable difficulties. This holds in particular when considering the necessity to construct datasets that thoroughly and adequately represent the highly heterogeneous nature of network traffic. These challenges cast doubts on the practicality of employing ML for designing NIDS.



\bibliographystyle{elsarticle-num} 
\bibliography{cas-refs}





\end{document}